\documentstyle[12pt,epsf,psfig]{article}
\newcommand{\bm}[1]{\mbox{\boldmath #1}}

\newcommand{\SR}{S_{\hbox{{\scriptsize Regge}}}}
\newcommand{\SLCS}{S_{\hbox{{\scriptsize LCS}}}}

\newcommand{\ZPR}{Z_{\hbox{{\scriptsize PR}}}}
\newcommand{\plaq}{\displaystyle \prod _{\partial \tilde{P}(l)} \hspace*{-6.4mm} \bigcirc \ U}
\newcommand{\sixj}[6]{
\renewcommand{\arraystretch}{1}
  \left\{
   \begin{array}{ccc}
    #1 & #2 & #3 \\
    #4 & #5 & #6
   \end{array}
\right\}}
\newcommand{\threej}[6]{
\renewcommand{\arraystretch}{1}
 \left(
  \begin{array}{ccc}
    #1 & #2 & #3 \\
    #4 & #5 & #6
  \end{array}
 \right)}
\renewcommand{\theequation}{\arabic{section}.\arabic{equation}}

\newcommand{\Str}{\hbox{Str}}

\newcommand{\1}{\bf\mbox{1}}

\newcommand{\C}{\widetilde{C}}
\newcommand{\del}{\partial}
\newcommand{\etab}{\overline{\eta}}

\newcommand{\st}{\widetilde{s}}
\newcommand{\scr}{\scriptsize}
\begin{document}
\renewcommand{\theequation}{\arabic {section}.\arabic{equation}}
\setlength{\baselineskip}{7mm}
\begin{titlepage}
\begin{flushright}
EPHOU-99-004 \\
March, 1999
\end{flushright}
 

\vspace{15mm}
\begin{center} 
{\Large Non-String Pursuit towards Unified Model on the Lattice}\\
\vspace{1cm}
{\bf {\sc Noboru Kawamoto}}\\
{\it{ Department of Physics, Hokkaido University }}\\
{\it{ Sapporo, 060-0810, Japan}}\\
\end{center}
\vspace{2cm}

\begin{abstract}
Non-standard overview on the possible formulation towards 
a unified model on the lattice is presented. 
It is based on the generalized gauge theory which is formulated 
by differential forms and thus expected to fit in a simplicial 
manifold.
We first review suggestive known results towards this direction. 
As a small step of concrete realization of the program, 
we propose a lattice Chern-Simons gravity theory which leads 
to the Chern-Simons gravity in the continuum limit via 
Ponzano-Regge model. 
We then summarize the quantization procedure of the generalized 
gauge theory and apply the formulation to the generalized topological
Yang-Mills action with instanton gauge fixing. 
We find $N=2$ super Yang-Mills theory with Dirac-K$\ddot{\hbox{a}}$hler 
fermions which are generated from ghosts via twisting mechanism. 
The Weinberg-Salam model is formulated by the generalized Yang-Mills 
action which includes Connes's non-commutative geometry formulation 
as a particular case. 
In the end a possible scenario to realize the program is 
proposed.
The formulations given here are by far incomplete towards the 
final goal yet include hopeful evidences. 
This summary of the overview is the extended version of the talk 
given at Nishinomiya-Yukawa memorial symposium 
(Nishinomiya-city Japan, Nov. 1998).
  
\end{abstract}

\vspace{0.5cm}
\noindent
To be published in the Proceedings of 13th Nishinomiya-Yukawa Memorial 
              Symposium, Nishinomiya city, Nov.  1998.\\
kawamoto@particle.sci.hokudai.ac.jp\\ 

\end{titlepage}

\setcounter{equation}{0}
\section{Introduction}
It is obviously the most challenging problem how to formulate the quantum 
gravity and the standard model in a unifying and constructive way. 
Towards a possible solution to this problem, the current trend is heading 
to the string related topics\cite{string1}\cite{string2}.
It is, however, not obvious that the string is the only possible formulation 
leading to the unified theory including quantum gravity. 
In fact, the two dimensional quantum gravity was formulated by a lattice 
gravity, the dynamical triangulation of random surface\cite{DT} which 
was analytically confirmed by Liouville theory\cite{KPZ}.
On the other hand the three dimensional Einstein gravity was successfully 
formulated by the Chern-Simons action even at the 
quantum level \cite{Witten1}.
There are thus other formulations of quantum gravity than the string 
related formulation.  

One of the important motivations that the super string could be the genuine 
formulation of unifying all the forces of gauge theories is 
that the super string may be able to control the divergences even with 
gravity, and thus the renormalizability and unitarity are natural 
consequences of the formulation. 
An alternative formulation to control the divergences would be the lattice 
formulation.

Suppose we aim to formulate a unified model, what could be 
the possible criteria to believe that it could be a realistic model.
Eventually the unified model should explain the origin of the 
following phenomenological parameters and characteristics: 
\begin{enumerate}
\item The group structure of the standard model: 
      $SU(3)\times SU(2)\times U(1)$.
\item The number of generation = 3. 
\item The pairing structure of quarks and leptons in the standard model: 
      Quarks interact strongly, weakly and electromagnetically while 
      leptons interact differently.
\item Our space-time is four dimension and Minkowskian. 
\item The quark and lepton masses, the mixing parameters; 
      Cabbibo-Kobayashi-Maskawa angles and possible lepton mixing angles.
\end{enumerate}

The fundamental unified theory should eventually propose a mean to evaluate 
the item 5 quantitatively. 
From the experiences of the lattice QCD it would be difficult to 
calculate the phenomenological quantities analytically from the fundamental 
theory. 
Instead we need to evaluate them numerically thus we need to formulate a 
constructive definition of a regularized unified theory including gravity. 

Concerning to the issues of quantum gravity it is more difficult to judge 
what could be the experimental evidences to confirm the quantum nature of 
the gravity. 
It could possibly be reflected to the large scale structure of the universe 
yet unconvincing. 
As we show later the quantum gravity in two dimensions can be 
well understood by the numerical simulations on the lattice which are 
confirmed analytically as well.
We expect that numerical method by lattice would be the only mean to evaluate 
the quantum nature of gravity even in higher dimensions. 

We thus believe that lattice theory is again a good candidate to fulfill 
our requirements for the quantitative unified theory.
In the above phenomenologically known results the first four items could be 
understood easier than the last one and could be related with the super 
symmetry. 

In this summary of overview we propose the generalized gauge theory 
which was proposed previously by the present author and Y.Watabiki
\cite{KW2} as a formulation towards a unified model on the lattice. 
In order to persuade the readers to accept the ideas and formulation 
I will collect the suggestive known results and include several recent 
investigations towards this direction and thus the summary 
is aimed to be self-contained. 
Our formulation is a non-standard approach towards the unified model 
and the formulation is not yet completed but there are several hopeful 
evidences that this approach may play an important role for 
the unified theory of the fundamental interactions. 

Here we list the contents of this summary towards a non-standard 
approach of a unified model on the lattice.
\begin{flushleft}
1. Introduction \\
2. Suggestive Known Results towards Unified Model on the Lattice \\
~~ 2.1 Fermionic Matter on the Lattice \\
~~ 2.2 Success of Two Dimensional Quantum Gravity on the Lattice \\
~~~~  2.2.1 Microscopic Description of Two Dimensional Random Surface \\
~~~~  2.2.2 Numerical Results on the Fractal Structure of Two Dimensional 
        Quantum \\  \hspace{1.6 cm} Gravity on the Lattice \\
~~ 2.3 Susskind Fermion---Staggered Fermion---Dirac K$\ddot{\hbox{a}}$hler
      Fermion on the Lattice \\
~~ 2.4  Chern-Simons Gravity and Ponzano-Regge Model \\
~~~~  2.4.1 Chern-Simons Gravity \\
~~~~  2.4.2 Ponzano-Regge Model \\
~~ 2.5 Four Dimensional Gravity on the Lattice \\
3. A Possible Formulation towards Gauge Gravity coupled Matter on the \\
       \hspace{0.5 cm} Simplicial Lattice Manifold  \\
~~  3.1 Possible Formulations towards Unified Model on the Lattice \\
~~  3.2 Generalized Gauge Theory \\
4. Gravity on the Lattice \\
~~ 4.1 First Step towards the Generalized Chern-Simons Actions 
     on the Lattice  \\
~~ 4.2 Lattice Chern-Simons Gravity \\
~~~~  4.2.1 Gauge Invariance on the Lattice \\
~~~~  4.2.2 Calculation of Partition Function  \\
~~~~~~  4.2.2.1 $e$ integration \\
~~~~~~  4.2.2.2 $U$ integration \\
~~~~  4.2.3 The Continuum Limit of the Lattice Chern-Simons Gravity \\
5. Quantization of Generalized Gauge Theory \\
6. Generalized Yang-Mills Theory \\
~~ 6.1 Generalized Topological Yang-Mills Theory \\
~~~~  6.1.1 Instanton Gauge Fixing of Topological Yang-Mills Model \\
~~~~  6.1.2 Twisted $N=2$ Super Yang-Mills Action with 
Dirac-K$\ddot{\hbox{a}}$hler Fermions \\
~~ 6.2 Weinberg-Salam Model from Generalized Gauge Theory \\
~~~~  6.2.1 Generalized Gauge Theory with Dirac-K$\ddot{\hbox{a}}$hler 
Fermions \\
~~~~  6.2.2 Weinberg-Salam Model from Generalized Gauge Theory as \\
            \hspace{1.6 cm} Non-Commutative Geometry Formulation \\
7. Possible Scenario and Conjectures for the Unified Model on the Lattice \\
\end{flushleft}

\setcounter{equation}{0}                
\section{Suggestive Known Results towards Unified Model on the Lattice}

\subsection{Fermionic Matter on the Lattice}

In considering the formulation towards the unified model on the lattice, 
I will first give a very suggestive and simple example to figure out 
how to find the possible model on the lattice. 
The first example is the two dimensional Ising model.
It is well known that the Ising model has a second order phase transition 
point and the model leads to a free fermion theory in a flat space 
at the phase transition point in the continuum limit. 
\begin{figure}
\begin{center}
 \begin{minipage}[b]{0.3\textwidth}
 \epsfxsize=\textwidth \epsfbox{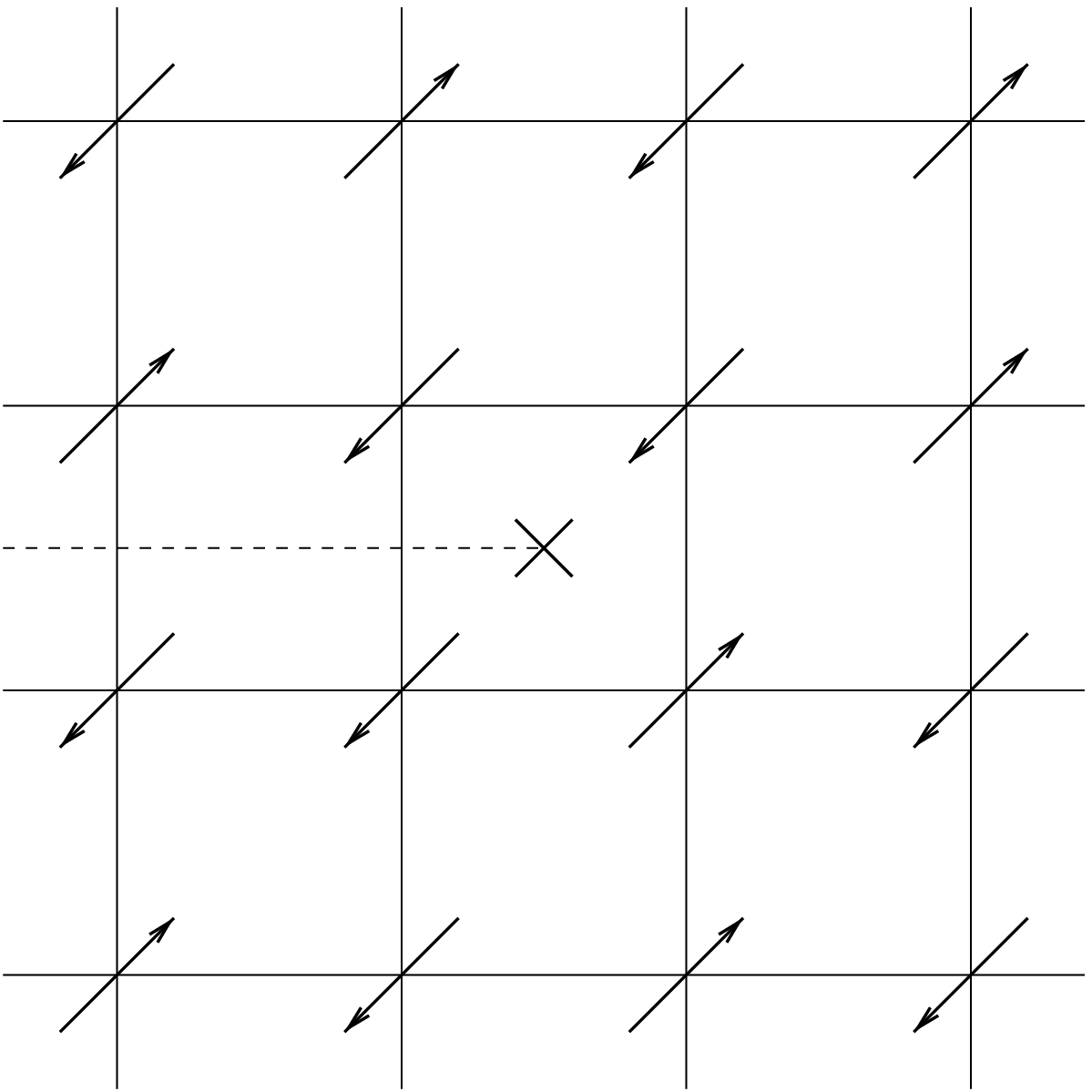} 
 \end{minipage}
\hspace{1 cm}
\begin{minipage}[b]{0.3\textwidth}
 \epsfxsize=\textwidth \epsfbox{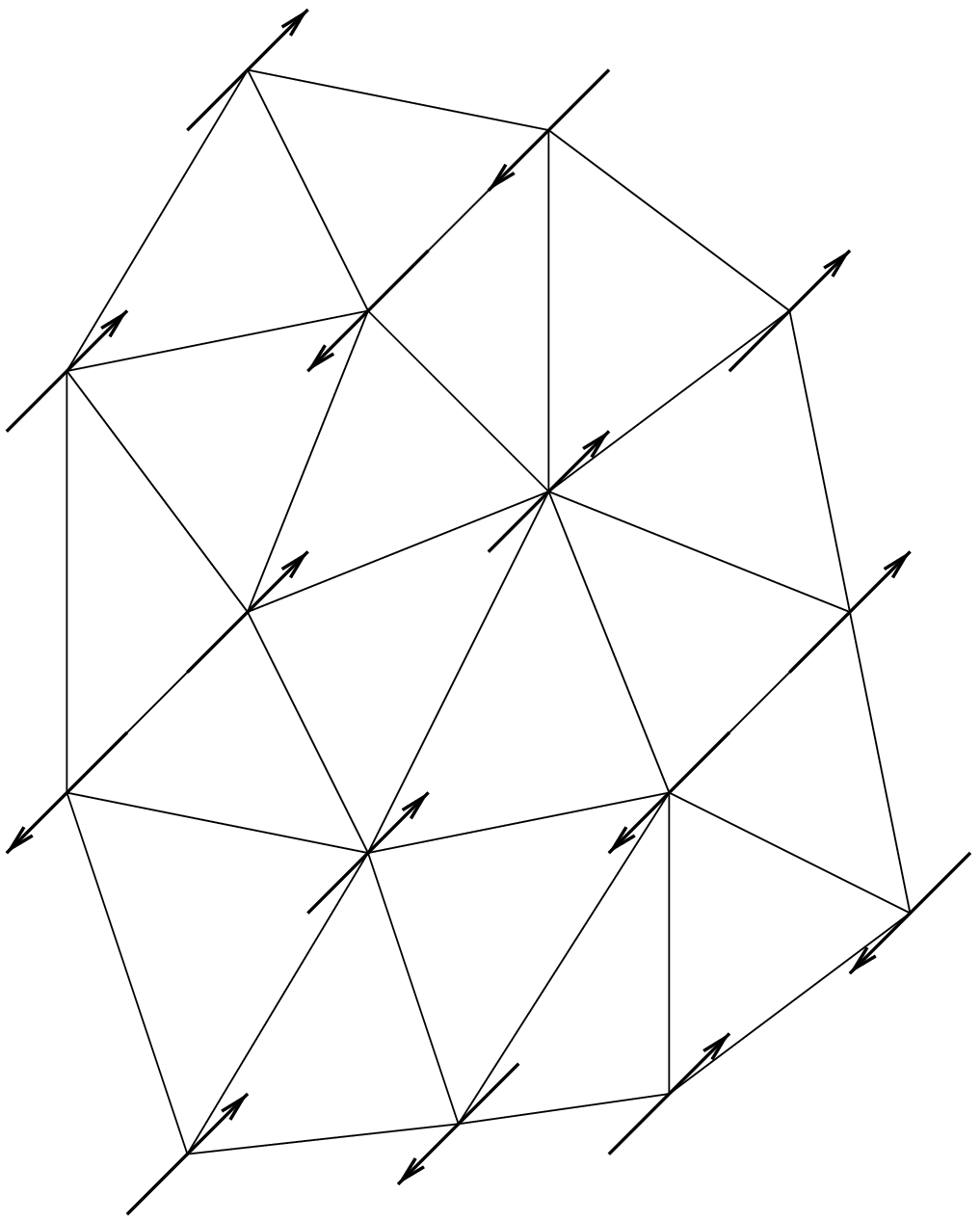} 
 \end{minipage}
\end{center}
 \caption{Ising spin on the square lattice and dynamically triangulated lattice}
 \label{fig:ising}
\end{figure}
In this system the lattice is the two dimensional square lattice and the 
matter field is sitting at the sites of the lattice and takes value 
$\pm 1$. 
What is surprising here is that the simple square lattice with the simplest 
matter at the site reproduces the fermionic matter in the continuum limit.
It is by now established that the Ising spin on the dynamically triangulated 
lattice reproduces the free fermion coupled to a gravitational 
background in the continuum limit. 
See fig.\ref{fig:ising}. 
This is a very symbolic example that matter fermion is generated 
via degrees of freedom of field sitting on sites and the curved 
space-time background is generated by the dynamically triangulated lattice 
in the continuum limit. 
In other words field theoretical matter and background gravitational 
field are essentially reproduced by the lattice formulation. 

It is important to recognize that there is a microscopic formulation 
to see how the fermionic degrees appear 
at the lattice level\cite{Ising1}\cite{Dotsenko}. 
Let me sketch the formulation.
The partition function of Ising model is
\begin{equation}
Z=\sum_{\sigma=\pm} \exp\{ \beta\sum_{<ij>}\sigma_i\sigma_j\},
\end{equation}
where $<ij>$ is a nearest neighbor pair of sites on a two dimensional square 
lattice.
We introduce the so called disorder parameter\cite{Kadanoff} 
\begin{equation}
 \mu_x = \prod^x_{-\infty} \exp {\{-2\beta \sigma\sigma'\}}, 
\end{equation} 
where the exponents in the product correspond to the links which are crossed 
by the dashed line starting from the dual site $x$ and ending at $-\infty$. 
We introduce the product of the disorder variable $\mu_x$ and order 
variables $\sigma_x$ 
\begin{eqnarray}
\begin{minipage}[c]{4.0cm}
 \epsfxsize=\textwidth \epsfbox{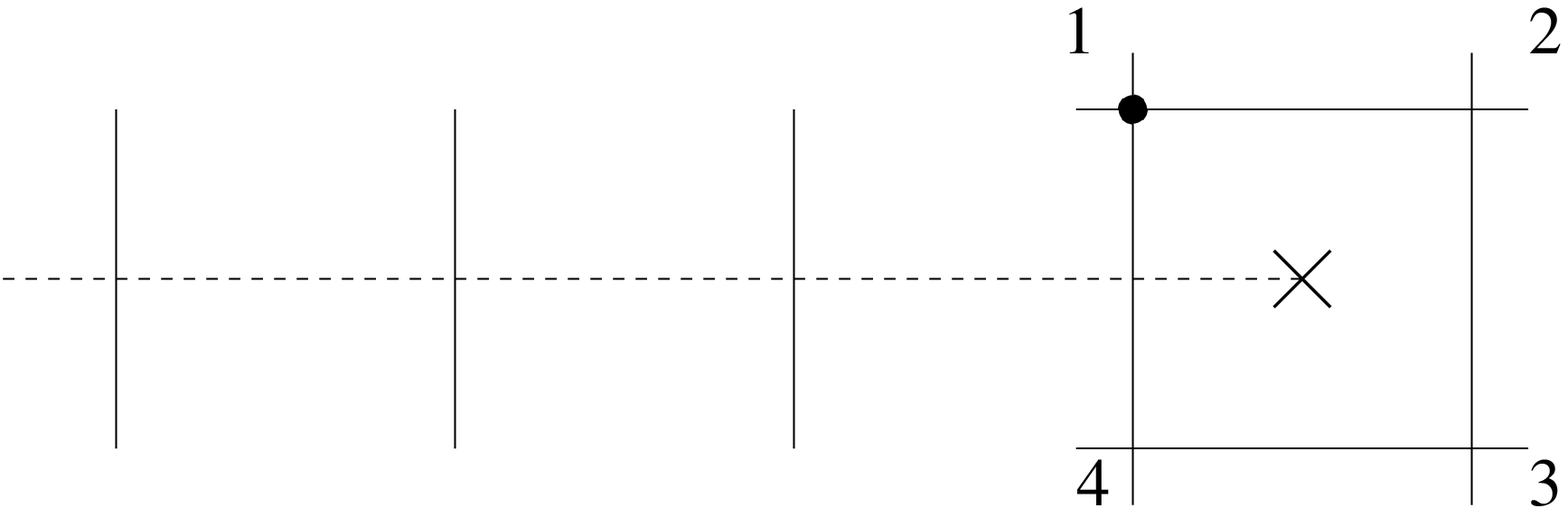} 
\end{minipage}
 ~~~~~~=~~&\mu_x\sigma_{x_1}~~\equiv~~ \chi^1(x) \\
\begin{minipage}[c]{4.0cm}
 \epsfxsize=\textwidth \epsfbox{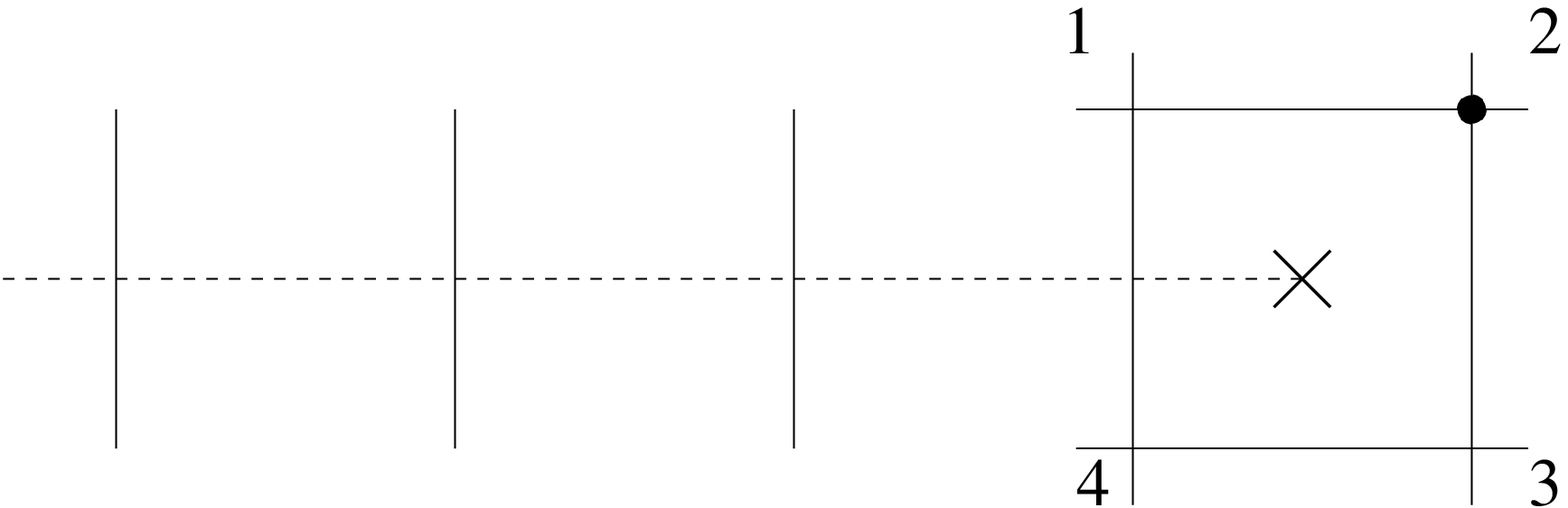} 
\end{minipage}
 ~~~~~~ =~~&\mu_x\sigma_{x_2}~~\equiv~~ \chi^2(x) \\
\begin{minipage}[c]{4.0cm}
 \epsfxsize=\textwidth \epsfbox{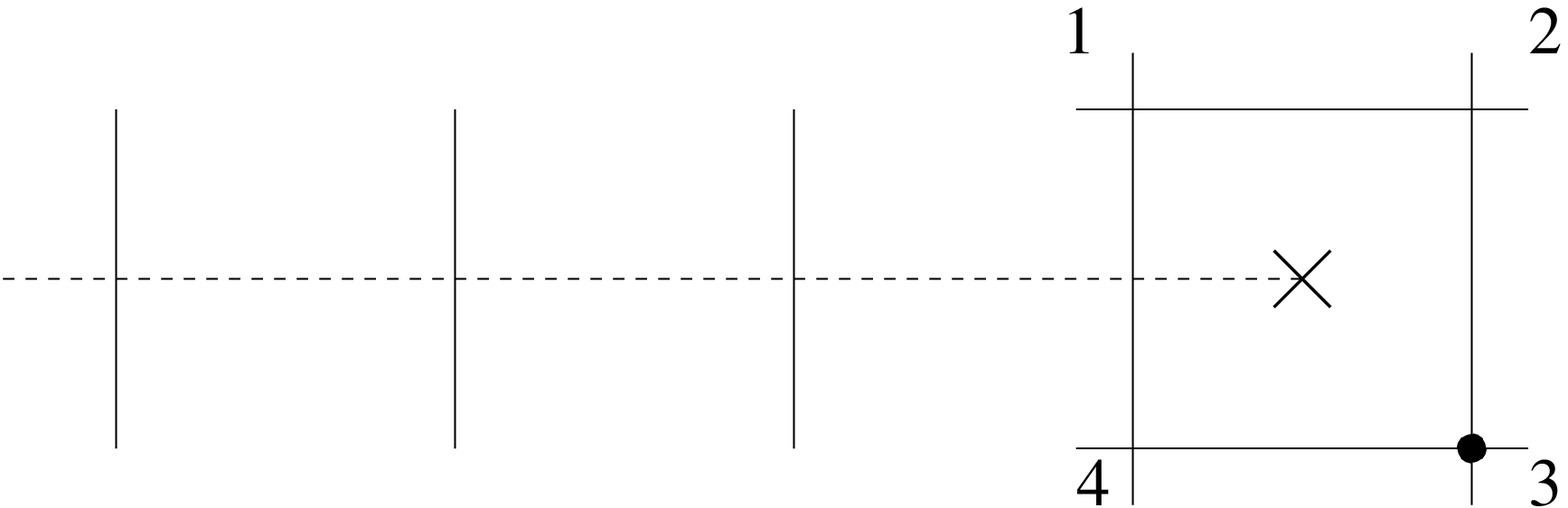} 
\end{minipage}
 ~~~~~~=~~&\mu_x\sigma_{x_3}~~\equiv~~ \chi^3(x) \\
\begin{minipage}[c]{4.0cm}
 \epsfxsize=\textwidth \epsfbox{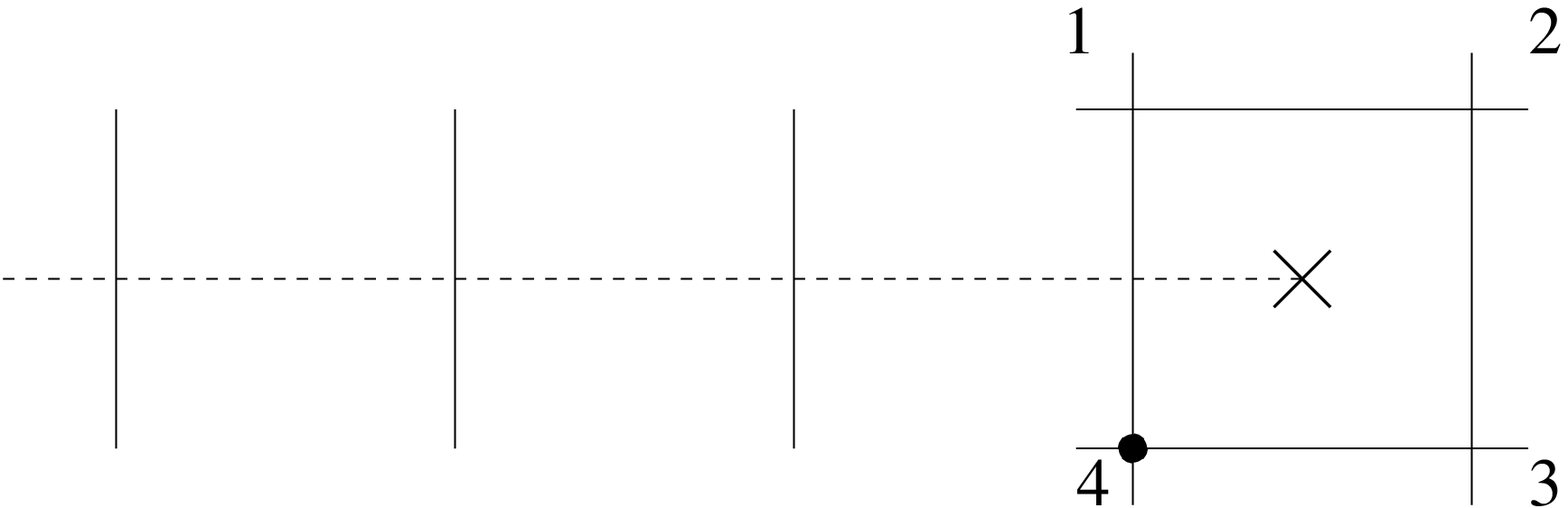} 
\end{minipage}
 ~~~~~~=~~&\mu_x\sigma_{x_4}~~\equiv~~ \chi^4(x), 
\end{eqnarray}
where $x_1$, $x_2$, $x_3$ and $x_4$ are original sites and surround the dual 
site $x$. 
Here we assume that the $\sigma\mu$ variables are always inside 
some correlation function 
$\chi^\alpha(x)=\mu_x\sigma_{x_\alpha}\sim<\mu_x\sigma_{x_\alpha}\cdots>$. 
Then the effect of the disorder variable is to flip the sign of 
$\beta\sigma\sigma'$ in the action along the dashed line.      

There is the following identity:
\begin{equation}
\sigma\exp \{-2\beta\sigma\sigma'\}=ch(2\beta)\sigma  -sh(2\beta)\sigma', 
\end{equation}
which combines with the disorder variable $\mu_x$ and leads to a graphical 
relation 
\begin{equation}
\begin{minipage}[c]{2.0cm}
 \epsfxsize=\textwidth \epsfbox{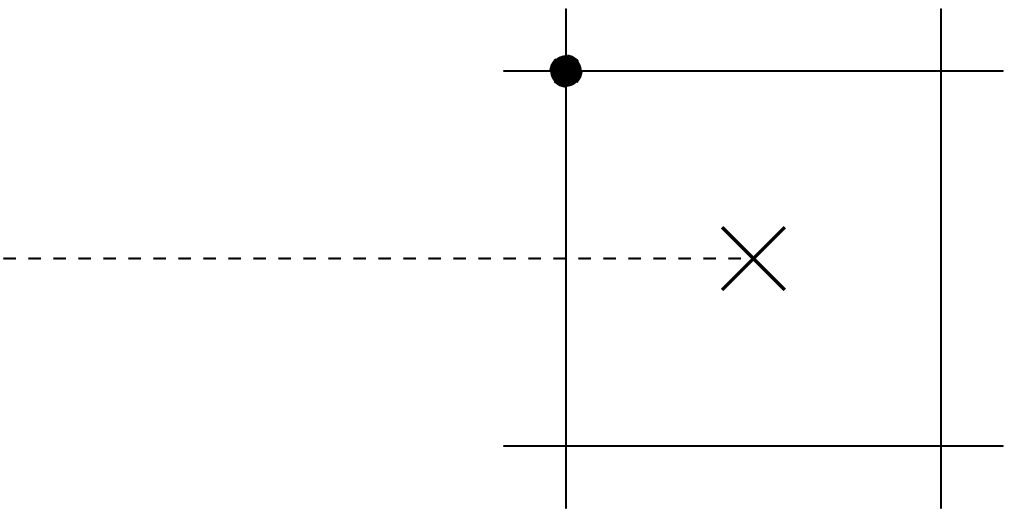} 
\end{minipage} 
    ~~=~~ch(2\beta)~
\begin{minipage}[c]{2.0cm}
 \epsfxsize=\textwidth \epsfbox{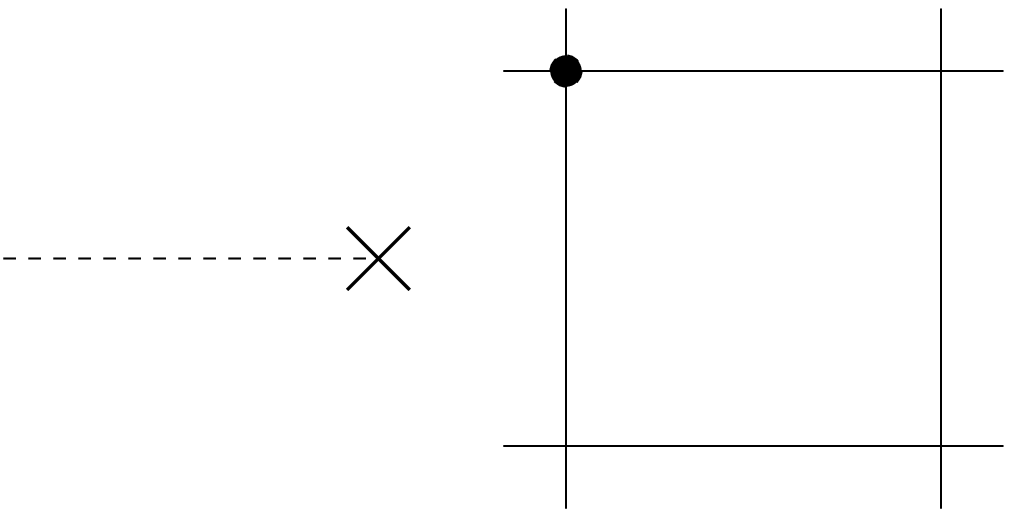} 
\end{minipage}
  ~-~sh(2\beta)
\begin{minipage}[c]{2.0cm}
 \epsfxsize=\textwidth \epsfbox{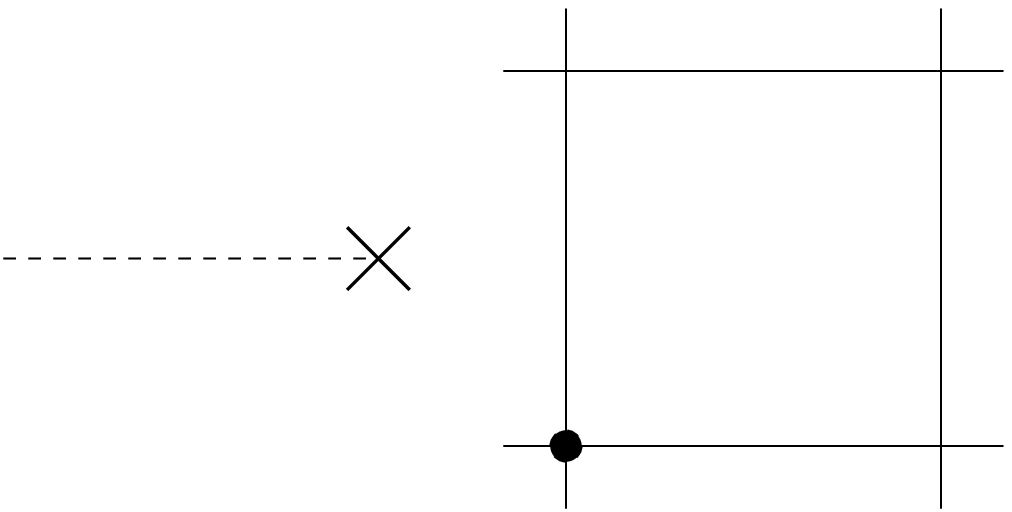} 
\end{minipage}
, 
\end{equation}
equivalently 
\begin{equation}
\chi^1(x)=ch(2\beta)\chi^2(x-\hat{1})  -sh(2\beta)\chi^3(x-\hat{1}),
\end{equation}
where $\hat{1}$ is a unit vector of 1 direction on the lattice. 
We can obtain similar relations for $\chi^2(x)$, $\chi^3(x)$ and 
$\chi^4(x)$, where we can use the invariance under a simultaneous change 
of $\sigma_n \rightarrow -\sigma_n$ and 
$\beta\sigma\sigma_n \rightarrow -\beta\sigma\sigma_n$ for an arbitrary 
given site $n$.

Redefining $\chi^\alpha(x)$, 
\begin{eqnarray}
\psi^1(x)&=\frac{1}{2}(\chi^2(x)+\chi^3(x)), ~~~ 
\psi^2(x)&=\frac{1}{2}(\chi^1(x)+\chi^2(x)), \\
\psi^3(x)&=\frac{1}{2}(-\chi^4(x)+\chi^1(x)), ~~~ 
\psi^4(x)&=\frac{1}{2}(-\chi^3(x)-\chi^4(x)),
\end{eqnarray}
we obtain
\begin{equation}
\left[\begin{array}{c}
             \psi^1(x)\\
             \psi^2(x)\\
             \psi^3(x)\\
             \psi^4(x)
      \end{array} \right]
             =e^{-2\beta}
\left[\begin{array}{cccc}
              1&1&0&-1\\
              1&1&1&0\\
              0&1&1&1\\
              -1&0&1&1
      \end{array}  \right]
\left[\begin{array}{c}\psi^1(x-\hat{1})\\
             \psi^2(x-\hat{2})\\
             \psi^3(x-\hat{3})\\
             \psi^4(x-\hat{4})
      \end{array} \right],             
\end{equation}
where the unit vectors are related $-\hat{3}=\hat{1}$ 
and $-\hat{4}=\hat{2}$.
This can be understood as a hopping relation of the
$\psi^\alpha$ fields on the lattice. 
Under the proper renormalization of the $\psi^\alpha$ fields, 
the hopping matrix is identified with the inverse propagator of 
fermion fields\cite{Ising1}. 

It is important to recognize that the microscopic description 
of the fields on the lattice is reflected to the field theoretic 
description of fermionic field in the continuum limit.

\begin{figure}
\begin{center}
 \begin{minipage}[b]{0.3\textwidth}
 \epsfxsize=\textwidth \epsfbox{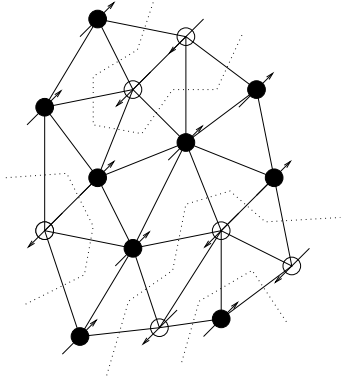} 
 \end{minipage}
 \end{center}
 \caption{Ising spin on a dynamically triangulated lattice}
 \label{fig:isinggr1}
\end{figure}

By now we know that the Ising spin located at sites of a dynamically 
triangulated lattice leads to a fermionic matter coupled to gravity 
in the continuum limit\cite{Migdal}\cite{Kazakov}.
In this case there is also a microscopic description of how fermionic 
degrees appear on the random lattice.
Let me briefly sketch the derivation. 
Here we don't follow to the matrix model formulation\cite{Kazakov} 
but give more direct derivation.  
We first note the standard observation concerning Ising model: 
the general configuration of spins can be considered as a set of 
alternating black drops ($\sigma=1$) and white bubbles ($\sigma=-1$).
The boundaries between drops and bubbles are drawn on the dual graph.
See fig. \ref{fig:isinggr1} .
The dual graph includes only three vertex on a dual site and thus 
coincides with Feynmann graphs of $\phi^3$ theory.
If we look at particular dual site, a boundary line may pass through 
the site or may not touch at all. 

We can show that the above mentioned property is reproduced by the 
following fermionic action on the lattice\cite{Migdal} 
\begin{equation}
Z=\int \prod_i d^2 \psi(i) exp(-S),
\end{equation}
with the action 
\begin{equation}
S=\frac{1}{2}\sum_i \bar{\psi}(i)\psi(i)-
\frac{1}{2}\sum_{<ij>}\bar{\psi}(i)K(i,j)\psi(j).
\end{equation}
We adopt the notation $\bar{\psi}_\beta=\psi_\alpha\epsilon_{\alpha\beta}$
and the normalization of Grassmann integral
\begin{equation}
\int d^2\psi \psi_\alpha\bar{\psi}_\beta = \delta_{\alpha\beta}.
\end{equation} 
There are two important properties to be satisfied by the action. 
(1) The boundary line does not possess particular direction thus the 
fermion must have Majorana nature. 
(2) The back-tracking motion is absent.
These properties are satisfied by the following constraints:
\begin{eqnarray}  
(1)&~~~~\bar{\psi}(i)K(i,j)\psi(j)&= \bar{\psi}(j)K(j,i)\psi(i) \\
(2)&~~~~K(i,j) K(j,i) &= 0.
\end{eqnarray}
The Majorana condition (1) can be written in a matrix form:
\begin{equation}  
\sigma_2 K(i,j)= -(\sigma_2 K(j,i))^T,
\end{equation}
where $(\sigma_2)_{\alpha\beta}=i\epsilon_{\alpha\beta}$ is the Pauli 
matrix and $T$ stands for transpose. 

The above two constraints will be satisfied, provided $K(i,j)$ has the form 
\begin{equation}
K(i,j)=A(ij)({\bf 1} + \sigma_1 n_1(ij) + \sigma_3 n_3(ij)),
\end{equation} 
where
\begin{equation}
A(ij)=A(ji),~~~n_1(ij)=-n_1(ji),~~~n_3(ij)=-n_3(ji),~~~
(n_1(ij))^2+(n_3(ij))^2=1.
\end{equation}

We can then guess a possible strategy how to get a field theory 
coupled gravity on the lattice.
The gravitational background is generated by the dynamical triangulation 
and the matter fields are generated by the fields on the simplex of 
the simplicial manifold.

\subsection{Success of Two Dimensional Quantum Gravity on the Lattice}

As we have seen in the previous section that Ising model on the 
two dimensional square lattice is equivalent to the free fermion 
theory of the flat space time in the continuum limit. 
On the other hand it is analytically known by now that 
Ising model on the randomly triangulated lattice is equivalent 
to the free fermion theory coupled to the two dimensional 
gravity\cite{Migdal}\cite{Kazakov}. 
These examples suggest that curved space time is generated by the 
random lattice and the matter fields are induced by some degrees of 
freedom on a simplex: for Ising model $\pm 1$ values on the sites 
(0-simplex). 

In two dimensions the relation between the lattice theory and the 
corresponding continuum field theory were analytically established 
with the help of conformal field theory. 
The central charge $c$ specifying a matter content of the continuum theory 
is a measure to differentiate different types of matrix models. 
The equivalence of a lattice model and the corresponding matrix model 
is established 
and the Liouville theory gave complementary understandings to the 
lattice theories\cite{KPZ}\cite{DT}\cite{DSlimit}.

In two dimensional quantum gravity conformal dimensions are predicted 
both by lattice models and the corresponding continuum theories and gave 
the same predictions. 
It is important to measure these physical quantities. 
Numerically it has been confirmed that two dimensional quantum space time 
coupled to $c=-2$ matter show clear fractal scaling and the fractal 
dimension is consistent with the theoretical value\cite{KKSW}\cite{KW1}.


\subsubsection{Microscopic Description of Two Dimensional 
Random Surface}

In general it is highly nontrivial how to relate a microscopic 
description of lattice theory and the corresponding continuum theory. 
In two dimensions it is possible to take a continuum limit of 
lattice theories for some particular cases. 
It has already been established that the matrix model is powerful 
tool to solve some lattice models in two dimensions and 
the procedure how to get the continuum limit is established\cite{DSlimit}. 
Here we give an example which has well defined microscopic 
description of $c=0$ lattice model and the continuum limit. 

We first derive a transfer matrix of two dimensional 
random surface on the lattice and take the continuum limit 
and derive the fractal structure of the $c=0$ random surface, 
a dynamically triangulated lattice without matter\cite{KKMW}.

\begin{figure}
\begin{center}
 \begin{minipage}[c]{0.3\textwidth}
 \epsfxsize=\textwidth \epsfbox{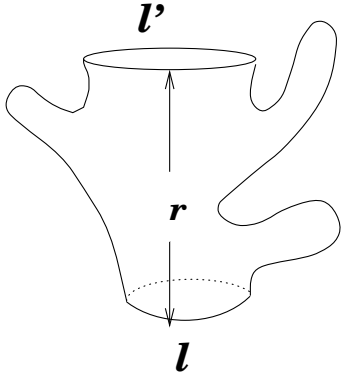} 
 \end{minipage}
\end{center}
\caption{ Composition law of cylinder amplitude } 
\label{fig:composition}
\end{figure}

\begin{figure}
\begin{center}
 \begin{minipage}[c]{0.3\textwidth}
 \epsfxsize=\textwidth \epsfbox{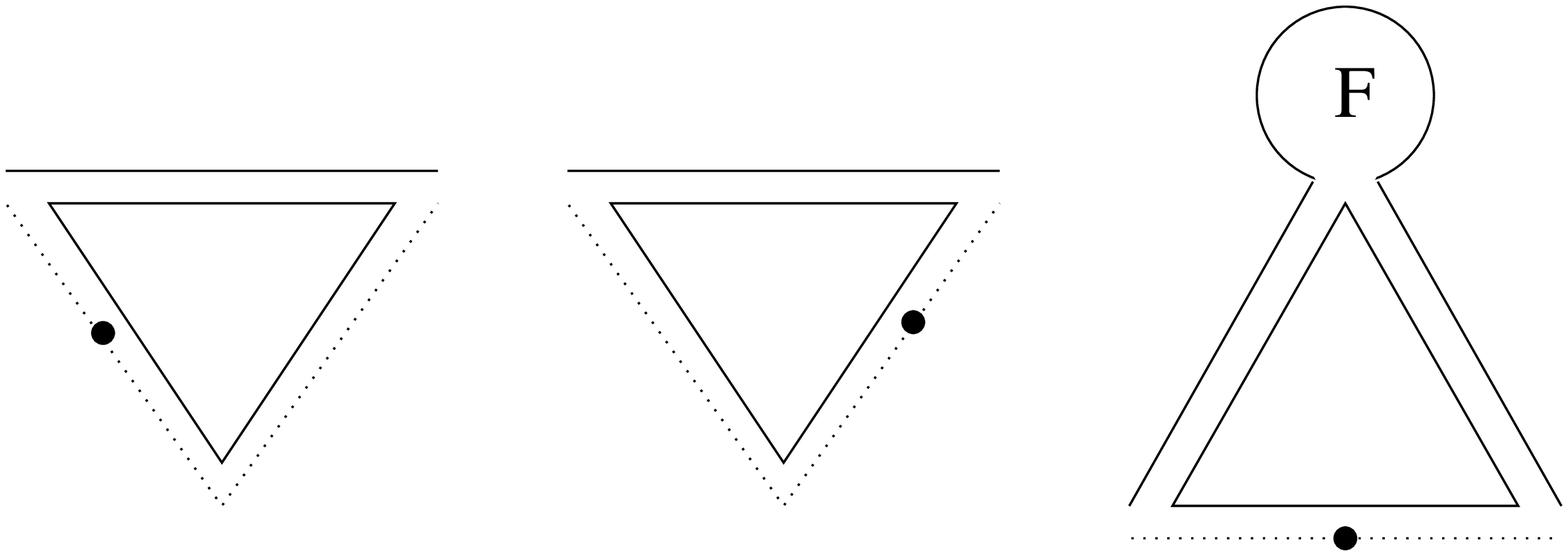} 
 \end{minipage}
\caption{ One step exit moves with a marked point with $F$ as disk amplitude }
\label{fig:exit moves}
 \begin{minipage}[c]{0.3\textwidth}
 \epsfxsize=\textwidth \epsfbox{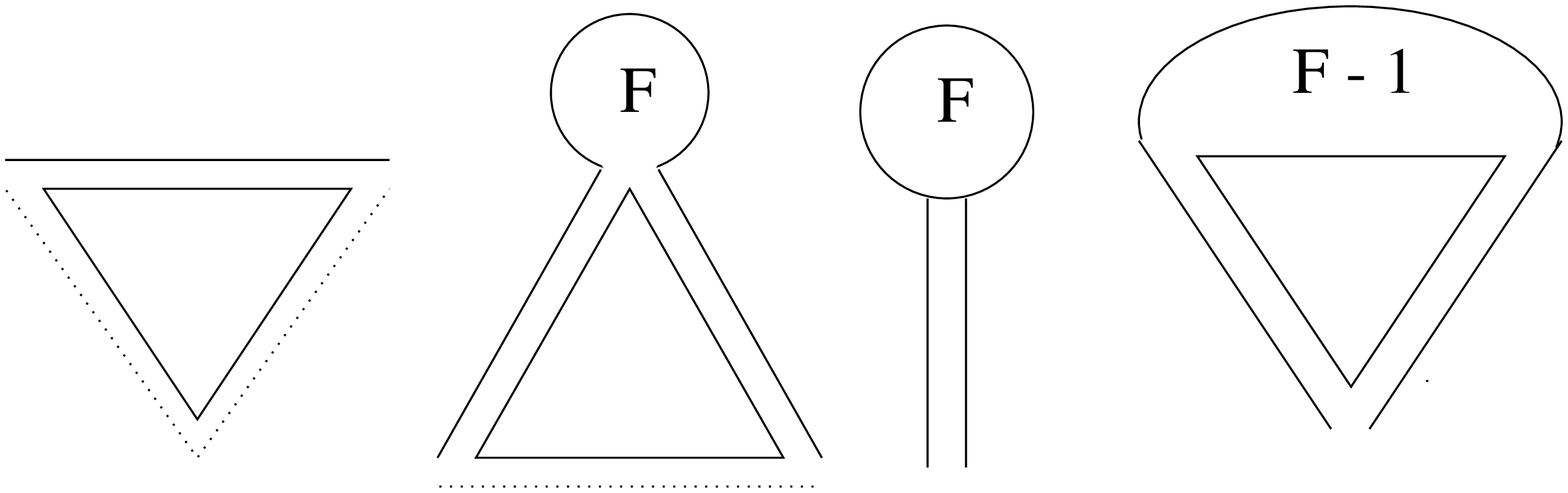} 
 \end{minipage}
\end{center}
 \caption{ One step entrance moves }
 \label{fig:entrance moves}
\end{figure}

We consider the cylinder amplitude $N(l,l';r;n)$ which counts a 
number of possible triangulations of random surface for a shape 
of cylinder with a entrance loop 
length $l$, a marked exit loop length $l'$, the geodesic distance 
of the two loops $r$ and the number of triangles $n$. 
See fig. \ref{fig:composition}. 
This quantity satisfies the following composition law:
\begin{equation}
N(l,l'';r_1+r_2;n_1+n_2) = \sum_{l'}N(l,l';r_1;n_1)N(l',l'';r_2;n_2).
\label{eqn:comp-law}
\end{equation}
We define the following quantity by summing up the triangles 
with a parameter $K$: 
\begin{equation}
N(l,l',r) \equiv \sum_{n=0}^\infty K^n N(l,l';r;n)\equiv 
       (\hat{N}(r))_{ll'},
\end{equation}
which satisfies the following relation 
\begin{equation}
       (\hat{N}(r))_{ll'}=((\hat{N}(1))^r)_{ll'}.
\end{equation}
We can thus claim that $(\hat{N}(1))_{ll'}$ is the transfer matrix of 
two dimensional surface. 
We now define the generating function of the transfer matrix by 
parametrizing $y$ and $y'$ for the entrance and exit loop length, 
respectively 
\begin{equation}
       \hat{N}(y,y';K)\equiv \sum_{l,l'=0}^\infty y^l(y')^{l'}N(l,l';1).
\end{equation}
It is surprising to recognize that the generating function of 
the transfer matrix can be obtained by a simple geometric series 
of the possible one steps 
\begin{eqnarray}
\hat{N}(y,y';K)&=& (2yy'^2K+y^2y'kF)\sum_{n=0}^\infty
                    [yy'2K+y2y'kF+y^2F+yK(F-1)]^n \nonumber \\
               &=& \frac{2yy'^2K+y^2y'kF}{1-yy'2K-y2y'kF-y^2F-yK(F-1)},     
\label{eqn:trans-gen}
\end{eqnarray}
where $F$ is a disk amplitude derived by Brezin, Itzykson, Parisi 
and Zuber\cite{BIPZ}.
The term in the parenthesis in the first line of 
(\ref{eqn:trans-gen}) is possible one step forward with marked 
point of the exit loop in the dynamical triangulation 
while the four terms in the square bracket is all the possible steps 
forward in the entrance loop. 
See fig.\ref{fig:exit moves} and \ref{fig:entrance moves}. 
It is interesting to note that the geometrical structure of the random 
surface is directly reflected to the analytic expression of the 
transfer matrix. 

For the $c=0$ random surface it is known how to take the continuum 
limit of random surface for the disk amplitude $F$ by \cite{BIPZ}. 
Here we simply point out that the continuum limit of the generating 
function of the transfer matrix can be taken in a well defined way.  
We can then obtain a continuum expression of the cylinder amplitude 
by solving differential equation. 
The details of this interesting example to obtain the continuum expression 
from the above lattice transfer matrix can be found in \cite{KKMW}.
By using the continuum cylinder amplitude we can derive interesting 
quantities which have direct relation with the fractal structure of 
the random surface and thus can be measured numerically. 

Let $\rho(L;R)dL$ be the continuum counterpart of the number of loops 
belonging to the boundary  
whose lengths lie between $L$ and $L+dL$. 
Here $L$ and $R$ are continuum counterparts of $l$ and $r$, respectively. 
The $\rho(L;R)dL$ turns out to be a scaling function in the continuum 
limit 
\begin{eqnarray}
 \rho(L;R)dL~=~ \Bigl(\frac{3}{7\sqrt{\pi}}\Bigr)
 \frac{1}{R^2}(x^{-5/2}+\frac{1}{2}x^{-3/2}+\frac{14}{3}x^{1/2})exp(-x)
\label{eqn:rho-scale}
\end{eqnarray} 
with $x=L/R^2$ as a scaling parameter.
It is very surprising that there is a scaling function for 
the quantum gravity.
This $\rho(L;R)dL$ can be compared with the numerically measured value.

\subsubsection{Numerical Results on the Fractal Structure of 
Two Dimensional Quantum Gravity on the Lattice}

It is important to ask what could be the observables of quantum gravity, 
which could be numerically measured. 
We have confirmed the fractal nature of the quantum gravity for 
$c=-2$ model numerically\cite{KKSW}, which is the first numerical 
confirmation of the fractal nature of quantum gravity. 
We have analyzed the following scalar fermion models corresponding to 
$c=-2$ model:
\begin{eqnarray}
S~&=&~\sum_{T}\int \prod_{i\in T} d \bar{\psi}_i d
\psi_i 
exp\{-\sum_{<ij>}(\bar{\psi}_i - \bar{\psi}_j)(\psi_i - \psi_j)\}   \cr
&=&~ \sum_{T} det \Delta (T) \cr   
&=&~ \frac{1}{N+2} T_{N+1} R_{N+1},
\label{eqn:c-2model}
\end{eqnarray} 
where all the possible triangulation $T$  are summed up. 
The scalar fermions are located on the dynamically triangulated lattice 
sites and thus $\Delta (T)$ is the lattice laplacian for the 
triangulation $T$.
$T_{N+1}$ and $R_{N+1}$ are the number of tree and rainbow 
diagrams with $N$ vertices and external lines, respectively. 
Tree Feynman diagrams of $\phi^3$ theory and planar triangulations have 
one to one correspondence. 
Here we fix the topology of the surface as a sphere and the number of 
triangles to be $N$.
$T_n$ and $R_n$ satisfy the same Schwinger-Dyson equation:
\begin{eqnarray}
T_n~=~\sum_{k=1}^{n-1} T_k T_{n-k},
\end{eqnarray} 
which leads to the solution $T_n=(2n-2)!/n!(n-1)!$ and 
$R_n$ has the same form.

Since we know the analytic expression of $T_n$ and $R_n$ 
we can reconstruct arbitrary triangulation of sphere topology 
by combining the tree and 
rainbow diagrams with the correct weight. 
The essential point of the $c=-2$ model is that it has the very simple 
relation (\ref{eqn:c-2model}) due to the first power of the lattice 
laplacian. 
Then $T_n$ and $R_n$ have very simple form and thus numerically 
huge number of triangulated surfaces can be generated. 

\begin{figure}
\begin{center}
 \begin{minipage}[b]{0.4\textwidth}
 \epsfxsize=\textwidth \epsfbox{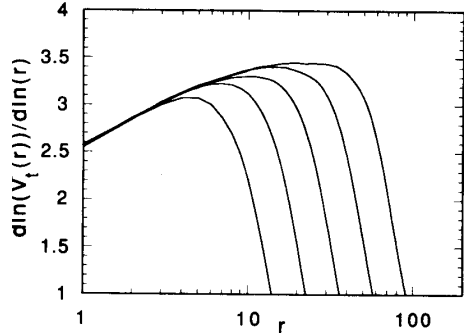} 
 \end{minipage}
 \end{center}
 \caption{Direct measurement of fractal dimension of $c=-2$ model}
 \label{fig:frexp1}
\end{figure}

We have used the formulation called "recursive sampling method" 
to generate the surface configuration numerically which was 
initiated by Agistein and Migdal for $c=0$ model \cite{AM}.
We have measured several quantities, in particular the 
number of triangles $V(r)$ within a geodesic distance $r$ parameterized by
\begin{equation}
V(r)~\equiv~<{\mbox {\rm number of triangles within r steps}}>~
\sim~r^\gamma.
\end{equation} 
We have found that the fractal dimension $\gamma$ approaches a constant 
value $\gamma \rightarrow 3.55$ with large $N (\rightarrow 5\times 10^6$).
This is the clear sign of the fractal scaling of two dimensional random 
surface\cite{KKSW}. 
See fig.\ref{fig:frexp1}

We have thus numerically confirmed that the fractal structure is the 
essential observable of two dimensional quantum gravity. 
In these numerical measurements we needed huge number of triangles to 
get saturated value of the fractal dimension. 

\begin{figure}
\begin{center}
 \begin{minipage}[a]{0.3\textwidth}
 \epsfxsize=\textwidth \epsfbox{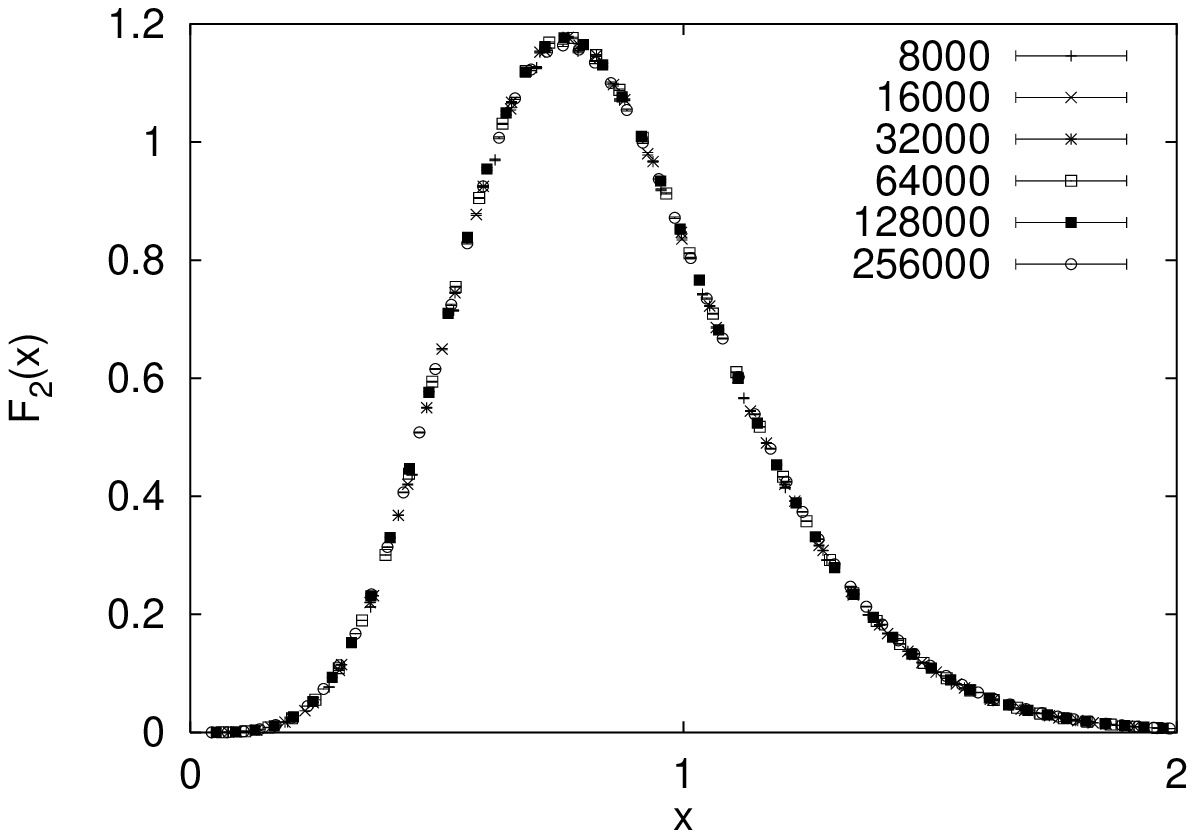} 
 \end{minipage}
\begin{minipage}[b]{0.3\textwidth}
 \epsfxsize=\textwidth \epsfbox{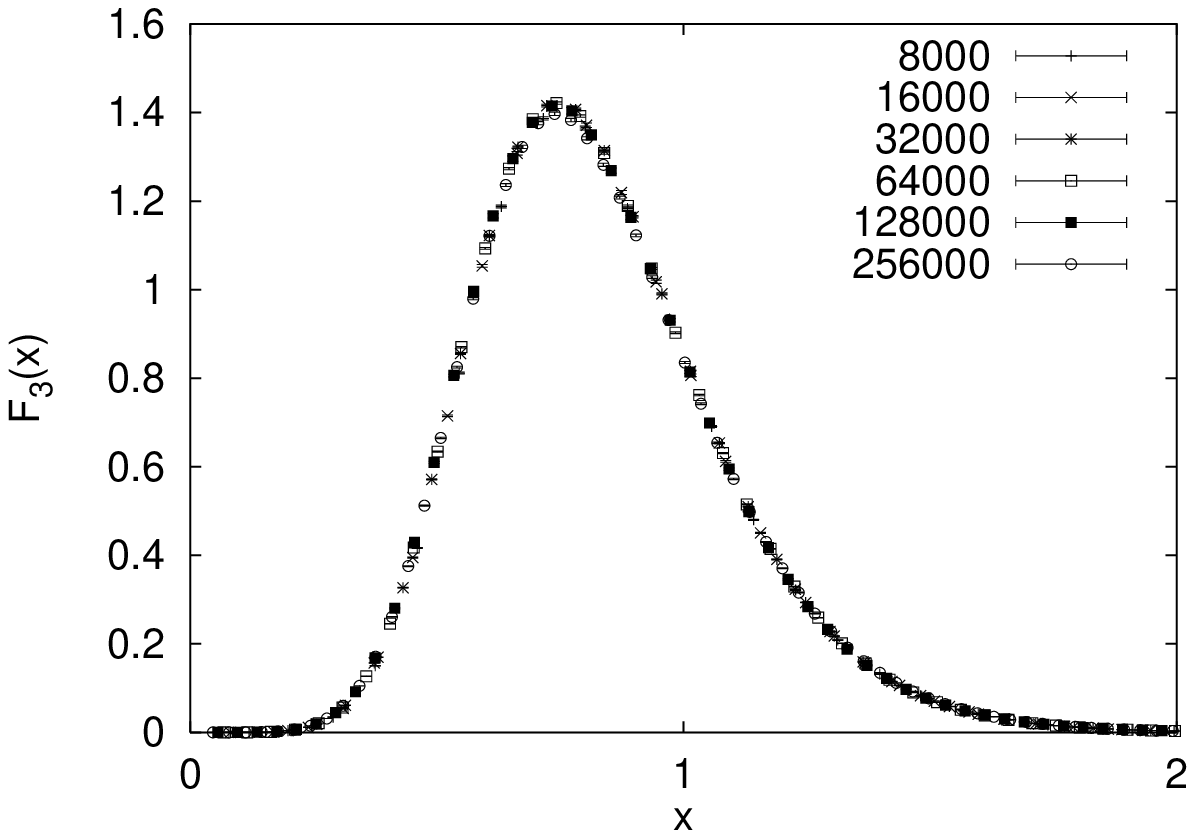} 
 \end{minipage}
\begin{minipage}[c]{0.3\textwidth}
 \epsfxsize=\textwidth \epsfbox{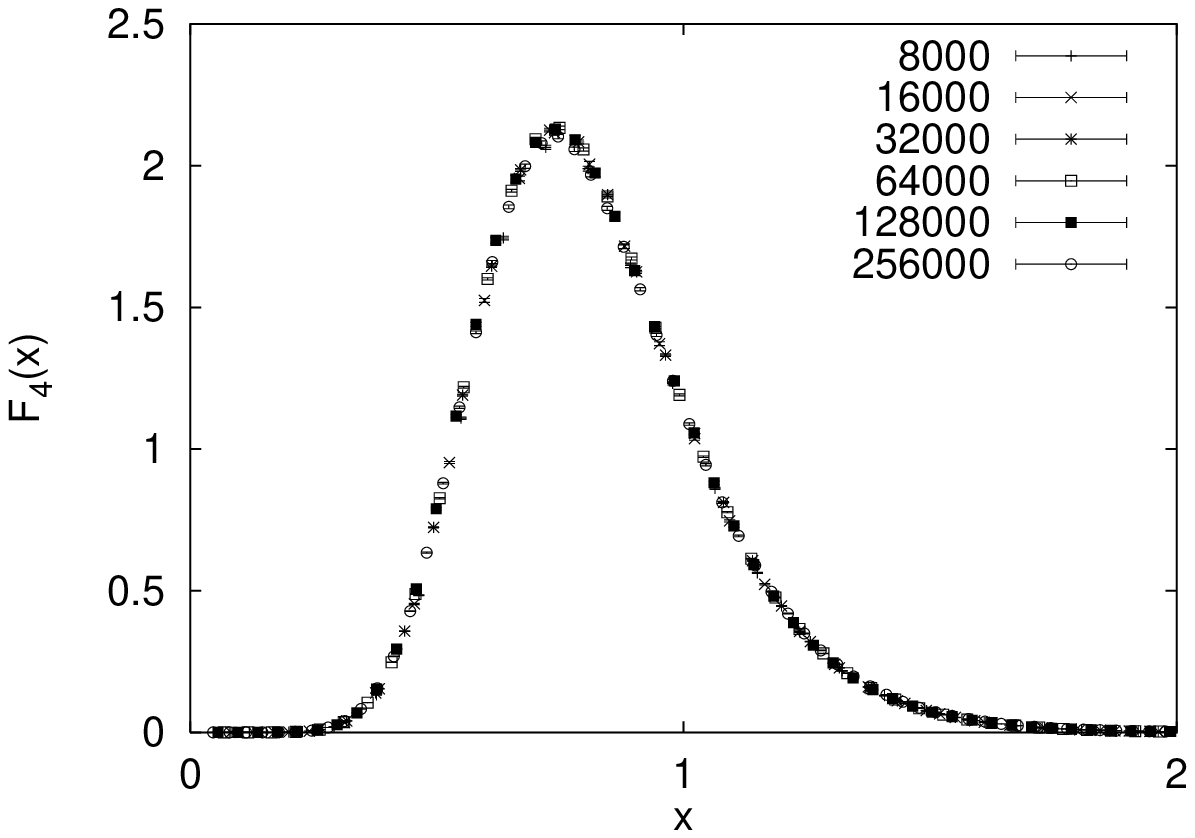} 
 \end{minipage}
\end{center}
 \caption{Finite size scaling of $F_2(y)$, $F_3(y)$ and $F_4(y)$ 
for the number of triangles given in figs.}
 \label{fig:finite scaling}
\end{figure}

Later we have used the formulation of finite size scaling methods 
applied to the quantum gravity and then we obtained 
very accurate value of the fractal dimension even with relatively 
smaller number of triangles but with huge number of 
configurations\cite{AAIJKWY}.
We define the following quantity: 
\begin{equation}
<L^n(r)> \equiv \sum_{l=1}^\infty l^n \rho(l,r),
\end{equation}
where $L(r)$ is the length of boundary loops located at $r$ steps from 
a marked point and $\rho(l,r)$ is the lattice counterpart of the quantity 
defined in (\ref{eqn:rho-scale}) for a given number of triangles $N$.
Finite size scaling formulation of the two dimensional quantum gravity 
predicts the following formula:
\begin{eqnarray}
<L^n(r)> &\sim& (N)^{2n/d_F}F_n(y) ~~~~~~ (n\geq 2)  \nonumber  \\
         y&=&\frac{r}{(N)^{1/d_F}},
\end{eqnarray} 
where $N$ is the number of triangles and $F(y)$ is a scaling function and 
$y$ is the scaling parameter. 
As we can see in figs.\ref{fig:finite scaling}, the data are beautifully 
scaling for $F_2$, $F_3$, $F_4$ with various numbers of 
triangles which are given in figures.

The numerical result for the fractal dimension of quantum space time 
coupled to $c=-2$ matter by comparing the above formula with the numerical 
distributions of $F$'s is 
$$
  d_F=3.56 \pm 0.04.
$$

On the other hand we gave the theoretical prediction of the fractal 
dimension of two dimensional space time coupled to $c$ matter by 
Liouville theory \cite{KW1}
\begin{eqnarray}
d_F&=&2\frac{\sqrt{25-c} + \sqrt{49-c}}{\sqrt{25-c} + \sqrt{1-c}} 
\nonumber \\
   &=& 3.561 ~~~~~~~~    (c=-2).
\end{eqnarray}
The numerical result and the theoretical data are perfectly consistent.

Here we have given an example, $c=-2$ model, of two dimensional quantum 
gravity which are well understood numerically and analytically.
We claim that the essence of the two dimensional quantum gravity 
is the fractal nature of the space time.

\subsection{Susskind Fermion---Staggered Fermion---Dirac 
K$\ddot{\hbox{a}}$hler Fermion on the Lattice}

If we want to put fermions on the lattice, there is the well known 
chiral fermion problem. 
The problem states that we cannot put chiral fermions on the lattice 
with avoiding species doublers\cite{nogo}. 
Or otherwise it is stated that the anomaly vanishes if we put fermions 
in a chiral invariant way\cite{Karsten-Smit}.
Wilson's proposal on this respect is to introduce non-chiral term in the 
lattice action and then each species doubler except for one gets a mass 
of order the inverse lattice constant and thus disappear in the 
continuum limit\cite{Wilson}. 
Then the chirality will be recovered and the 
remaining unique species generates the necessary anomaly 
in the continuum limit. 
In fact there is a clear chiral invariant line on the 
hopping parameter-coupling constant plane where the quak mass and 
the pion mass vanishes in the Wilson's lattice QCD 
formulation\cite{Kawamoto}. 

There is an alternative approach on this problem. 
The origin of the species doublers are related to the simplicial 
nature of the manifold. 
Susskind proposed an idea to collect these species doublers to 
construct Dirac fermions\cite{Susskind}. 
For example in four dimensional square lattice there are $2^4=16$ 
corners of Brillouin zone in the momentum space and each doubler 
corresponds to a component of the Dirac spinors thus leads to 
four flavor copies of Dirac fermions. 

One of the essential points here is to recognize that the species 
doublers in the configuration space correspond to points located at 
the center of the simplexes of the fundamental hyper cube.  
This is related to do with the fact that the Susskind fermion 
formulation is the flat space version of the Dirac-K$\ddot{\hbox{a}}$hler 
fermion formulation\cite{DK}\cite{BJ} which is formulated by 
differential forms and 
thus independent of the particular choice of lattice shape and thus 
can be identified as a fermion formulation on the simplicial manifold.

Here we explain this situation by starting from the naive fermion 
formulation. 
If we naively discretize the Dirac equation on the square lattice 
we obtain
\begin{equation}  
S_F=\frac{1}{2}\sum_{x,\hat{\mu}}[\overline{\psi}(x)\gamma_\mu\psi
    (x+\hat{\mu})-\overline{\psi}(x+\hat{\mu})\gamma_\mu\psi(x)],
\label{eqn:naive-ferm}
\end{equation}
where $x$ denotes the lattice site with integer coordinates 
$x_1,x_2,\cdots,x_d$ in units of lattice spacing which is taken to be 1.
$\hat{\mu}$ is the unit vector along the $\mu$th direction.
The $\gamma$-matrices satisfy the following Clifford algebra:
\begin{equation}  
\{\gamma_\mu,\gamma_\nu\}=-2\delta_{\mu\nu},
\label{eqn:clifford-alg}
\end{equation}
with the conventions $\gamma_\mu^\dagger=-\gamma_\mu$ in $d$-dimension.
We proposed the following transformation which relates the naive 
fermion formulation and Susskind fermion formulation \cite{KS}: 
\begin{equation}  
\psi(x)=A(x)\chi(x),~~~~~~~~~~~\overline{\psi}(x)=
\overline{\chi}(x)A^\dagger(x),
\label{eqn:KS-trans}      
\end{equation}
where 
\begin{equation}
A(x)=\gamma_1^{x_1}\gamma_2^{x_2}\cdots\gamma_d^{x_d}.
\label{eqn:KS-mat}
\end{equation}
Then the naive fermion formulation of Dirac action leads to
\begin{equation}  
S_F=-\frac{1}{2}\sum_{x,\hat{\mu},\alpha}\eta_\mu(x)
[\overline{\chi}^\alpha(x)\chi^\alpha(x+\hat{\mu})+
\overline{\chi}^\alpha(x+\hat{\mu})\chi^\alpha(x)], 
\label{eqn:stag-ferm}
\end{equation}
with 
\begin{equation}
\eta_\mu(x)=(-1)^{x_1+x_2+\cdots+x_{\mu-1}}.
\label{eqn:KS-mat}
\end{equation}
This formulation is called staggered fermion formulation in lattice QCD. 
The $\gamma$ matrices in the naive fermion formulation have now 
disappeared in the staggered fermion formulation. 
Instead we get $C\equiv 2^{[d/2]}$ copies of spinors.

We now consider $d$-dimensional hypercube defined on the lattice, 
with its origin at the site $2y$ and corners 
\begin{equation}
    x_\mu=2y_\mu + \eta_\mu, ~~~~~\eta_\mu=0~ \hbox{or}~ 1,~~~
    \mu=1,\cdots,d.\nonumber 
\label{eqn:double-size}
\end{equation}
Thus $\eta$ labels the set of $C^2$ $d$-dimensional vectors which 
point on the corners of the hypercube.
We introduce the following notations for the convenience:
\begin{eqnarray} 
\chi(2y+\eta)&\equiv& (-1)^y\chi_\eta(y) \nonumber \\
\overline\chi(2y+\eta)&\equiv& (-1)^y\overline\chi_\eta(y) \nonumber \\
         (-1)^y&=&(-1)^{y_1+\cdots +y_d}.
\end{eqnarray} 
Then $S_F$ can be written in the following form\cite{Saclay}:
\begin{equation}  
S_F=\sum_{x,\hat{\mu}}\sum_{\eta,\eta'}[\overline{\chi}_\eta(y)
    \Gamma^\mu_{\eta,\eta'}\Delta_\mu\chi_{\eta'}(y) + 
    \overline{\chi}_\eta(y)\widetilde{\Gamma}^\mu_{\eta,\eta'}
    \delta_\mu\chi_{\eta'}(y)],
\label{eqn:double-space-action}
\end{equation}
where the first and second lattice derivatives are
\begin{eqnarray} 
\Delta_\mu f(y) &\equiv& \frac{1}{4}[f(y+\hat{\mu})-f(y-\hat{\mu})]
\nonumber \\
\delta_\mu f(y) &\equiv& \frac{1}{4}[f(y+\hat{\mu})-2f(y)+f(y-\hat{\mu})].
\end{eqnarray} 
$\Gamma^\mu$ and $\Gamma^{\mu'}$ are defined by
\begin{eqnarray} 
\Gamma^\mu_{\eta,\eta'}&\equiv&\frac{1}{C}\hbox{tr}
(\Gamma_\eta^\dagger\gamma_\mu\Gamma_{\eta'}), \nonumber \\
\widetilde{\Gamma}^\mu_{\eta,\eta'} &\equiv& \frac{1}{C}\hbox{tr}
(\Gamma_\eta^\dagger\gamma_\mu\Gamma_{\eta'})
(\delta_{\eta',\eta-\hat{\mu}}-\delta_{\eta',\eta+\hat{\mu}}),
\label{eqn:Gamma-def} 
\end{eqnarray} 
with 
\begin{equation}
\Gamma_\eta=\gamma_1^{\eta_1}\gamma_2^{\eta_2}\cdots\gamma_d^{\eta_d}.
\label{eqn:gamma-prod}
\end{equation}
Introducing the following notation for the fermionic fields to get 
rid of the factor of trace in (\ref{eqn:Gamma-def}):
\begin{eqnarray} 
\chi_\eta(y)&=&\sqrt{C}\sum_{\alpha,a}\Gamma^{*\alpha a}_\eta
                           q^{\alpha a}(y) \nonumber \\
\overline{\chi}_\eta(y)&=&\sqrt{C}\sum_{\alpha,a}\overline{q}^{\alpha a}(y)
                           \Gamma^{\alpha a}_\eta,
\label{eqn:newferm-def} 
\end{eqnarray} 
we obtain the final form of Dirac-K$\ddot{\hbox{a}}$hler fermion 
action on the flat space 
\begin{equation}  
S_F=2^d\sum_{y,\mu}[\overline{q}(y)(\gamma_\mu\otimes{\bf 1})
                    \Delta_\mu q(y) + 
                    \overline{q}(y)(\gamma^\dagger_5\otimes t_\mu^\dagger
                    t_5^\dagger)
                    \delta_\mu q(y)].
\end{equation}
The fermionic part of this action can be interpreted as Susskind fermions 
while this action has bosonic counterparts as well.
The bosonic part including second derivative is higher order 
than the fermionic part with respect to the lattice constant. 
In the fermion bilinears the first matrix acts on Greek indices 
interpreted as Dirac indices, the second one on Latin indices interpreted 
as flavor indices. $t_\mu$ stands for the matrix $\gamma^*_\mu$ when 
acting on flavor space, and $\gamma_5=\gamma_1\gamma_2\cdots \gamma_d$.

\subsection{ Chern-Simons Gravity and Ponzano-Regge Model}
\subsubsection{Chern-Simons Gravity}
In this subsection we explain formulations of three dimensional gravity.
We first summarize the Chern-Simons gravity formulated by 
Witten\cite{Witten1}.
We choose the gauge group as Euclidean version of three dimensional 
Poincare group $ISO(3)$.
Then we define one form gauge field and zero form gauge parameter as 
\renewcommand{\arraystretch}{1.5}
\begin{equation}
\begin{array}{rcl}
  A_\mu&=&e^a_\mu P_a + \omega^a_\mu J_a, \\
 v  &=& \rho^a P_a + \tau^a J_a,
\end{array}
\end{equation}
where $e^a_\mu$ and $\omega^a_\mu$ are dreibein and spin connection, 
respectively, and $\rho$ and $\tau$ are the corresponding gauge parameters. 
The momentum generator $P_a$ and the angular momentum generator $J_a$ 
of $ISO(3)$ satisfy 
\begin{equation}
 [J_a,J_b] = \epsilon_{abc} J^c, ~~~
 [J_a,P_b] = \epsilon_{abc} P^c, ~~~
 [P_a,P_b] = 0.
\end{equation} 
Using the invariant quadratic form which is special 
in three dimensions, we can define the inner product 
\begin{equation}
 \langle J_a,P_b \rangle = \delta_{ab},
 ~~\langle J_a,J_b \rangle = \langle P_a,P_b \rangle = 0.
\end{equation} 
We then obtain Einstein-Hilbert action of three 
dimensional gravity from Chern-Simons action   
\begin{equation}
 \int \Bigl\langle AdA + \frac{2}{3}A^3 \Bigr\rangle = 
 \int \epsilon^{\mu\nu\rho}e_{\mu a}F^a_{\nu\rho}~d^3x,
\end{equation}  
where 
\begin{equation}
 F^a_{\mu\nu}= \partial_\mu\omega^a_\nu - 
               \partial_\nu\omega^a_\mu +
               \epsilon^a_{~bc}\omega^b_\mu\omega^c_\nu.\label{CSCurvature}
\end{equation} 

The component wise gauge transformation of 
$\delta A_\mu = - D_\mu v$ is given by 
\begin{equation}
\begin{array}{rcl}
 \delta e^a_\mu &=& -D_\mu\rho^a -\epsilon^{abc}e_{\mu b}\tau_c,\\
 \delta \omega^a_\mu &=& -D_\mu\tau^a.\label{CS gauge transformation}
\end{array}
\end{equation} 
At this stage it is important to recognize that 
the local Lorentz transformation is generated by the 
gauge parameter $\tau$
\begin{equation}
\begin{array}{rcl}
 \delta e^a_\mu &=&  -\epsilon^{abc}e_{\mu b}\tau_c,\\
 \delta \omega^a_\mu &=& -D_\mu\tau^a, \label{LLGT}
\end{array}
\end{equation}
while the gauge transformation of diffeomorphism is generated 
by the gauge parameter $\rho$
\begin{equation}
\begin{array}{rcl}
 \delta e^a_\mu &=& -D_\mu\rho^a, \\
 \delta \omega^a_\mu &=& 0.\label{DGT}
\end{array}
\end{equation}

Three dimensional Einstein gravity is thus elegantly 
formulated by Chern-Simons action. 
This is essentially related to the fact that 
the three dimensional Einstein gravity does not include
dynamical graviton and thus can be formulated by the 
topological Chern-Simons action.
The equivalence of the above action and Einstein-
Hilbert action is, however, valid only if the dreibein $e^a_\mu$ 
is invertible.
The quantization and perturbative renormalizability 
around the nonphysical classical background 
$e^a_\mu = 0$ 
is the natural consequence of the formulation.

\subsubsection{Ponzano-Regge Model}

Ponzano and Regge noticed that angular momenta of 6-$j$ symbol 
can be identified as link lengths of a tetrahedron\cite{Ponzano-Regge}.
\begin{figure}
\begin{center}
 \begin{minipage}[b]{0.4\textwidth}
 \epsfxsize=\textwidth \epsfbox{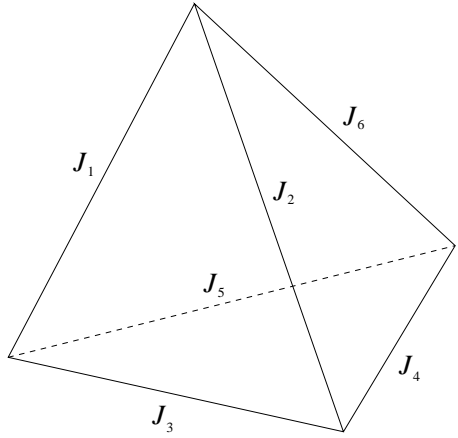} 
 \end{minipage}
\end{center}
 \caption{tetrahedron with angular momenta on the links}
 \label{fig:colored tetra}
\end{figure}
In particular they showed the following approximate relation: 
\begin{equation}
 (-1)^{\sum_{i=1}^{6} J_{i}} 
  \sixj{J_1}{J_2}{J_3}{J_4}{J_5}{J_6}
  \sim 
  \frac{1}{\sqrt{12 \pi V}} \cos \left( \SR + \frac{\pi}{4} \right) 
  \quad (\hbox{all } J_i \gg  1) ,
  \label{P-R formula}
\end{equation}
where $\SR$ is the Regge action of Regge calculus\cite{Regge} 
for a tetrahedron having link length $J_k$ $(k=1\sim 6)$ 
which correspond to the angular momentum of the corresponding 6-$j$ symbol 
and $V$ is the volume of the tetrahedron.

Based on this observation they proposed the following partition function: 
\begin{equation}
 \ZPR = \lim_{\lambda \rightarrow \infty} \sum_{J \leq \lambda} 
  \prod_{\hbox{\tiny vertices}} \Lambda (\lambda) ^{-1} 
  \prod_{\hbox{\tiny edges}} (2J+1) 
  \prod_{\hbox{\tiny tetrahedra}} (-1)^{\sum J_i} 
    \sixj{J_1}{J_2}{J_3}{J_4}{J_5}{J_6}
\label{ZofPR}.
\end{equation}
Thus the partition function $\ZPR$ is 
the product of the partition function of each tetrahedron which 
reproduces the cosine of the Regge action in contrast with 
the exponential of the Regge action in Regge calculus.
There is an argument about the origin of the cosine, that right 
and left handed contributions of the general coordinate frames 
contribute separately and thus the summation of the exponential 
with the different sign factor for the Regge action appears.
It is thus natural to expect that this 
action leads to a gravity action.

\begin{figure}
\begin{center}
 \begin{minipage}[b]{0.4\textwidth}
 \epsfxsize=\textwidth \epsfbox{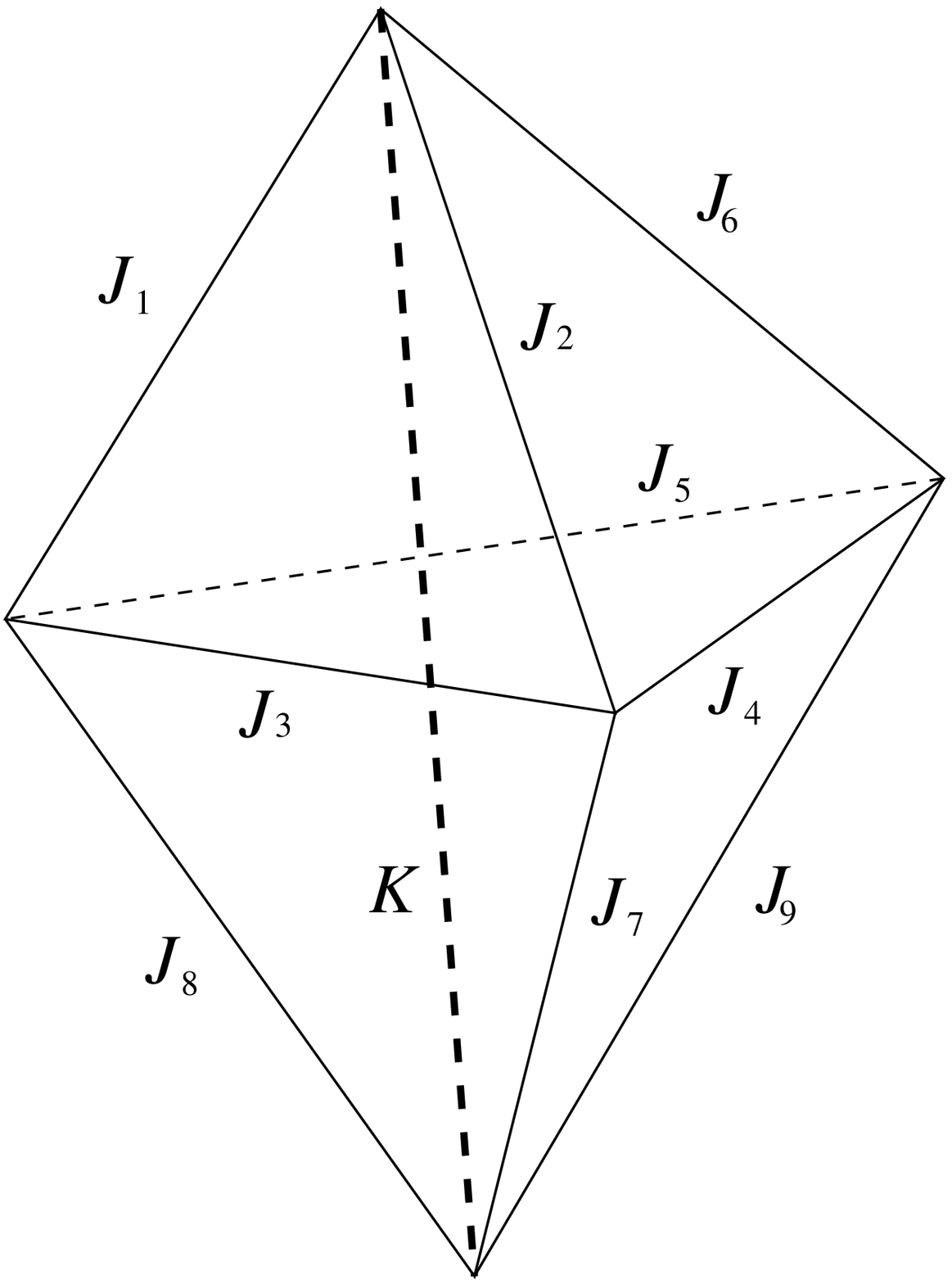} 
 \end{minipage}
 \hspace*{2cm}
  \begin{minipage}[t]{0.4\textwidth}
 \epsfxsize=\textwidth \epsfbox{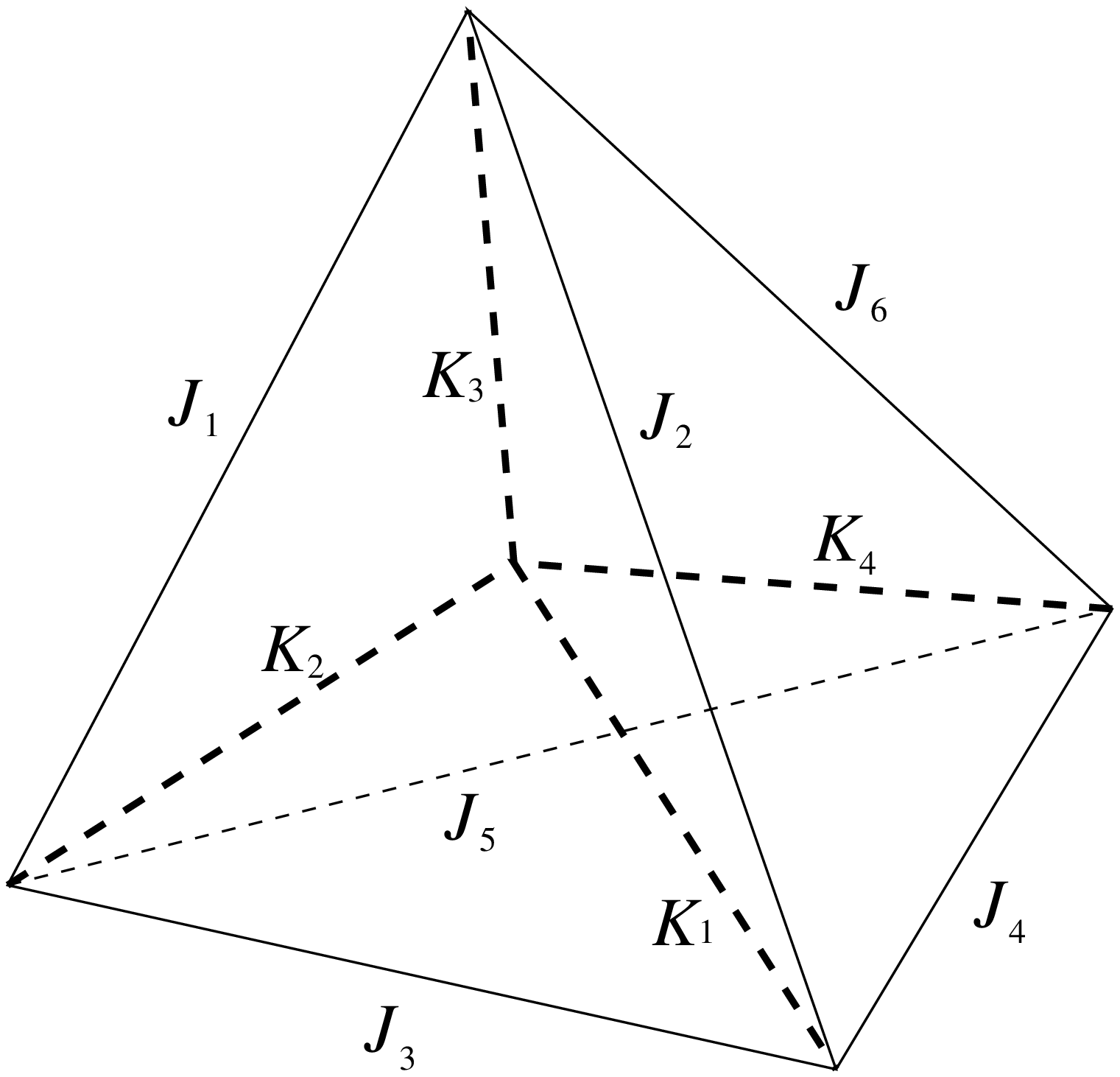} 
 \end{minipage}
\end{center}
 \caption{2-3 move and 1-4 move}
 \label{fig:Alexander move}
\end{figure}

Important characteristic of the Ponzano-Regge action is that 
it has a topological nature on a simplicial manifold. 
The action is invariant under the following 2-3 and 1-4 
Alexander moves. 
The 2-3 and 1-4 moves are related to the following 6-$j$ relations:
\begin{eqnarray}
 && \sum_{K} (-1)^{K+\sum_{i=1}^{9}J_i} (2K+1) 
  \sixj{J_1}{J_8}{K}{J_7}{J_2}{J_3} \sixj{J_7}{J_2}{K}{J_6}{J_9}{J_4}
  \sixj{J_6}{J_9}{K}{J_8}{J_1}{J_5} 
    \nonumber \\
 &=& \sixj{J_3}{J_4}{J_5}{J_6}{J_1}{J_2} \sixj{J_3}{J_4}{J_5}{J_9}{J_8}{J_7},
  \label{2-3move}
\end{eqnarray}
and 
\begin{eqnarray}
&&  \sum_{K_i} \left[ \prod_{i=1}^{4} (2K_i + 1) \right]
 (-1)^{\sum K_i} \Lambda (\lambda)^{-1} 
  \sixj{J_1}{J_2}{J_3}{K_1}{K_2}{K_3}
  \sixj{J_4}{J_6}{J_2}{K_3}{K_1}{K_4} \nonumber \\
&& \hspace{2cm} \times
  \sixj{J_3}{J_4}{J_5}{K_4}{K_2}{K_1}
  \sixj{J_1}{J_5}{J_6}{K_4}{K_3}{K_2} 
= (-1)^{\sum J_i} \sixj{J_1}{J_2}{J_3}{J_4}{J_5}{J_6}.
\label{1-4move}
\end{eqnarray}
The geometrical correspondence of 2-3 and 1-4 moves with 
two tetrahedra into three tetrahedra and one tetrahedron 
into four tetrahedra is obvious from 
fig.\ref{fig:Alexander move}.
In the formula of 1-4 move there appears the following infinite sum 
which is then introduced as a regularization factor in the denominator 
with a cutoff $\lambda$:  
\begin{eqnarray}
 \Lambda (\lambda)
&=& \frac{1}{2 J_1 + 1}  
\!\!\!\!\!\!\!\!\!\!
  \sum_{ 
  \mbox{ \scriptsize $
  \begin{array}{c} 
   K_2,K_3 \leq \lambda,\\ 
   |K_2 - K_3| \leq J_1 \leq K_2 + K_3
  \end{array} $}
   }
\!\!\!\!\!\!\!\!\!\!
(2 K_2 + 1)(2 K_3 + 1) \nonumber  \\
&=& \sum_{J=0}^{\lambda} (2J+1)^2  
\sim \frac{4 \lambda^3}{3} ~~ (\lambda \rightarrow \infty)
\label{Lambda}.
\end{eqnarray}
It is known that these two Alexander moves reproduce any three dimensional 
simplicial manifold.
Thus the partition function $\ZPR$ is invariant 
under the variation of metric and is expected to be topological.  

There is a beautiful generalization of this model to avoid the above 
cut off dependence by introducing the quantum group formulation as a 
regularization by Turaev and Viro\cite{Turaev-Viro}. 
Their proposal triggered number of later investigations for three 
dimensional lattice 
gravity\cite{Ooguri-Sasakura}\cite{Boulatov}\cite{3dgravity}. 
Turaev, Ooguri-Sasakura and Boulatov suggested the equivalence of the 
Chern-Simons gravity and the Ponzano-Regge 
model\cite{Turaev-Viro}\cite{Ooguri-Sasakura}\cite{Boulatov}.    

In the later section we show that the continuum limit of the lattice 
Ponzano-Regge model leads to the Chern-Simons gravity by explicitly 
constructing a lattice gauge gravity model.

\subsection{Four Dimensional Gravity on the Lattice }

In analogy with the formulation of three dimensional gravity 
by Chern-Simons action and 6-$j$ symbols of the Ponzano-Regge model, 
it is possible to formulate 
a four dimensional topological gravity by $BF$ theory and 15-$j$ symbols
\cite{Ooguri}.
There are variants of the similar four dimensional models developed later
\cite{FDgravity}\cite{Smolin}.

Ooguri proposed the following model on the four dimensional simplicial 
manifold composed of four dimensional fundamental simplexes:
\begin{equation}
Z=\sum_C\frac{1}{N_{\hbox{\tiny sym}}}\lambda^{N_4(C)}Z_C,
\end{equation}
where the summation $\sum_C$ is over oriented four dimensional symplicial 
manifolds, and 
$N_{\hbox{\tiny sym}}$ is a symmetric factor and $N_4(C)$ is a number of 
4-simplexes in $C$. 
The partition function for a given $C$ is
\begin{eqnarray}
Z_C&=&\sum_J\prod_{t:\hbox{\tiny triangles}}(2J_t+1)\prod_{\hbox{\tiny tetrahedra}}
               \{6j\}\prod_{\hbox{\tiny 4-simplexes}}\{15j\} \\
   &=&\sum_J\sum_{m,n}\prod_{t:\hbox{triangles}}(2J_t+1)
               \prod_{T:\hbox{\tiny tetrahedra}} \int dU_T 
               D^{J_1,T}(U_T)\cdots D^{J_4,T}(U_T),           
\label{eqn:15j}
\end{eqnarray}
where the first line is the symbolic presentation of the geometrical 
nature of the four dimensional fundamental simplex which has ten triangles, 
each of which corresponds to 3-$j$ symbol, and ten 3-$j$ symbols lead 
15-$j$ symbol,
which is analogous to the situation where 6 angular momenta were assigned 
to the links of tetrahedron and four triangles of a tetrahedron 
carry four 3-$j$ symbols which leads to a 6-$j$ symbol in three dimensions. 
By using the orthonormality of the 6-$j$ symbols one can show that the 
results of this summation is independent of the choice of splitting of the 
vertices and depends only on the structure of the complex $C$, where 
extra 6-$j$ symbols are necessary to keep this invariance. 
This is again analogous to the Alexander move invariance of the 
Ponzano-Regge model.

The second line of eq.(\ref{eqn:15j}) comes from the fact that the 15-$j$ 
and 6-$j$ symbols are written down by 3-$j$ symbols and 
the products of 3-$j$ symbols can be replaced by the products of 
$D$ functions by using the following relations:
\begin{eqnarray}
 \int DU D^{J_1}_{m_1n_1}(U) D^{J_2}_{m_2n_2}(U&)& D^{J_3}_{m_3n_3}(U)   
= \threej{J_1}{J_2}{J_3}{m_1}{m_2}{m_3}\threej{J_1}{J_2}{J_3}{n_1}{n_2}{n_3},
\label{eqn:3d3j}\\
 \sixj{J_1}{J_2}{J_3}{J_4}{J_5}{J_6} 
= \sum_{\hbox{\tiny{all $m_i$}}}
 (-&1&)^{\sum_i (J_i - m_i)}
 \threej{J_1}{J_2}{J_3}{-m_1}{-m_2}{-m_3}\nonumber \\
 &&\hspace{-2cm} \times 
 \threej{J_1}{J_5}{J_6}{m_1}{-m_5}{m_6} 
 \threej{J_4}{J_2}{J_6}{m_4}{m_2}{-m_6}
 \threej{J_4}{J_5}{J_3}{-m_4}{m_5}{m_3},
\label{eqn:6j3j}
\end{eqnarray}
where the 3-$j$ symbols carry $m_i$ suffices which correspond to the third 
components of the angular momentum $J_i$.
Each tetrahedron $T$ carries a group element $U_T$, and 
$J_{1,T},\cdots J_{4,T}$ are the spins on the links dual to the four 
triangles of $T$. 
The matrix elements $D^{J}$'s are multiplied around triangles in $C$.
Each triangle $t$ is shared by a finite number of tetrahedra; 
$T_1{(t)},T_2{(t)},\cdots,T_{n_t}{(t)}$.
We can perform the sum over the spin-$J_t$ on the link dual to $t$ using
the formula
\begin{equation} 
 \sum_{J_t}(2J_t+1)\hbox{Tr}\left[ D^{J_t}(U_{T_1}^{(t)})
                   \cdots D^{J_t}(U_{T_{n_t}}^{(t)})\right] 
               = \delta(U_{T_1}^{(t)}\cdots U_{T_{n_t}}^{(t)},1).
\label{eqn:flat-con}
\end{equation}
After carrying out $dU$ integration we obtain 
the condition that the holonomy around the triangle $t$ is 
trivial, the lattice version of flat connection condition.

The above result suggests that $Z_C$ is related to the partition function of 
the $BF$ model which is defined for an oriented manifold $M$ as 
\begin{equation} 
Z_{BF} = \int DB~DA \hbox{exp}(i\int_M <B,dA+[A,A]>),
\end{equation} 
where $A$ is a connection one-form on the manifold $M$, while $B$ 
is a Lie algebra valued two-form and $<,>$ is the invariant bilinear 
form on the Lie algebra. 
In this so called $BF$ action\cite{BF} with $F\equiv dA+[A,A]$, we can consider 
$B$ as Lagrange multiplier then we obtain the flat connection condition 
which is equivalent to the lattice version obtained from 
eq.(\ref{eqn:flat-con}). 
This action is the four dimensional counterpart of the Chern-Simons action 
in the sense that the flat connection condition is the equation of 
motion. 
We can thus naively expect that the 15-$j$ model leads to the $BF$ theory 
in the continuum limit. 
This is similar to the situation that the Ponzano-Regge model leads to 
the Chern-Simons gravity in the continuum limit which we are going 
to show in the later section.

\setcounter{equation}{0}
\section{A Possible Formulation towards Gauge Gravity coupled to Matter 
on the Simplicial Lattice Manifold}

\subsection{A Possible Formulation towards Unified Model on the Lattice}

From the known results which we have collected in the last subsections, 
I intend to figure out my personal view on what could be the necessary 
formulation to realize the possible formulations towards unified model 
on the lattice. 

I found that the example of Ising spin on the square lattice and dynamically 
triangulated lattice are the most convincing examples of the following 
picture: a regularized 
gravitational background can be generated by the dynamical triangulation 
and the matter field fermions can be generated by some degrees of 
freedom of fields located on the simplexes of a simplicial manifold. 
Lattice models of conformal field theories give many more varieties of 
the similar examples where matter central charge $c$ takes different 
values for different type of lattice models. 
One may wonder it would be special case of two dimensional field 
theory due to the conformal invariance on the second order 
phase transition point of two dimensional lattice models.
I find it is very natural to expect to have the similar mechanism working 
even in higher dimensions.
It would be, however, very difficult to solve four dimensional 
lattice models analytically while it was possible to solve two 
dimensional model analytically and derive the corresponding conformal 
invariant continuum models. 

As we have seen in the analyses of the fractal structure of quantum 
gravity, numerical simulation played an important role. 
In the two dimensional example of $c=-2$ model theoretical and numerical 
analyses on the fractal dimension were perfectly consistent. 
Since we don't expect the solvability in higher dimensional lattice 
gravity models with matter, numerical simulations may play an important 
role again. 
In considering the present situation, the lattice QCD would be a good example 
to find analogy in the lattice gravity formulation. 
Lattice QCD is not solvable but is perfectly correct formulation 
of QCD even in non perturbative regime. 
The Lorentz invariance is lost but the gauge 
invariance was strictly kept in the finite lattice regime. 

We need to find a gravitational counterpart of the lattice QCD 
formulation. 
In three dimensions Chern-Simons gravity is formulated as a gauge 
theory. 
It would be thus very instructive to find lattice gravity formulation 
of Chern-Simons gravity.
As we show in the next section it is possible to formulate the Chern-Simons 
gravity in analogy with Wilson's lattice QCD formulation. 
What is surprising there is that the continuum limit of the Chern-Simons 
gravity can be considered analytically with the help of Ponzano-Regge 
model.  

\begin{figure}
\begin{center}
 \begin{minipage}[b]{0.4\textwidth}
 \epsfxsize=\textwidth \epsfbox{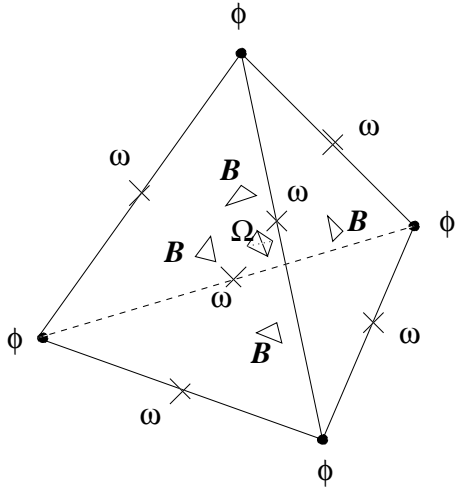} 
 \end{minipage}
\end{center}
 \caption{Differential forms on the symplices}
 \label{fig:glatgr}
\end{figure}

We find it is instructive that the three dimensional gravity is formulated 
by gauge theory of one form with Chern-Simons action. 
If one can formulate a gauge theory in terms of forms, the general coordinate 
invariance is automatic. 
On the other hand it is also very natural to expect that the gravity theory 
can very well be formulated by a gauge theory.
If we intend to formulate a lattice gravity theory as a gauge theory 
we need to formulate gauge theory in terms of differential forms.
Furthermore differential forms have very natural correspondence with 
simplexes on the simplicial manifold. 
As we can see in fig.\ref{fig:glatgr} the fields of differential forms are 
naturally put on the simplexes. 
We thus propose that the generalized gauge theory which are formulated 
by differential forms would be a good candidate to formulate lattice 
gravity theory even with matter fermions, possibly 
Dirac-K$\ddot{\hbox{a}}$hler fermions, which are also formulated 
by differential forms\cite{DK}\cite{BJ}.

\subsection{Generalized Gauge Theory}

The generalized Chern-Simons actions, which were proposed by 
the present author and Watabiki about ten years ago, is a generalization 
of the ordinary 
three-dimensional Chern-Simons theory into arbitrary dimensions~\cite{KW2}.
We summarize the results in this section.
The essential point of the generalization is to extend a one-form gauge field 
and zero-form gauge parameter to a quaternion valued 
generalized gauge field and gauge parameter which contain forms of all 
possible degrees.
Correspondingly the standard gauge symmetry is extended to much higher 
topological symmetry.
These generalizations are formulated in such a way that the generalized 
actions have the same algebraic structure as the ordinary three-dimensional 
Chern-Simons action. 

Since this generalized Chern-Simons action is formulated completely parallel 
to the ordinary gauge theory, the generalization can be extended further 
to the topological Yang-Mills action and generalized Yang-Mills actions of 
arbitrary dimensions.  

In the most general form, a generalized gauge field $\cal{A}$ and a gauge 
parameter $\cal{V}$ are defined by the following component form:
\begin{eqnarray}
 {\cal A} & = & {\mbox{\bf 1}}\psi + {\mbox{\bf i}} \hat{\psi} +  
            {\mbox{\bf j}} A + {\mbox{\bf k}} \hat{A}, 
\label{eqn:ggf} \\
 {\cal V} & = & {\mbox{\bf 1}} \hat{a} + {\mbox{\bf i}} a +  
            {\mbox{\bf j}} \hat{\alpha} + {\mbox{\bf k}} \alpha, 
\label{eqn:ggp}
\end{eqnarray}
where $( \psi, \alpha )$, $( \hat{\psi}, \hat{\alpha} )$, 
$( A, a )$ and  $( \hat{A},\hat{a} )$ are direct sums of 
fermionic odd forms, fermionic even forms, bosonic odd forms 
and bosonic even forms, respectively, and they take values on a gauge algebra.
The bold face symbols ${\bf 1}$, ${\bf i}$,  ${\bf j}$ and ${\bf k}$ 
satisfy the algebra
\begin{equation}
\begin{array}{c}
{\bf 1}^2={\bf 1}, \quad {\bf i}^2=\epsilon_1 {\bf 1}, \quad 
{\bf j}^2=\epsilon_2 {\bf 1}, \quad {\bf k}^2=-\epsilon_1 \epsilon_2 {\bf 1},
\\
{\bf i}{\bf j}=-{\bf j}{\bf i}={\bf k}, \quad 
{\bf j}{\bf k}=-{\bf k}{\bf j}=-\epsilon_2 {\bf i}, \quad 
{\bf k}{\bf i}=-{\bf i}{\bf k}=-\epsilon_1 {\bf j},
\end{array}
\end{equation}
where $(\epsilon_1,\epsilon_2)$ takes the value $(-1,-1), (-1,+1), (+1,-1)$ or 
$(+1,+1)$.
Throughout this paper we adopt the convention 
$(\epsilon_1,\epsilon_2)=(-1,-1)$ unless otherwise stated, 
then the above algebra corresponds to the quaternion algebra.
The following graded Lie algebra can be adopted as a gauge algebra:
\begin{eqnarray}
\left[ T_a , T_b \right] &=& f^c_{ab} T_c, \nonumber\\
\left[ T_a , \Sigma_\beta \right] &=& g^\gamma_{a\beta} \Sigma_\gamma, \\
\left\{ \Sigma_\alpha , \Sigma_\beta \right\} &=& h^c_{\alpha\beta} T_c, 
\nonumber
\end{eqnarray}
where all the structure constants are subject to consistency conditions 
which follow from the graded Jacobi identities.
If we choose $\Sigma_\alpha=T_a$ especially, this algebra reduces to 
$T_a T_b = k^c_{ab}T_c$ which is closed under multiplication.
A specific example of such algebra is realized by Clifford algebra~\cite{KW3}.
The components of the gauge field $\cal{A}$ and parameter $\cal{V}$ are 
assigned to the elements of the gauge algebra in a specific way:
\begin{equation}
 \begin{array}{rclrclrclrcl}
 A          &=& T_a A^a, \quad 
 \hat{\psi} &=& T_a \hat{\psi}^a, \quad 
 \psi       &=& \Sigma_\alpha \psi^\alpha, \quad 
 \hat{A}    &=& \Sigma_\alpha \hat{A}^\alpha, \\ 
 \hat{a}    &=& T_a \hat{a}^a, \quad
 \alpha     &=& T_a \alpha^a, \quad
 \hat{\alpha}&=&\Sigma_\alpha \hat{\alpha}^\alpha, \quad 
 a          &=& \Sigma_\alpha a^\alpha.
 \end{array}
\end{equation}
An element having the same type of component expansion as $\cal{A}$ 
and $\cal{V}$ belong to $\Lambda_-$ and $\Lambda_+$ class, respectively, and 
these elements fulfill 
the following $Z_2$ grading structure:
$$
[\lambda_+ , \lambda_+] \in \Lambda_+, \qquad
[\lambda_+ , \lambda_-] \in \Lambda_-, \qquad
\{\lambda_- , \lambda_-\} \in \Lambda_+, 
$$
where $\lambda_+ \in \Lambda_+$ and $\lambda_- \in \Lambda_-$.
The elements of $\Lambda_-$ and $\Lambda_+$ can be regarded as generalizations 
of odd forms and even forms, respectively.
In particular the generalized exterior derivative which belongs to 
$\Lambda_-$ is given by
\begin{equation}
Q={\bf j}d, 
\end{equation}
and the following relations similar to the ordinary differential algebra hold:
$$
\{Q , \lambda_-\}=Q\lambda_-, \qquad
[Q , \lambda_+]=Q\lambda_+, \qquad
Q^2=0,
$$
where $\lambda_+ \in \Lambda_+$ and $\lambda_- \in \Lambda_-$.
To construct the generalized Chern-Simons actions, 
we need to introduce two kinds of traces 
\begin{equation}
 \begin{array}{rcl}
 \mbox{Tr}[T_a , \cdots]=0, 
   &{\qquad}& \mbox{Tr}[\Sigma_\alpha , \cdots]=0, \\
 \mbox{Str}[T_a , \cdots]=0, 
   &{\qquad}& \mbox{Str}\{\Sigma_\alpha , \cdots\}=0, \\  
 \end{array}                                                
\label{eqn:gtr}
\end{equation}
where $(\cdots)$ in the commutators or the anticommutators denotes a product 
of generators. In particular $(\cdots)$ should include an odd number
of $\Sigma_\alpha$'s in the last eq. of (\ref{eqn:gtr}).
$\mbox{Tr}$ is the usual trace while $\mbox{Str}$ is the super trace 
satisfying the above relations. 
These definitions of the traces are crucial 
to show that the generalized gauge theory action can be invariant under 
the generalized gauge transformation presented bellow.

As we have seen in the above the generalized gauge field ${\cal A}$ and 
parameter ${\cal V}$, and the generalized differential operator $Q$ 
play the same role as the one-form gauge field and zero-form gauge 
parameter and differential operator of the usual gauge theory, respectively.
We can then construct generalized actions in terms of these generalized 
quantities. 
We first define a generalized curvature 
\begin{equation}
 {\cal F} \equiv \{Q+{\cal A},Q+{\cal A}\} = Q{\cal A} + {\cal A}^2. 
 \label{eqn:gcurv} 
\end{equation}

We can construct generalized actions of Chern-Simons form, topological 
Yang-Mills form and Yang-Mills form.
They have the standard forms with respect to the generalized quantities
\begin{eqnarray}
S_{GCS}&=&\int_M \mbox{Tr}^*
        \left( \frac{1}{2}{\cal A}Q{\cal A}+\frac{1}{3}{\cal A}^3 \right),
         \nonumber \\
S_{GYM}&=&\int_M \mbox{Tr}^*
        \left( {\cal F} {\cal F} \right),
	\label{eqn:GAS} \\
S_{GYM}&=&\int_M \mbox{Tr}^*
        \left( {\cal F} *{\cal F} \right),\nonumber
\end{eqnarray}
where $*$ is generalized Hodge star defined later. 
The product of the generalized fields should be understood as wedge 
product.
The generalized trace $\mbox{Tr}^*$ have several types: 
$\mbox{Tr}^*=\mbox{Tr}_{\bf q}$ or $\mbox{Tr}^*=\mbox{Str}_{\bf q}$.
$\mbox{Tr}_{\bf q}(\cdots)$ and $\mbox{Str}_{\bf q}(\cdots) \ \ 
({\bf q}={\bf 1}, {\bf i}, {\bf j}, {\bf k})$ are defined so as to pick up 
only the coefficients of ${\bf q}$ from $(\cdots)$ and take the traces defined 
by eq.(\ref{eqn:gtr}).
The reason why we obtain the four different types of action is related to 
the fact that the generalized actions in the trace are quaternion valued 
and thus we should pick up particular combination in order to keep the 
generalized gauge invariance. 

For example the generalized Chern-Simons action 
$\frac{1}{2}{\cal A}Q{\cal A}+\frac{1}{3}{\cal A}^3 $, 
belongs to $\Lambda_-$ class and thus possesses the 
four different component types, the same types as in ${\cal A}$ of 
(\ref{eqn:ggf}).  
For example the ${\bf k}$-th component of the generalized Chern-Simons action 
is even dimensional bosonic component and thus the action 
\begin{equation}
 \displaystyle{S^e_{GCS}= \int_M \mbox{Tr}_{\bf k} 
        \left( \frac{1}{2}{\cal A}Q{\cal A}+\frac{1}{3}{\cal A}^3 \right)}
	\label{eqn:egcs}
\end{equation}
is even dimensional bosonic action while ${\bf j}$-th component of the 
generalized Chern-Simons action is odd dimensional bosonic component and 
thus the action 
\begin{equation}
 \displaystyle{S^o_{GCS}=\int_M \mbox{Str}_{\bf j}
        \left( \frac{1}{2}{\cal A}Q{\cal A}+\frac{1}{3}{\cal A}^3 \right)}
	\label{eqn:ogcs}
\end{equation}
is odd dimensional bosonic action. 
It is also possible to obtain fermionic generalized Chern-Simons actions 
by taking ${\bf 1}$-th and ${\bf i}$-th components. 
We then need to pick up $d$-form terms to obtain $d$ dimensional actions 
defined on a $d$-dimensional manifold $M$.
We can thus construct any dimensional generalized Chern-Simons actions.

On the other hand the generalized topological Yang-Mills action 
${\cal F} {\cal F}$ belongs to $\Lambda_+$ class and thus possesses the 
four different component types, the same types as in ${\cal V}$ of 
(\ref{eqn:ggp}).
For example the ${\bf 1}$-th component of the generalized topological 
Yang-Mills action is even dimensional bosonic component and thus the action 
\begin{equation}
 \displaystyle{S_{GYM}=\int_M \mbox{Str}_{\bf 1}
        \left( {\cal F} {\cal F} \right)} \label{eqn:gyma}
\end{equation}
is even dimensional bosonic action.
We can similarly construct odd dimensional bosonic action by taking 
${\bf i}$-th component of the generalized topological Yang-Mills action.

Important fact is that these generalized Chern-Simons actions and the 
topological Yang-Mills actions are invariant under the following 
generalized gauge transformations:
\begin{equation}
\delta {\cal A}=[Q+{\cal A} , {\cal V} ], 
\label{eqn:ggt}
\end{equation}
where ${\cal V}$ is the generalized gauge parameter defined by 
eq.(\ref{eqn:ggp}).
It should be noted that this symmetry is much larger than the usual gauge 
symmetry, in fact topological symmetry, since the gauge parameter ${\cal V}$ 
contains as many gauge parameters as gauge fields of various forms in 
${\cal A}$.

There is another suggestive topological nature due to the parallel 
construction with the standard gauge theory. 
In ordinary gauge theory the integral for the $n$-th power of the 
trace of curvature is called Chern character and has topological 
nature. 
In the generalized gauge theory it is possible to define generalized Chern 
character which is expected to have topological nature related to 
\lq\lq generalized index theorem"  
\begin{eqnarray} 
 {\mbox{Str}}_{\bf\mbox{1}}({\cal{F}}^n)
 &=&{\mbox{Str}}_{\bf\mbox{1}}(Q\Omega_{2n-1}), 
\label{eqn:bepo} \\
 {\mbox{Tr}}_{\bf\mbox{i}}({\cal{F}}^n)
 &=&{\mbox{Tr}}_{\bf\mbox{i}}(Q\Omega_{2n-1}), 
\label{eqn:bopo}
\end{eqnarray}
where $\Omega_{2n-1}$ is the \lq\lq generalized" Chern-Simons forms. 
Especially, for $n=2$ case in (\ref{eqn:bepo}), we obtain 
the topological Yang-Mills type action of (\ref{eqn:gyma}) related to 
the generalized Chern-Simons action with one dimension lower 
on an even-dimensional manifold $M$,
\begin{equation}
 \int_M{\mbox{Str}}_{\bf\mbox{1}}{\cal F}^2
 =\int_M{\mbox{Str}}_{\bf\mbox{1}}\Bigg( Q\Big({\cal{A}}Q{\cal{A}}
 +\frac{2}{3}{\cal A}^3\Big)\Bigg),  \label{eqn:2cc}
\end{equation}
which has the same form of the standard relation. 

In the case of the generalized Yang-Mills action the story is different.
In order to define the generalized Yang-Mills action we need to define 
the dual of the generalized curvature by defining Hodge star operation 
which breaks the gauge symmetry of the higher forms as will be explained 
later.
Correspondingly the generalized Yang-Mills action is not invariant 
under the generalized gauge transformation (\ref{eqn:ggt}).
If we, however, restrict to use the zero form gauge parameter then 
the action recovers gauge invariance. 
It is, however, important to recognize that the generalized Yang-Mills 
formulation is essentially the general realization of the 
non-commutative geometry formulation of gauge theory \`{a} la 
Connes\cite{Connes}. 
We can thus formulate the Weinberg-Salam model by our generalized Yang-Mills 
action which will be explained later.

\setcounter{equation}{0}
\section{Gravity on the Lattice}

\subsection{First Step towards the Generalized Chern-Simons Actions 
on the Lattice }    \label{First Step}

Here we first show the concrete expressions of generalized Chern-Simons 
actions in two, three and four dimensions. 
For generalized gauge fields and parameters we introduce the following 
notations:
\begin{eqnarray}
 {\cal A} & = &  {\mbox{\bf j}} A + {\mbox{\bf k}} \hat{A}   \nonumber \\ 
 &\equiv& {\mbox{\bf j}}(\omega + \Omega) + 
 {\mbox{\bf k}}( \phi + B  + \epsilon_1 H) \label{eqn:def-gf} \\
{\cal V} & = & {\mbox{\bf 1}} \hat{a} + {\mbox{\bf i}} a \nonumber \\ 
 &\equiv& {\mbox{\bf 1}}(v + \epsilon_1 b + V) + 
 {\mbox{\bf i}}( u +  U), \label{eqn:def-gp}
\end{eqnarray} 
where we have omitted the fermionic gauge fields $\psi, \hat{\psi}$ 
and gauge parameters $\alpha, \hat{\alpha}$ for simplicity.
Here zero, one, two, three, four form gauge fields and gauge parameters 
are denoted as 
$\phi, \omega, B, \Omega, H$ and $v, u, b, U, V$, respectively.

By substituting eq.(\ref{eqn:def-gf}) into $S^e_{GCS}$ and $S^o_{GCS}$ of 
eq.(\ref{eqn:egcs}) and eq.(\ref{eqn:ogcs}) and picking up two, three 
and four form part of the action, we can obtain more explicit 
expressions of our generalized Chern-Simons 
actions in two, three and four dimensions.
\begin{eqnarray}
S_2 &=& - \int \mbox{Tr}\{\phi(d\omega + \omega^2) + \phi^2 B\}, 
\label{S_2} \\
S_3 &=& - \int \mbox{Str}\{\frac{1}{2}\omega d\omega + \frac{1}{3}\omega^3 -
              \phi(d B + [\omega,B]) + \phi^2 \Omega \},\label{S_3}    \\
S_4 &=& - \int \mbox{Tr}\{ B(d\omega + \omega^2) + \phi(d \Omega + 
               \{\omega, \Omega\}) + 
             \phi B^2 + \phi^2 H \}. \label{S_4}
\end{eqnarray} 
These actions are invariant under the following gauge transformations:
\begin{eqnarray}
\delta \phi &=& [\phi,v],                      \nonumber \\
\delta \omega &=& dv + [\omega,v] - \{\phi,u\},    \nonumber \\
\delta B &=& du + \{\omega,u\} + [B,v] + [\phi,b],      \nonumber \\
\delta \Omega &=& db + [\omega,b] + [\Omega,v] - \{B,u\} + 
\{\phi,U\},\nonumber \\
\delta H &=& -dU-\{\omega,U\}+\{\Omega,u\}+[H,v]+[B,b]+[\phi,V],  
\label{eqn:com-ggt}
\end{eqnarray} 
which are obtained by substituting eqs.(\ref{eqn:def-gf}) and 
(\ref{eqn:def-gp})into the gneralized gauge transformations (\ref{eqn:ggt}).  

Equations of motion derived from the actions $S_2$, $S_3$ and $S_4$ are 
\begin{eqnarray}
\phi^2  \ &=& \ 0  ,         \nonumber \\
d \phi \, + \, [ \, \omega , \phi \, ] \ &=& \ 0  ,          \nonumber \\
d \omega \, + \, \omega^2 \, + \,  \{ \phi , B \} \ &=& \ 0  ,     \nonumber \\
d B \, + \, [ \, \omega , B \, ] \, + \, [ \, \Omega , \phi \, ] \ &=& \ 0 ,
 \nonumber \\
d \Omega \, + \, \{ \omega , \Omega \} 
          \, + \, B^2 \, + \, \{ \phi , H \} \ &=& \ 0  ,  
\label{eqn:com-eqmotions}
\end{eqnarray}
which are component-wise expressions of the vanishing generalized 
curvature ${\cal F}=0$ as equations of motion of the generalized Chern-Simons 
action, obtained by substituting (\ref{eqn:def-gf}) into (\ref{eqn:gcurv}). 

Classically we have already shown that gravity theories can be formulated 
by using the generalized Chern-Simons actions of two, three and four 
dimensions. 
In two dimensions $SL(2,R)$ topological gravity formulated by 
Verlinde and Verlinde\cite{Verlinde-Verlinde} can be formulated 
as a local gauge theory of the generalized Chern-Simons action by using 
Clifford algebra\cite{KW3}. 
In three dimensions the gravity theory formulated by generalized Chern-Simons 
action with three dimensional Super Poincar\`{e} algebra\cite{Yoshida} 
leads to the standard Chern-Simons gravity 
of Einstein-Hilbert action formulated by Witten\cite{Witten1}.   
Thus classically three dimensional Einstein gravity can be formulated by 
both the standard Chern-Simons action and the generalized Chern-Simons action. 
In four dimensions a topological conformal gravity can be formulated by 
the generalized Chern-Simons action with Conformal 
algebra\cite{KW3}.
In these analyses the zero form gauge field, as a classical solution of 
the equation motion, played a role of 
\lq\lq order parameter" of gravity phase.   
In other words the above mentioned gravity theories are realized 
when the zero form part of the classical solution vanishes.

Classically gravity theories in two, three and four dimensions are formulated 
by the generalized Chern-Simons actions. 
We have recently shown that the quantization of the generalized Chern-Simons 
actions can be carried out\cite{KSTU}. 
It turned out that the quantization of the generalized Chern-Simons action 
is highly nontrivial due to the infinite reducibility of the theories.
We will briefly explain the formulation of the quantization later.
We thus know how to quantize the above mentioned gravity theories.   

Since these gravity theories are formulated by the generalized 
Chern-Simons actions we naively expect that we may be able to formulate these 
gravity theories on the simplicial lattice manifold by simply putting 
$n$-form gauge fields on the $n$-simplexes of the manifold. 
The story is not so simple as we show the three dimensional example 
in the following.

\subsection{Lattice Chern-Simons Gravity }  \label{Action}

Hereafter we concentrate on the three dimensional gravity since the 
formulation of the three dimensional Einstein gravity is established by the 
standard Chern-Simons action. 
Firstly it should be noted that the generalized Chern-Simons action 
includes the standard Chern-Simons action as a part of the full action. 

To make the story simpler we consider how to formulate the standard 
Chern-Simons action, which is the part of the 
generalized Chern-Simons action composed of one form only, 
on the simplicial manifold.  

We consider a three-dimensional piece-wise linear simplicial manifold 
which is composed of tetrahedra.
In 3-dimensional Regge calculus it is known that curvature is concentrated 
on the links of tetrahedra\cite{Regge}.
We intend to formulate a lattice gravity theory in terms of gauge variables, 
dreibein $e$ and spin connection $\omega$.
In analogy with the lattice gauge theory where link variables surrounding a 
plaquette induce a gauge curvature, We have proposed 
the formulation 
that dual link variables $U(\tilde{l}) = e^{\omega(\tilde{l})}$ 
located at the boundary of a dual plaquette $\tilde P$ 
($\tilde{l} \in \partial\tilde{P}(l)$) 
associated to an original link $l$ 
induce the curvature of the gravity theory 
\cite{Kawamoto-Nielsen}\cite{Dadda}.
It was further pointed out that the dreibein $e^a(l)$ is located on the 
original link $l$.
In order to avoid sign complications we consider a Euclidean version of 
three-dimensional local 
Lorentz group $SO(3)$ but exclude $SU(2)$ case.
Here we explicitly construct a lattice Chern-Simons gravity by extending our 
previous formulation. 
The details of the formulation can be found in \cite{KNS}.

\begin{figure}[t]
\begin{center}
 \begin{minipage}[c]{0.4\textwidth}
 \epsfxsize=\textwidth \epsfbox{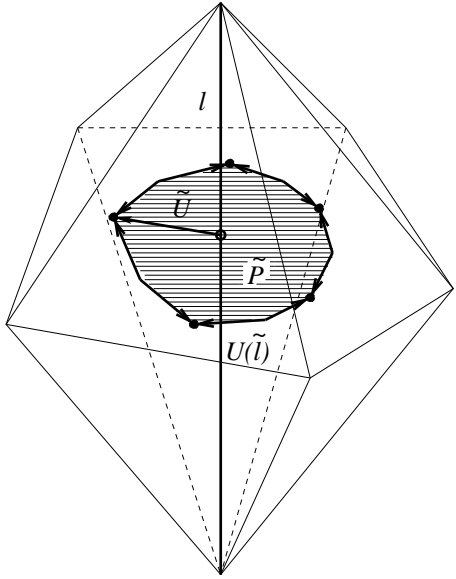} 
 \end{minipage}
\end{center}
 \caption{dual link variables on $\partial \tilde{P}$}
 \label{fig:action}
\end{figure}

Here we slightly modify the formulation given above in order that each 
tetrahedron gets independent contribution to the partition function 
and at the same time the orientability could be naturally accommodated.  
We divide the dual link, which connects the centers of neighboring 
tetrahedra, into two links by the center of mass of the common triangle 
of the neighboring tetrahedra.
We may keep to use the terminology of dual plaquette and dual link 
even for those modified plaquettes and links. 
Correspondingly we put different link variables $U$ for the doubled dual links.
We then assign the directions of $U$-links inward for each tetrahedron 
as shown in Fig.\ref{fig:action}. 

Using these variables,
we consider the following lattice version of $ISO(3)$ Chern-Simons gravity 
action on the simplicial manifold,
\begin{equation}
 \SLCS = \sum_{l} \epsilon_{abc} e^{a}(l)
  \Bigl[ \ln \plaq \Bigr]^{bc}, \label{LCS1} 
\end{equation}
where,
$\partial \tilde{P}(l)$ is a boundary of the $\tilde{P}(l)$,
which is a (dual) plaquette around the link $l$,
and $\displaystyle \plaq $ denotes the product of $U(\tilde{l})$ 
along $\partial \tilde{P}(l)$.
We define the ``curvature'' $F^{ab}(l)$ of the link $l$
by the following equation,
\begin{equation}
 \Bigl[ \plaq \Bigr]^{ab} \equiv \Bigl[ e^{F(l)} \Bigr]^{ab}.
\end{equation}
The leading term of $F$ with respect to the lattice unit is the 
ordinary curvature
$d\omega + \omega \wedge \omega$
similar to the ordinary lattice gauge theory.

Classically the Chern-Simons action impose a torsion free condition as an 
equation of motion. 
The torsion free nature is lost at the quantum level since we integrate 
out the dreibein and spin connection.
We now introduce the following vanishing holonomy constraint which relates 
the dreibein and 
spin connection even at the quantum level:
\begin{equation}
 \Bigl[ \plaq \Bigr]^{ab} e^b = e^a. \label{constraint}
\end{equation}
The dreibein $e^a$ associated to a original link may be parallel transported around 
the boundary of the dual plaquette $\partial \tilde{P}(l)$ to the original 
location and yet the direction of the dreibein should not be changed.
We may interpret this constraint as a gauge fixing condition 
of gauge diffeomorphism symmetry which we will explain 
later.
Due to the constraint the group $SO(3)$ becomes ``effectively abelian'', 
i.e. the direction of the rotation associated with the curvature is 
parallel to that of $e^a$.
This can be seen as follows: we can reduce the above constraint to
the following one:
\begin{equation}
  F^{ab} e^b = 0, \label{constraint2}
\end{equation}
hence $F^a \equiv \frac{1}{2} \epsilon ^{abc} F^{bc}$ is parallel
to $e^a$: $e^a \propto F^a$.

Here we should reconsider the constraint (\ref{constraint}). 
Firstly it should be noted that the $\displaystyle\Bigl[ \plaq \Bigr]^{ab}$ 
is an 
element of $SO(3)$ and thus the eigenvalue equation of this element always has 
eigenvalue +1. 
Thus the number of the independent constraints in eq.(\ref{constraint})   
is not three but two.
Taking into account the parallel and anti-parallel nature of $e^a$ and $F^a$ 
in the constraint, we can rewrite the correct constraint equation  
\begin{equation}
\frac{e^3}{|e|} 
\left[ \prod_{a=1}^{2}
 \delta \left( \frac{F^a}{|F|} + \frac{e^a}{|e|} \right)
+ \prod_{a=1}^{2}
\delta \left( \frac{F^a}{|F|} - \frac{e^a}{|e|} \right) \right], 
\label{constraint3}
\end{equation}
where $|e|$ and $|F|$ are length of $e^a$ and $F^a$, respectively.
The coefficient factor $\frac{e^3}{|e|}$ is necessary to keep the 
rotational invariance of the constraint relation, which 
can be easily checked by polar 
coordinate expression of the constraint relation.

Now we show that discreteness of the length of the dreibein $|e|$ 
comes out as a natural consequence of the specific choice of the lattice 
gauge gravity action.
We first introduce the following normalized matrix $I$,
\begin{equation}
 I \equiv I^{a} J_{a} , 
\quad  I^{a} \equiv \frac{F^{a}}{\sqrt{F^{a}F_{a}}},
\end{equation}
here $[J_{a}]_{bc} = i \epsilon_{abc}$ is the generator of $SO(3)$.
This matrix satisfies the following relation,
\begin{equation}
 e^{i \theta I} = 1 - I^2 (1-\cos \theta) + i I \sin \theta,
\end{equation}
then
\begin{equation}
 e^{i 2 \pi n I} = 1, \quad n \in \bm{Z} \label{periodicity1}.
\end{equation}
Using the above relation 
and $F^a \propto e^a$ by the constraint (\ref{constraint}),
we find that our lattice Chern-Simons action $\SLCS$ has 
the following ambiguity:
\begin{eqnarray*}
 \SLCS
  &=& \sum _{l} \epsilon _{abc} e^{a}(l)  
  \left[ \ln e^{F(l)} \right] ^{bc} \\
  &=& \sum _{l} \epsilon _{abc} e^{a}(l)  
  \left[ \ln e^{F(l) + i 2 \pi n I} \right] ^{bc} \\
  &=& \sum _{l} \bigl[ 2 e^{a}(l) F_a (l) + 4 \pi n |e(l)| \bigr] \\
  &=& \SLCS + \sum_{l} 4 \pi n |e(l)|,
\label{ambiguity}
\end{eqnarray*}
here $|e|$ is the length of $e^a$, $|e| \equiv \sqrt{e_a e^a}$.
This ambiguity leads to an ambiguity in the partition 
function
\begin{equation}
 Z = \int {\cal D}U {\cal D}e ~ e^{i S_{\hbox{\tiny LCS}}}
   = \int {\cal D}U {\cal D}e ~
   e^{ i S_{\hbox{\tiny LCS}} + i \sum_{l} 4 \pi n |e|}.
\end{equation}
Imposing the single valuedness of 
$e^{i S_{\hbox{\tiny LCS}}}$, we obtain the constraint 
that $\sum_{l} 2 |e(l)| $ should be integer, or equivalently $|e(l)|$ should 
be half integer.

\subsubsection{Gauge Invariance on the Lattice} \label{Gauge Invariance}

The gauge transformations of the continuum Chern-Simons gravity have 
been given by (\ref{CS gauge transformation}) which includes 
the local Lorentz gauge transformation (\ref{LLGT}) and 
the gauge transformation of diffeomorphism (\ref{DGT}).
We first note that  
the dreibein and the curvature defined in (\ref{CSCurvature}) 
transform adjointly under the local Lorentz gauge transformation  
\begin{equation}
\begin{array}{rcl}
 \delta e^a_\mu &=  -\epsilon^{abc}e_{\mu b}\tau_c,\\
 \delta F^a_{\mu\nu} &= -\epsilon^{abc}F^b_{\mu\nu}\tau_c. \label{LLGT2}
\end{array}
\end{equation} 

We consider that the lattice version of the local Lorentz gauge parameters
are sitting on the dual sites and the middle of the original links, 
the same point of the dreibein. 
For simplicity we consider here in this section that the 
dual link is not divided into two dual links by the center 
of original triangle.
Then the dual link variable $U(\tilde{l}) = e^{\omega(\tilde{l})}$ 
transforms under the lattice local Lorentz transformation as 
\begin{equation}
U(\tilde{l}) \rightarrow V^{-1} U(\tilde{l}) V', \label{LLLspin}
\end{equation}
where the gauge parameters $V$ and $V'$ are elements of $SO(3)$ 
and located at the end points of dual link $\tilde{l}$.
Defining the matrix form of the dreibein by 
$E^{cb}_\mu(l)=\epsilon^{abc}e_{a\mu}(l)$, 
we can rewrite the lattice Chern-Simons action (\ref{LCS1}) by 
\begin{equation}
 \SLCS = \sum_{l} \hbox{Tr} ( E(l)F(l)), \label{LCSA} 
\end{equation}
where $\displaystyle F(l)^{ab}=\Bigl[ \ln \plaq \Bigr]^{ab}$. 

Corresponding to the continuum local Lorentz transformation, we 
can define the lattice version of local Lorentz transformation of
$E(l)$ and $F(l)$ according to (\ref{LLLspin}) 
\begin{equation}
 \begin{array}{ccc}
     E(l) & \rightarrow & V^{-1}E(l) V,  \\  
     F(l) & \rightarrow & V^{-1}F(l) V.
\end{array}
\label{LLLGT}
\end{equation}
It is obvious that the lattice Chern-Simons action (\ref{LCSA}) 
is invariant under the lattice local Lorentz transformation.


The continuum Chern-Simons gravity action is invariant under the 
gauge transformation of diffeomorphism (\ref{DGT}) which transforms 
dreibein $e^a_\mu$ but not spin connection $\omega^a_\mu$. 
We can show that the lattice Chern-Simons action is invariant under 
the lattice version of gauge diffeomorphism by formulating the 
lattice version of Bianchi identity\cite{KNS}.

We now point out that the constraint (\ref{constraint}) or equivalently 
(\ref{constraint2}) breaks the lattice gauge diffeomorphism while 
the lattice Chern-Simons action itself is invariant, as is shown above.  
The lattice dreibein is transformed but the lattice curvature is not 
transformed under the lattice gauge transformation of the diffeomorphism. 
The precise expression of the constraint (\ref{constraint3}) tells us that 
the dreibein $e^a$ can be rotated by using two gauge parameters of the 
gauge transformation of diffeomorphism to be parallel or anti-parallel to 
the curvature $F^a$. 
The length of the dreibein is discretized and thus the third gauge parameter 
can be exhausted. 
In this sense we can identify the equivalent constraint,
(\ref{constraint}), (\ref{constraint2}), (\ref{constraint3}) as
a gauge fixing condition of 
the lattice gauge transformation of diffeomorphism.

\subsubsection{Calculation of Partition Function} \label{Integration}

In the previous section we have found that the length of dreibein 
is discretized to half integer for $SO(3)$.
Taking into account the discreteness of the dreibein, 
the total partition function leads 
\begin{eqnarray}
 Z &=& \int {\cal D}U  \prod_{l} Z_{l}, \\ 
 Z_{l} &=& \int d^3 e ~\frac{e^3}{|e|}
  \left[ \prod_{a=1}^{2}
   \delta \left( \frac{F^a}{|F|} + \frac{e^a}{|e|} \right)
  + \prod_{a=1}^{2}
  \delta \left( \frac{F^a}{|F|} - \frac{e^a}{|e|} \right)
  \right] \nonumber \\ 
 && \hspace*{1cm} \times
  \frac{1}{2}\sum_{J=0}^{\infty} \delta \left( |e| - \frac{J}{2} \right)
  e^{2 i e^a F^a},
\end{eqnarray}
where $Z_{l}$ is the partition function
associated with a link $l$.

{\bf 4.2.2.1 $e$ integration}\\

Due to the rotational invariance of the constraints, we can take 
$e^3$ as the third direction of local Lorentz frame without loss of 
generality.
We can then evaluate $e^a$ integral of $Z_{l}$ immediately
thanks to the delta functions
\begin{eqnarray*}
 Z_{l}
  &=& \int d^3 e ~ |e|^2 \frac{e^3}{|e|}
  \left[ \prod_{a=1}^{2}
   \delta \left( e^a + |e| \frac{F^a}{|F|} \right)
  + \prod_{a=1}^{2}
  \delta \left( e^a - |e| \frac{F^a}{|F|}\right)
  \right] 
  \frac{1}{2}\sum_{J} \delta \left( |e| - \frac{J}{2} \right)
  e^{2 i e^a F^a} \\
  &=& \frac{1}{2}\sum_{J} \left( \frac{J}{2} \right)^2
  \left( e^{2i\frac{J}{2} |F|} + e^{-2i\frac{J}{2} |F|} \right) \\
  &=& \sum_{J} \frac{1}{4} J^2 \cos(J|F|).
\end{eqnarray*}
Using the following formula for the character $\chi_J$
of the spin-$J$ representation of $SO(3)$,
\begin{equation}
 \chi _{J} (e^{i \theta^a J_a})
  = \chi _{J} (|\theta|)
  = \frac{\sin\left( (2J+1) \frac{|\theta|}{2} \right)}
  {\sin \left(\frac{|\theta|}{2} \right)},
\end{equation}
where $|\theta|$ is the length of $\theta^a$,
we find 
\begin{eqnarray}
 \chi _{J}(|F|) - \chi _{J-1}(|F|)
 &=& 2 \cos(J|F|). 
\end{eqnarray}
Hence we can naively calculate the link partition function,
\begin{eqnarray*}
 Z_{l}
  &=& \sum_{J=1}^{\infty} \frac{1}{8} J^2 (\chi _{J} - \chi _{J-1}) \\
  &=& -\frac{1}{8}
      \sum_{J=0}^{\infty} (2J+1) \chi _{J}.
\end{eqnarray*}
This calculation is not precise, because 
the summation is not convergent.
We need to show that there is a regularization procedure which leads to 
a validity of the above calculation after the regularization.

It is possible to give the similar formulation by using Heat Kernel 
regularization. 
Then the action leads
\begin{eqnarray}
 Z_{l} &=& 
  - \frac{1}{8} \sum_{J} (2J+1)\chi _J e^{-J(J+1)t}.
\end{eqnarray}
This regularization factor $e^{-J(J+1)t}$ breaks
the Alexander move invariance of the partition function but 
it will be recovered at the end of the calculation when 
we take the limit $t \rightarrow 0$. 

 {\bf 4.2.2.2 $U$ integration}\\

After $e^a$ integration and dividing the unimportant constant factor 
$\prod_{l} (-1/8)$, the partition function leads 
\begin{equation}
 Z = \int {\cal D}U \prod_{l} \sum_{J=0}^{\infty} \left(2J+1\right)
  \chi_{J}\Bigl(\plaq \Bigr) ~ e^{-J(J+1)t}, 
\end{equation}
where we take $t\rightarrow 0$ limit in the end of calculation.
We now carry out $DU$ integration of this partition function. 
Thanks to the character of the partition function, $DU$ integration is 
straightforward. 
We show that the Ponzano-Regge partition function will be reproduced after 
$DU$ integration with 6-$j$ symbols together with correct coefficients and 
sign factors. 

Before getting into the details we figure out how 6-$j$ symbols appear. 
The character in the partition function is a product of $D$-function
around the boundary of dual plaquette associated to a original link. 
Each tetrahedron has six original links and there are two dual links which 
is a part of a product on the boundary of the dual plaquette associated 
to each original link.
In other words three dual links associated to a $DU$ integration thrust into 
each triangle from the center of the tetrahedron. 
Therefore twelve dual links are associated to a tetrahedron.
Each $DU$ integration of the product of three $D$-function reproduces two 
3-$j$ symbols, thus we get eight 3-$j$ symbols for each tetrahedron. 
Four out of eight 3-$j$ symbols lead to a 6-$j$ symbol and the rest of four 
3-$j$ symbols lead to give a trivial factor together with the 3-$j$ symbols 
from the neighboring tetrahedra.

We first note that the character appearing in the partition function is 
a product of $D$-functions
\begin{equation}
 \chi_J(|F|) = \chi_J \Bigl(\plaq \Bigr)
             = D^J_{m_1m_2}(U_1)D^J_{m_2m_3}(U_2)\cdots 
               D^J_{m_km_1}(U_k),
\end{equation}
where $U_i$ is a dual link variables on the boundary of dual plaquette 
$\tilde P(l)$ associated to a link $l$ and $m_i$ is the third component of 
angular momentum $J$ which is assigned to the link $l$. 
As we have already pointed out that the direction of $U_i$ for each link 
is defined inward for each tetrahedron. 
On the other hand the direction of the loop composed of the product of 
dual links associated to the link $l$ can be chosen arbitrarily. 
Therefore some of $U_i$ in the above $D$-functions are $U_i^\dagger$. 
If the original link $l$ is a link of a particular tetrahedron, 
two $D$-functions out of the above product are located inside the 
tetrahedron.

\begin{figure}
\begin{center}
 \begin{minipage}[b]{0.4\textwidth}
 \epsfxsize=\textwidth \epsfbox{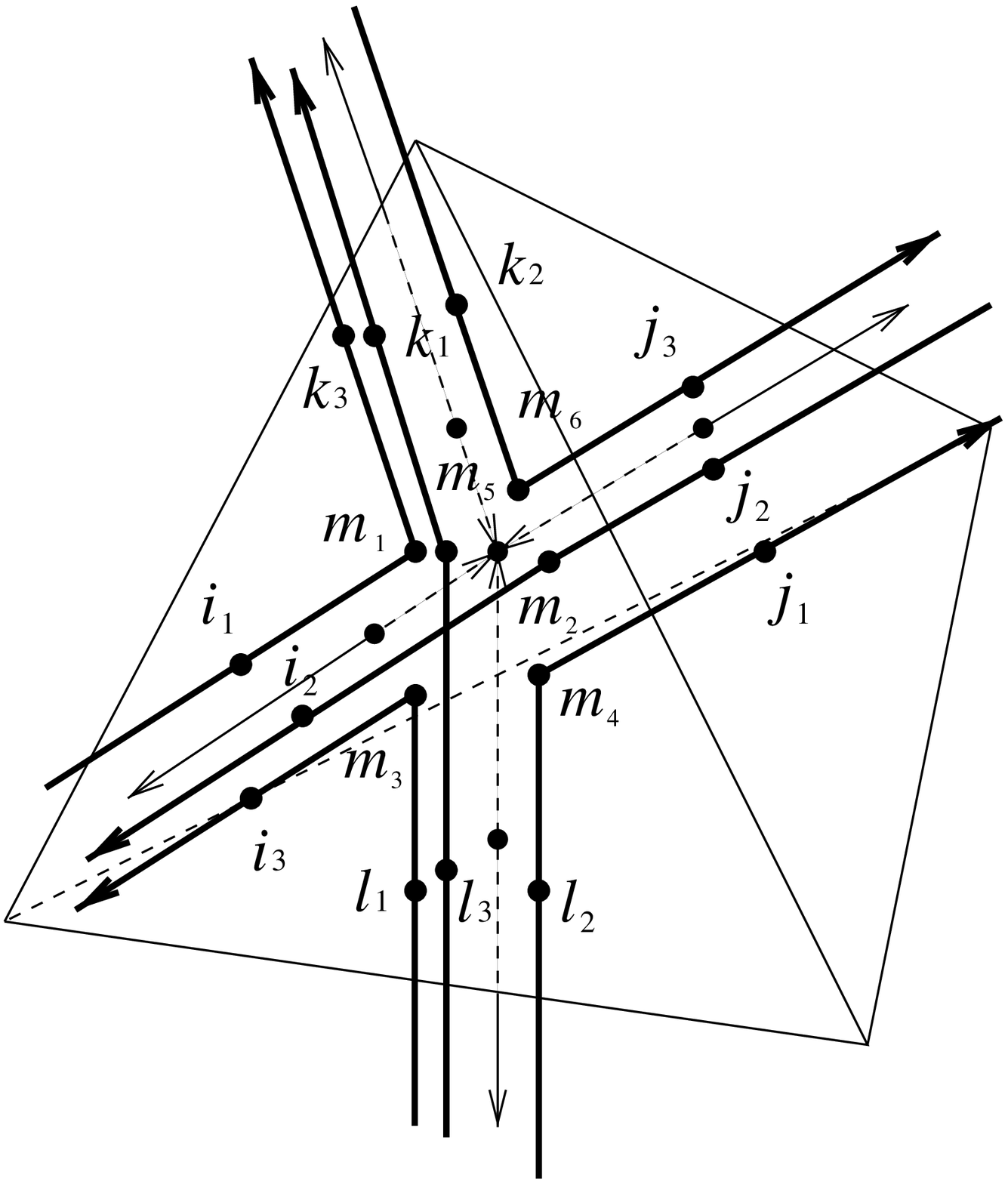} 
 \end{minipage}
 \hspace*{5mm}
 \begin{minipage}[b]{0.4\textwidth}
 \epsfxsize=\textwidth \epsfbox{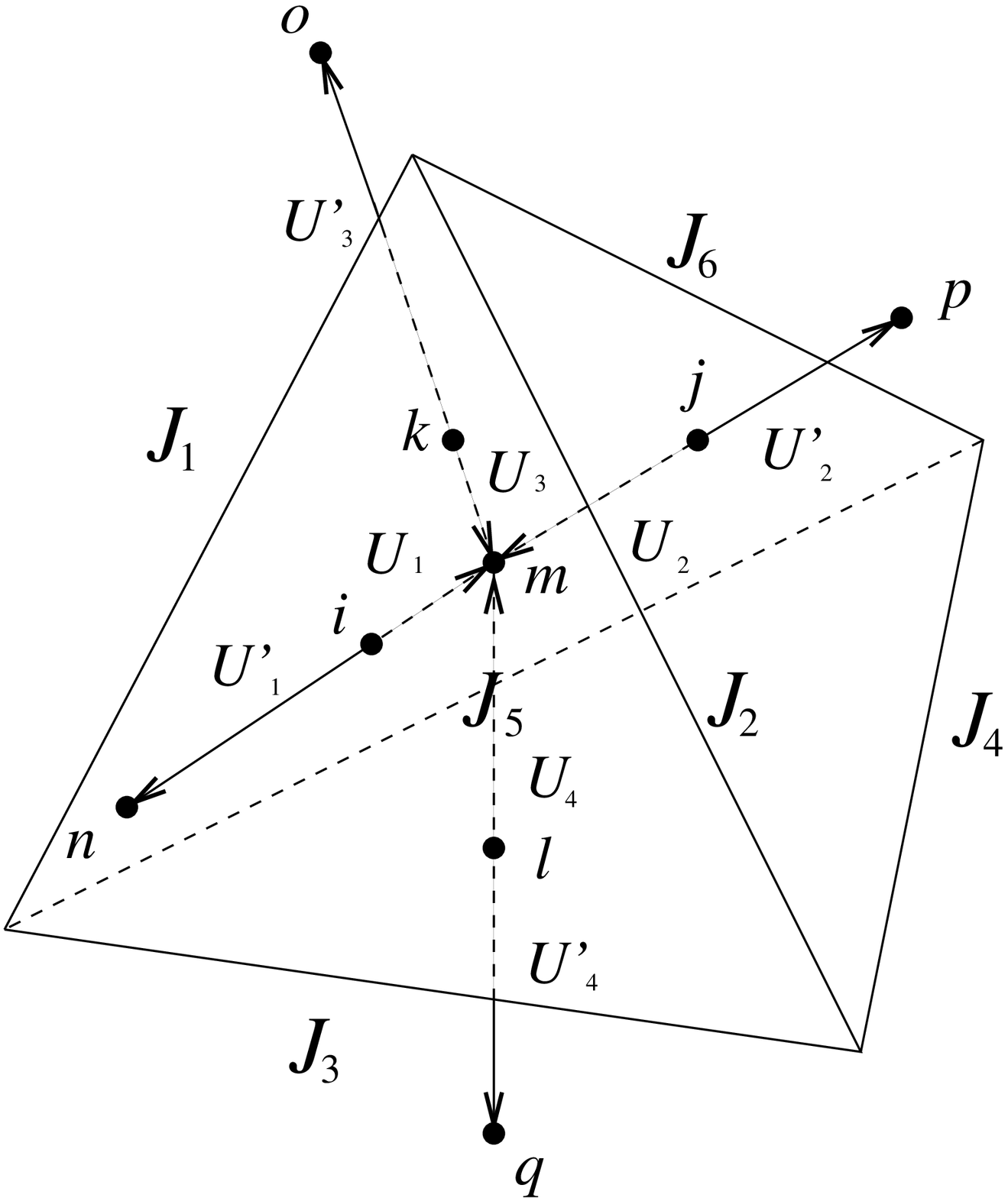} 
 \end{minipage}
\end{center}
 \caption{dual links related to neighboring tetrahedra and the orientability}
 \label{fig:U-int}
\end{figure}

We now choose a particular situation which is shown in Fig.\ref{fig:U-int}. 
The twelve $D$-functions associated to this particular tetrahedron are  
\begin{eqnarray*}
&& I_{U_1U_2U_3U_4} = \int \prod_{i=1}^{4} DU_i ~
  D^{J_1}_{i_1m_1}(U_1) D^{J_1}_{m_1k_3}(U_3^{\dagger}) \cdot
  D^{J_2}_{j_2m_2}(U_2) D^{J_2}_{m_2i_2}(U_1^{\dagger}) \\
&& \quad\quad\quad\quad \times
  D^{J_3}_{l_1m_3}(U_4) D^{J_3}_{m_3i_3}(U_1^{\dagger}) \cdot
  D^{J_4}_{l_2m_4}(U_4) D^{J_4}_{m_4j_1}(U_2^{\dagger}) \\
&& \quad\quad\quad\quad \times
  D^{J_5}_{l_3m_5}(U_4) D^{J_5}_{m_5k_1}(U_3^{\dagger}) \cdot
  D^{J_6}_{k_2m_6}(U_3) D^{J_6}_{m_6j_3}(U_2^{\dagger}).
\end{eqnarray*}
We pick up the $D$-functions associated to $DU_1$ integration 
\begin{equation}
 I_{U_1}= (-)^{i_2-m_2+i_3-m_3}
 \int DU_1 D^{J_1}_{i_1m_1}(U_1)D^{J_2}_{-i_2-m_2}(U_1)
               D^{J_3}_{-i_3-m_3}(U_1), 
\end{equation}
where we have used the following formula to rewrite only with $U_1$ variable:
\begin{equation}
 D^I_{mn} (U^{\dagger}) =
 D^{I *}_{nm} (U) = (-)^{n-m} D^I_{-n-m} (U) 
  \label{D conjugate}.
\end{equation}
We can now use the formula relating the integration of
three $D$-functions and two 3-$j$ symbols given in (\ref{eqn:3d3j}) 
and obtain 
\begin{equation}
  I_{U_1}= (-)^{i_2-m_2+i_3-m_3} \threej{J_1}{J_2}{J_3}{m_1}{-m_2}{-m_3}
                        \threej{J_1}{J_2}{J_3}{i_1}{-i_2}{-i_3}. 
\end{equation}
After carrying out $DU_2DU_3DU_4$ integration, we obtain
\begin{eqnarray*}
I_{U_1U_2U_3U_4}
&=&
  (-)^{i_2-m_2+i_3-m_3} \threej{J_1}{J_2}{J_3}{m_1}{-m_2}{-m_3}
                        \threej{J_1}{J_2}{J_3}{i_1}{-i_2}{-i_3} \\
&\times&  (-)^{j_1-m_4+j_3-m_6} \threej{J_4}{J_2}{J_6}{-m_4}{m_2}{-m_6}
                        \threej{J_4}{J_2}{J_6}{-j_1}{j_2}{-j_3} \\
&\times&  (-)^{k_3-m_1+k_1-m_5} \threej{J_1}{J_5}{J_6}{-m_1}{-m_5}{m_6}
                        \threej{J_1}{J_5}{J_6}{-k_3}{-k_1}{k_2} \\
&\times&  \hspace{3cm}        \threej{J_4}{J_5}{J_3}{m_4}{m_5}{m_3}
                        \threej{J_4}{J_5}{J_3}{l_2}{l_3}{l_1}.
\end{eqnarray*}
We now use the formula given in (\ref{eqn:6j3j}) which relates 6-$j$ symbol and 
four 3-$j$ symbols which carry $m_i$ suffices associated to 
the center of the tetrahedron.
 We then find 6-$j$ symbols after $DU_1DU_2DU_3DU_4$ integration
\begin{eqnarray}
I_{U_1U_2U_3U_4}
&=&
  (-)^{\sum_{i=1}^{6} J_i} \sixj{J_1}{J_2}{J_3}{J_4}{J_5}{J_6} \nonumber \\
&\times&
  (-)^{i_2+i_3} \threej{J_1}{J_2}{J_3}{i_1}{-i_2}{-i_3} 
  (-)^{j_3+j_1} \threej{J_4}{J_2}{J_6}{-j_1}{j_2}{-j_3} \nonumber \\
&\times&
  (-)^{k_3+k_1} \threej{J_1}{J_5}{J_6}{-k_3}{-k_1}{k_2}
                 \threej{J_4}{J_5}{J_3}{l_2}{l_3}{l_1}.
                 \label{6jwith3j}
\end{eqnarray}

Here we are considering $SO(3)$ case then the third component of the 
angular momentum $m_i$ is integer and thus we can use the relation 
$(-)^{m_i}=(-)^{-m_i}$.
We now look at the rest of the 3-$j$ symbols in 
eq.(\ref{6jwith3j}) which carry the 
suffices $i,j,k,l$. 
As we can see from Fig.\ref{fig:U-int}
that $DU_1$ integration reproduces two 
3-$j$ symbols and one of them associated to the suffices $m_k$ is 
absorbed to reproduce the 6-$j$ symbol and the other 3-$j$ symbol carrying the 
suffices $i_k$ could be combined with another 3-$j$ symbol obtained from 
$DU'_1$ integrations of the neighboring tetrahedron. 
Those 3-$j$ symbols are associated to the boundary triangle of the 
two neighboring tetrahedra carrying suffix $i_k$. 
In this particular case of Fig.\ref{fig:U-int} we obtain the following two 3-$j$ symbols
\begin{eqnarray*}
I^b_{i}
=
  \sum_{i_1i_2i_3}(-)^{i_1+i_2+i_3} 
   \threej{J_1}{J_2}{J_3}{i_1}{-i_2}{-i_3} 
   \threej{J_1}{J_2}{J_3}{i_1}{-i_2}{-i_3}. 
\end{eqnarray*}
Since the three angular momentum vectors $J_1,J_2,J_3$ construct the 
boundary triangle, the third components satisfy the relation 
$i_1-i_2-i_3=0$.
Using the following formula:
\begin{equation}
 \sum_{m_1m_2m_3}
 \threej{J_1}{J_2}{J_3}{m_1}{m_2}{m_3}
 \threej{J_1}{J_2}{J_3}{m_1}{m_2}{m_3} = 1, 
\end{equation}
and noting $(-)^{i_1+i_2+i_3}=(-)^{i_1-i_2-i_3}=1$ for $SO(3)$ case, 
these two 3-$j$ symbols lead to a trivial factor. 
Thus all the terms together lead to the Ponzano-Regge partition function.

\subsubsection{The Continuum Limit of the Lattice Chern-Simons Gravity }

We have explicitly shown that the partition function of the $ISO(3)$ 
lattice Chern-Simons action exactly coincides with the Ponzano-Regge model 
after the integration of the dreibein and the dual link variables.
The discreteness of the length of the dreibein is the natural consequence 
of the logarithm form in the lattice Chern-Simons action.
On the simplicial lattice manifold constructed from tetrahedra, 
the drei beins are located on the original links while the lattice version of 
the spin connection, the dual link variables are located on the dual links.

Since the Ponzano-Regge model is invariant under the 2-3 and 1-4 Alexander 
moves, the partition function is invariant under how the three dimensional 
space is divided into small pieces by tetrahedra.
It is natural to expect that the partition function is invariant in the 
continuum limit and thus the lattice Chern-Simons action leads to the 
continuum Chern-Simons action. 

In the $ISO(3)$ lattice Chern-Simons action there are 6 gauge parameters. 
Two gauge parameters of the lattice gauge diffeomorphism can be used to 
rotate the dreibein $e^a$ to be parallel or anti-parallel to the curvature 
$F^a$ and the remaining one gauge parameter of the lattice gauge 
diffeomorphism can be exhausted to make the length of the dreibein discrete. 
There remain three gauge parameters of the lattice local Lorentz gauge 
symmetry, which are expected to convert into the three vector parameters 
of general coordinate diffeomorphism symmetry. 
There are two reasons to expect this scenario. 
Firstly the lattice action coincides with the Ponzano-Regge model 
which is Alexander move invariant and is thus expected to be metric 
independent. 
In fact the lattice Chern-Simons action in the continuum limit is metric 
independent since it is composed of one form.  
Secondly the general coordinate transformation of diffeomorphism and the 
local Lorentz transformation are on shell equivalent in the continuum 
$ISO(3)$ Chern-Simons gravity\cite{Witten1}.  

It is interesting to recognize that the drei bein and spin connection are 
located on the original and dual links, respectively. 
This geometrical dual nature will be the reflection of the algebraic 
dual nature of the $ISO(3)$ Chern-Simons gravity where the abelian momentum 
generators $P_a$ 
and the angular momentum generator $J_a$ are algebraically dual to 
each other\cite{Ashtekar-Romano}.

\setcounter{equation}{0}
\section{Quantization of Generalized Gauge Theory}

Since the generalized gauge theory has nontrivial algebraic structures, 
we need to understand the origin of these algebras. 
For example the quaternion structure plays a very crucial role to treat 
the generalized fields and parameters in a uniform way.
We expected that the quaternion structure will be essentially related 
with the algebraic structure of the quantization of the theories. 
This expectation has turned out to be true.
We found several other interesting features of the generalized gauge theory 
in the quantization procedure.
Here in this section we briefly summarize how to quantize the generalized 
Chern-Simons action\cite{KSTU}. 
In particular we concentrate on the quantization of even dimensional 
generalized Chern-Simons actions. 
To be concrete for the notations we focus on the four dimensional case here.
The details of the quantization of two dimensional model is given in 
\cite{KSTU}.
The extension to odd dimensions goes quite parallel to the even dimensional 
case. 

We consider to quantize the even dimensional generalized Chern-Simons 
action 
\begin{equation}
 \displaystyle{S^e_{GCS} = \int_M \mbox{Tr}_{\bf k}  
        \left( \frac{1}{2}{\cal A}Q{\cal A}+\frac{1}{3}{\cal A}^3 \right)},
	\label{eqn:egcs2}
\end{equation}
where the explicit component wise notations for the classical gauge fields 
and gauge parameters are given in (\ref{eqn:def-gf}) and (\ref{eqn:def-gp}).

An important characteristic of the generalized Chern-Simons action is 
the infinite reducibility of the action which can be stated as follows. 
We extend the generalized gauge field and parameter to accommodate the 
ghost and the ghost for ghost $\cdots$ as
\begin{eqnarray}
 {\cal{V}}_{2n} & = & {\mbox{\bf j}} \left(u_{2n} + 
                    U_{2n}\right) 
                    + {\mbox{\bf k}} \left( v_{2n} +  
                    b_{2n} + V_{2n}\right)  
                     \  \in \Lambda_{-},                
\label{eqn:Ve4}\\
 {\cal{V}}_{2n+1} & = &  {\mbox{\bf 1}} \left( v_{2n+1} + 
                b_{2n+1} +  V_{2n+1}\right)  
               - {\mbox{\bf i}}\, \left(u_{2n+1} + 
                 U_{2n+1} \right) \  \in \Lambda_{+}, 
\label{eqn:Vo4}\\
                & &  \hspace{9cm} n = 0, 1, 2, \cdots, \nonumber
\end{eqnarray}
where ${\cal V}_0 \equiv {\cal A}$ and ${\cal V}_1 \equiv {\cal V}$ 
and the parameters with $n\ (>1)$ correspond the $n$-th reducibility 
parameters. 
The infinite reducibility of the generalized Chern-Simons actions 
can be seen by the transformation of ${\cal{V}}_n$
\begin{equation}
  \delta {\cal{V}}_n  = 
             (-)^n [ \ Q + {\cal{A}} \ , \ {\cal{V}}_{n+1} \ ]_{(-)^{n+1}},
                            \qquad  n = 0,1,2,\cdots 
\label{eqn:ir}
\end{equation}
which satisfies the on-shell relation
\begin{eqnarray}
  \delta(\delta{\cal{V}}_n) & = &
  \delta {\cal{V}}_n\Bigr\vert_{{\cal V}_{n+1}\rightarrow 
                                   {\cal V}_{n+1}+\delta{\cal V}_{n+1}} 
                                   - \delta {\cal{V}}_n \nonumber \\
             & = & 0,                     
\label{eqn:irr}
\end{eqnarray}
where we have used the equation of motion of generalized Chern-Simons 
action.
In the above equations, $[ \ , \ ]_{(-)^n}$ is a commutator for odd $n$ 
and an anticommutator for even $n$.
Since the transformation (\ref{eqn:ir}) for $n=0$ represents the gauge 
transformation, eq.(\ref{eqn:irr}) implies that the gauge transformation
is infinitely reducible.

In order to quantize this kind of system we need to introduce infinite 
series of ghosts as generalized fields 
\begin{eqnarray}
& & C_n, \ \ C_{n\mu}, \ \ \frac{1}{2!}C_{n\mu\nu}, \ \ 
      \frac{1}{3!}C_{n\mu\nu\rho}, \ \ \frac{1}{4!}C_{n\mu\nu\rho\sigma},
\label{eqn:4d-cnc} \\
& & \hspace{8cm} n=0, \pm 1, \pm 2, \cdots, \pm\infty, \nonumber 
\end{eqnarray}
where the index $n$ indicates the ghost number of the fields and 
the fields with even (odd) ghost number are bosonic (fermionic).
The fields with ghost number 0 are the classical fields
$$
C_0=\phi, \ \ C_{0\mu}=\omega_\mu, \ \ 
  C_{0\mu\nu}=B_{\mu\nu}, \ \ 
      C_{0\mu\nu\rho}=\Omega_{\mu\nu\rho}, \ \ 
         C_{0\mu\nu\rho\sigma}=H_{\mu\nu\rho\sigma}.
$$
Then we redefine a generalized gauge field
\begin{eqnarray}
\widetilde{{\cal A}} &=& {\mbox{\bf 1}}\psi + {\mbox{\bf i}} \hat{\psi} +  
            {\mbox{\bf j}} A + {\mbox{\bf k}} \hat{A} 
            \ \in \Lambda_-, 
\label{eqn:4d-ggf} \\
&&\psi  =  \sum_{n = -\infty}^{\infty} \Big(C_{2n+1 \mu} dx^{\mu}+
                \frac{1}{3!}C_{2n+1\mu\nu\rho}
                 dx^\mu\wedge dx^\nu\wedge dx^\rho \Big),
                                                             \nonumber \\
&&\hat{\psi} =  \sum_{n = -\infty}^{\infty}  \Big( C_{2n+1} + 
             \frac{1}{2!} C_{2n+1 \mu \nu} dx^{\mu}\wedge dx^{\nu} +
             \frac{1}{4!}C_{2n+1 \mu\nu\rho\sigma}
                    dx^\mu\wedge dx^\nu\wedge dx^\rho\wedge dx^\sigma\Big),
                                                              \nonumber \\
&&A  =  \sum_{n = -\infty}^{\infty} \Big(C_{2n \mu} dx^{\mu}+
                \frac{1}{3!}C_{2n\mu\nu\rho}
                      dx^\mu\wedge dx^\nu\wedge dx^\rho \Big),
                                                             \nonumber \\
&&\hat{A} =  \sum_{n = -\infty}^{\infty}  \Big( C_{2n} + 
             \frac{1}{2!} C_{2n \mu \nu} dx^{\mu}\wedge dx^{\nu} +
             \frac{1}{4!}C_{2n \mu\nu\rho\sigma}
                      dx^\mu\wedge dx^\nu\wedge dx^\rho\wedge dx^\sigma\Big),
                                                              \nonumber 
\end{eqnarray}
where we have explicitly shown the differential form dependence.

Using the above definitions of generalized gauge fields, we 
define the following extended generalized action which has 
the same form as the original one 
\begin{equation}
\displaystyle{S_{min}=
        \int  \ {\mbox{Tr}}^0_{\mbox{\bf {k}}} 
             \left( \frac {1}{2}\widetilde{{\cal A}}Q\widetilde{{\cal A}} 
                + \frac {1}{3} \widetilde{{\cal A}}^3 \right)},
\label{eqn:mina2}
\end{equation}
where the upper index 0 on ${\mbox{Tr}}$ indicates to pick 
up only the part with ghost number 0.
We can then show that this action satisfies the master equation 
of antibracket formalism \`{a} la Batalin and Vilkovisky\cite{BV}. 

In the construction of Batalin and Vilkovisky, ghosts, ghosts for ghosts$\cdots$ 
and the corresponding antifields are introduced according to the reducibility 
of the theory.
We denote a minimal set of fields by $\Phi^A$ which include classical fields 
and ghost fields, and the corresponding  antifields by $\Phi_A^{\ast}$.
If a field has  ghost number $n$, its antifield has ghost number $-n-1$.
Then a minimal action is obtained by solving the master equation
\begin{eqnarray}
  (S_{min}(\Phi , \Phi^*),S_{min}(\Phi , \Phi^*)) &=& 0,
\label{eqn:me} \\
  (X,Y)&=&{\partial_r X \over \partial \Phi^A}
          {\partial_l Y \over \partial \Phi_A^*}
          -{\partial_r X \over \partial \Phi_A^*}
          {\partial_l Y \over \partial \Phi^A},
\label{eqn:ab}
\end{eqnarray}
with the boundary conditions
\begin{eqnarray}
  S_{min} \Bigr\vert_{\Phi^{\ast}_A = 0} & = & S_0, 
\label{eqn:bbc} \\
  \frac{\partial S_{min}}{\partial \Phi^{\ast}_{a_n}} 
                             \Bigr\vert_{\Phi^{\ast}_A = 0} 
           & = & Z^{a_n}_{a_{n+1}} \Phi^{a_{n+1}}, 
\label{eqn:bbd} \hspace{2cm} n = 0,1,2,\cdots, 
\end{eqnarray}
where $S_0$ is the classical action and $Z^{a_n}_{a_{n+1}} \Phi^{a_{n+1}}$ 
represents the $n$-th reducibility transformation where the reducibility 
parameters are replaced by the corresponding ghost fields.

The BRST transformation and the nilpotency of the transformation can be 
shown by using generalized fields 
\begin{eqnarray*}
 \delta_{\lambda} \widetilde{{\cal A}} 
   &\equiv& s \widetilde{{\cal A}} \lambda 
    = -\widetilde{{\cal F}} \ {\mbox{\bf i}} \lambda, \\
 s^2\widetilde{{\cal A}}\lambda_2\lambda_1
 &\equiv& \delta_{\lambda_2} \delta_{\lambda_1} \widetilde{{\cal A}} 
   =  - [ \ Q + \widetilde{{\cal A}} \ , \ \widetilde{{\cal F}} \ ] 
                               \lambda_2 \lambda_1 = 0,
\end{eqnarray*}
where the last relation holds due to the generalized Bianchi identity. 
We can then show that $S_{min}$ satisfies the master equation 
$$
  \delta_{\lambda} S_{min} 
    = (S_{min},S_{min})_{\lambda,{\bf k}}
    = ( S_{min} , S_{min} ) \cdot \lambda = 0,
$$
where $(~,~)$ is the original antibracket defined by 
(\ref{eqn:me}) with the following identifications of antifields:
\begin{eqnarray}
\frac{1}{4!}\epsilon^{\mu\nu\rho\sigma}C_{-2n+1 \mu\nu\rho\sigma}
                            &=&C^*_{2(n-1)}, \hspace{1.5cm}
\frac{1}{4!}\epsilon^{\mu\nu\rho\sigma}C_{-2n \mu\nu\rho\sigma}
                            =-C^*_{2n-1}, \nonumber \\
\frac{1}{3!}\epsilon^{\mu\nu\rho\sigma}C_{-2n+1 \nu\rho\sigma}
                            &=&C^{\mu*}_{2(n-1)}, \hspace{1.5cm}
\frac{1}{3!}\epsilon^{\mu\nu\rho\sigma}C_{-2n \nu\rho\sigma}
                            =C^{\mu*}_{2n-1}, \nonumber \\
\frac{1}{2!}\epsilon^{\mu\nu\rho\sigma}C_{-2n+1 \rho\sigma}
                            &=&-C^{\mu\nu*}_{2(n-1)}, \hspace{1.5cm}
\frac{1}{2!}\epsilon^{\mu\nu\rho\sigma}C_{-2n \rho\sigma}
                            =-C^{\mu\nu*}_{2n-1}, 
\label{eqn:4d-idantif} \\
\epsilon^{\mu\nu\rho\sigma}C_{-2n+1 \sigma}
                            &=&C^{\mu\nu\rho*}_{2(n-1)}, \hspace{1.5cm}
\epsilon^{\mu\nu\rho\sigma}C_{-2n \sigma}
                            =C^{\mu\nu\rho*}_{2n-1}, \nonumber \\
\epsilon^{\mu\nu\rho\sigma}C_{-2n+1}
                            &=&C^{\mu\nu\rho\sigma*}_{2(n-1)}, \hspace{1.5cm}
\epsilon^{\mu\nu\rho\sigma}C_{-2n }
                            =-C^{\mu\nu\rho\sigma*}_{2n-1}. \nonumber 
\end{eqnarray}

In order to fix gauge completely we need to introduce the 
non-minimal action with the proper choice of gauge fermion.
We can then eliminate the antifields via the chosen gauge fermion. 
We just mention here that all these procedures can be completed 
properly\cite{KSTU}. 
We can thus complete the quantization of the generalized 
Chern-Simons action in even dimensions. 

\setcounter{equation}{0}
\section{Generalized Yang-Mills Theory}

Interesting applications of the generalized gauge theory would be to 
find the realistic correspondence with the known formulations. 
Needless to say that the Yang-Mills action has been playing a crucial role 
to describe the three out of four fundamental forces of our nature and may 
play again an important role in describing unified theory of all the four 
fundamental forces including gravity.

Here in this section we concretely show that the generalized Yang-Mills 
action is fundamentally related to the super symmetry and interpretation of 
matter fermions\cite{Kawamoto-Tsukioka} and non-commutative geometry 
formulation of gauge theory\cite{KTU}.

\subsection{Generalized Topological Yang-Mills Theory}

In this subsection we analyze the two dimensional version of the 
generalized topological Yang-Mills action 
and show that the instanton gauge fixing of the generalized topological 
Yang-Mills action leads to a $N=2$ super Yang-Mills theory together with the 
twisting procedure.
In four dimensions Witten showed that a topological gauge theory having 
$N=2$ super symmetry leads to the $N=2$ super Yang-Mills theory with the 
twisting procedure\cite{Witten2}. 
Later it has been pointed out that the $N=2$ super Yang-Mills theory 
with the twisting procedure can be derived from the topological Yang-Mills 
action with instanton gauge fixing\cite{Baulieu-Singer}. 
This subject has been intensively investigated \cite{top-twist} 
in particular the twisting mechanism has been cleared up in two 
dimensions with the help of super conformal field theory\cite{Eguchi-Yang}. 

Our formulation of this section\cite{Kawamoto-Tsukioka} is the two 
dimensional realization of the 
known four dimensional scenario and can be extended to arbitrary dimensions.  
 
To make the formulation concrete and simpler we specify to the two 
dimensional case. 
As we have already mentioned that the action we consider satisfies the 
following well known relation: 
\begin{equation}
 \int_M{\mbox{Str}}_{\bf\mbox{1}}{\cal F}^2_0
 =\int_M{\mbox{Str}}_{\bf\mbox{1}}\Bigg( Q\Big({\cal{A}}_0Q{\cal{A}}_0
 +\frac{2}{3}{\cal A}_0^3\Big)\Bigg),  \label{eq:gym2}
\end{equation}
where ${\cal A}_0$ and ${\cal F}_0$ are the two dimensional counter part
of the generalized classical gauge field and curvature. 
More explicitly they are given by 
\begin{eqnarray}
 {\cal{A}}_0 &=& {\bf\mbox{j}}\omega 
             + {\bf\mbox{k}}\Big( \phi 
                             +B \Big)
             \qquad\in\Lambda_-, \\
 {\cal{F}}_{0} &=& Q{\cal{A}}_0 + {\cal{A}}_0^2 \\
               &=& -{\bf\mbox{1}}\Big( d\omega + \omega^2 + \{ \phi, B \}
                                       +\phi^2 \Big) 
                +{\bf\mbox{i}}\Big(d\phi+[\omega, \phi] \Big) 
	       \qquad\in\Lambda_+.
\end{eqnarray}

Due to the topological nature of the action, i.e., the action vanishes 
identically if the two dimensional manifold $M$ does not have boundary, 
the action has so called shift symmetry. 
In other words the action is invariant under the gauge transformation 
of an arbitrary function ${\cal{E}}_0$.   
Thus the gauge transformation of the generalized topological Yang-Mills 
action has the following form:
\begin{equation}
 \delta {\cal A}_{0} = [Q+{\cal{A}}_0, {\cal{V}}_0] 
                       +{\cal{E}}_0, 
\label{eqn:ty-gt}
\end{equation}
where ${\cal{V}}_0$ is the generalized gauge parameter 
\begin{equation}
 {\cal{V}}_0 = {\bf\mbox{1}}\Big(v 
                             +b \Big) 
              +{\bf\mbox{i}}u \qquad\in\Lambda_+, 
\end{equation}
while ${\cal{E}}_0$ is a new gauge parameter of the shift symmetry 
and is given by 
\begin{equation}
 {\cal{E}}_0 = {\bf\mbox{j}}\xi^{(1)}
             + {\bf\mbox{k}}\Big(\xi^{(0)} 
                          +\xi^{(2)}\Big)
             \qquad\in\Lambda_-, 
\end{equation}
where $\xi^{(0)},\xi^{(1)}$ and $\xi^{(2)}$ are zero-, one-  and two 
form bosonic gauge parameters, respectively. 
At this stage the readers may wonder why we should keep the original gauge 
transformation in the gauge transformation of ${\cal A}_{0}$ in  (\ref{eqn:ty-gt}) 
since the gauge parameter of the shift 
symmetry ${\cal{E}}_0$ is an arbitrary function and could absorb the 
change of the original gauge transformation.
It will turn out later that both of the shift symmetry and the original 
symmetry are essential to induce 
$N=2$ super symmetry with matter fermions via twisting procedure.  

The generalized topological action has the following obvious reducibility: 
\begin{equation}
\begin{array}{rcl}
 {\cal{V}}_0 &=& {\cal{V}}_1, \\
 {\cal{E}}_0 &=& -[Q+{\cal{A}}_0, {\cal{V}}_1]. 
 \label{eqn:ty-reducibility} 
\end{array}
\label{eq:reducibility}
\end{equation}
Thus this system is a first stage reducible system in the terminology of 
Batalin and Vilkovisky. 
Correspondingly we need to introduce ghost fields $C^{(n)}$ and 
$\widetilde{C}^{(n)}$ with respect to 
the gauge parameters ${\cal{V}}_0$ and ${\cal{E}}_0$, and 
ghost for ghost field $\eta^{(n)}$ with respect to ${\cal{V}}_1$
, respectively.
Here the suffix $(n)$ with $n=0,1,2$ denotes the form degree. 

We then redefine the generalized gauge field by introducing the ghost field 
$C^{(n)}$
\begin{equation}
 {\cal{A}} = {\bf\mbox{1}}C^{(1)} 
              +{\bf\mbox{i}}\Big(C^{(0)}  +C^{(2)}\Big) 
           +{\bf\mbox{j}}\omega  +{\bf\mbox{k}}\Big(\phi  + B\Big)
             \qquad\in\Lambda_-.
\label{eqn:gfa}
\end{equation}
We need to introduce another generalized field to accommodate the ghost of 
the shift symmetry $\widetilde{C}^{(n)}$ and ghost for ghost $\eta^{(n)}$
\begin{equation}
 {\cal{C}} = {\bf\mbox{1}}\Big(\eta^{(0)} + \eta^{(2)} \Big)
              +{\bf\mbox{i}}\eta^{(1)}
              +{\bf\mbox{j}}\Big(\widetilde{C}^{(0)} + 
	      \widetilde{C}^{(2)}\Big)
              +{\bf\mbox{k}}\widetilde{C}^{(1)} \qquad\in\Lambda_+. 
\label{eqn:gfc}
\end{equation}
Here ${\cal{C}}$ belongs $\Lambda_+$ and could be identified as a part of 
generalized curvature later.

Furthermore we extend the differential operator $Q$ 
by introducing the BRST operator $s$ as a fermionic zero-form 
\begin{equation}
 {\cal{Q}} \equiv Q+Q_B  = {\bf\mbox{j}}d+{\bf\mbox{i}}s, \qquad\in\Lambda_-.
\end{equation}
It should be noted 
that $s$ commutes with $d$, $i.e.$ $[d, s]=0$ and $s^2=0$.
This operator satisfies the nilpotency property 
because of quaternionic structures 
\begin{equation}
{\cal{Q}}^2 = 0.
\label{eq:np}
\end{equation}
The following graded Leibnitz rule acting on generalized gauge fields 
${\cal{A}}, {\cal{B}}\in\Lambda_\pm$ can be derived: 
\begin{equation}
 {\cal{Q}}({\cal{A}}{\cal{B}})
 =({\cal{Q}}{\cal{A}}){\cal{B}}
    +(-)^{|{\cal{A}}|}{\cal{A}}({\cal{Q}}{\cal{B}}),
\label{eqn:lr} 
\end{equation}
where $|{\cal{A}}|=0$ for ${\cal{A}}\in\Lambda_+$ and 
$|{\cal{A}}|=1$ for ${\cal{A}}\in\Lambda_-$. 

We can now define the generalized curvature by using the redefined 
gauge field 
\begin{eqnarray}
 {\cal{F}} &=& {\cal{Q}}{\cal{A}}+{\cal{A}}^2 \nonumber \\
           &=& {\cal{F}}_0 + {\cal{C}}, 
\label{eqn:gecu}
\end{eqnarray}
where the second relation is imposed to relate the BRST transformation 
of the components fields. 
The nilpotency of the BRST transformation is assured by the 
Bianchi identity of the generalized field 
\begin{equation}
 {\cal{Q}}{\cal{F}}+[{\cal{A}}, {\cal{F}}]=0.
\label{eqn:geBi}
\end{equation}
The component wise expressions of BRST transformation can be read from 
(\ref{eqn:gecu}) and (\ref{eqn:geBi}).

\subsubsection{Instanton Gauge Fixing of Topological Yang-Mills Model}

We next introduce a particular model to carry out explicit analyses. 
In particular we choose the Clifford algebra as a graded Lie 
algebra, which closes under the multiplication and is simplest 
nontrivial example. 
More specifically we take the following two dimensional antihermitian 
Euclidean Clifford algebra:
\begin{equation}
 \begin{array}{rcl}
 T^a &:& 1, \quad \gamma_5, \\
 \Sigma^\alpha &:& \gamma^a, 
 \end{array}
\end{equation}
where $\gamma^a=(i\sigma^1, i\sigma^2)$, 
which satisfy $\{ \gamma^a, \gamma^b\}=-2\delta^{ab}$ 
and $\displaystyle\gamma_5=\frac{1}{2}\epsilon_{ab}\gamma^a\gamma^b
      =-i\sigma^3$ with $\epsilon_{12}=1$. 
A grading generator can be identified by $\gamma_5$ and then we define 
the super trace 
$$
{\mbox{Str}}(\cdots)=\hbox{Tr}(\gamma_5\cdots).
$$ 

The two dimensional generalized topological Yang-Mills action leads 
\begin{eqnarray}
S_0&=&\frac{1}{2}\int_M{\mbox{Str}}_{\bf\mbox{1}}{\cal F}_0^2, \nonumber \\
   &=&\int_M d^2x\Big(\epsilon^{\mu\nu}F_{\mu\nu}|\phi|^2
                              +\epsilon^{\mu\nu}\epsilon^{ab}(D_\mu\phi)_a
                              (D_\nu\phi)_b \Big)        \nonumber \\
   &=&\int_M d^2x\epsilon^{\mu\nu}\partial_\mu
                 \Big(2\omega_\nu|\phi|^2+
		 \epsilon^{ab}\phi_a\del_\nu\phi_b\Big),			\label{eqn:2ty}
\end{eqnarray}
where the scalar part of the one-form field $\omega_{\mu s}$ 
and the two form field $B_{a\mu\nu}$ do not appear in the action. 
It can be seen from the above relation that the square of the generalized 
curvature is related to the one dimensional generalized Chern-Simons 
action with the particular choice of the graded Lie algebra.

We can now find out a two dimensional instanton relation of our generalized 
topological Yang-Mills action by imposing self- (anti-self-) dual condition 
\begin{equation}
*{\cal F}_0=\pm{\cal F}_0.
\label{eqn:ic}
\end{equation}
In the present model we take the following duality relation for the gauge 
algebra and the quaternions 
\begin{eqnarray}
 *1 = -\gamma_5, ~~~
 {*}\gamma^a = -{\epsilon^a}_b\gamma^b, ~~~
 {*}\gamma_5 = -1, \nonumber\\ 
 {*}{\bf\mbox 1}={\bf\mbox 1}, ~~~~~~~~~~~~~~~
 {*}{\bf\mbox i}=-{\bf\mbox i}.
\end{eqnarray}
We can then find the following minimal condition of the action leading 
to instanton relations: 
\begin{eqnarray}
 & &\pm\frac{1}{2}\int\Str_{\1}{\cal F}_0\wedge{\cal F}_0
     +\frac{1}{2}\int\Str_{\1}{\cal F}_0\wedge*{\cal F}_0         \nonumber \\
 &=& \frac{1}{4}\int\Str_{\1}
     \Big({\cal{F}}_0\pm*{\cal F}_0\Big)\wedge
                     *\Big({\cal F}_0\pm*{\cal F}_0\Big)  \nonumber \\ 
 &=&\int d^2x\Bigg(\Big(\frac{1}{2}
                            \epsilon^{\mu\nu}F_{\mu\nu}\pm|\phi|^2\Big)
                            \Big(\frac{1}{2}
                            \epsilon^{\rho\sigma}F_{\rho\sigma}\pm|\phi|^2\Big)
                                                             \nonumber \\ 
 & &\qquad\quad            +\frac{1}{2}\Big((D_\mu\phi)_a
                                            \pm{\epsilon_\mu}^\nu
                                                   {\epsilon_a}^b
                                                   (D_\nu\phi)_b\Big)
                                       \Big((D^\mu\phi)^a
                                        \pm{\epsilon^\mu}_\rho
                                                   {\epsilon^a}_c
                                                   (D^\rho\phi)^c\Big)\Bigg).
\label{eqn:tpi}
\end{eqnarray}
Then the instanton condition is the absolute minima of the generalized 
Yang-Mills action 
\begin{eqnarray}
 \frac{1}{2}\epsilon^{\mu\nu}F_{\mu\nu}-|\phi|^2 &=& 0, 
                                                       \label{eq:sdgf1} \\
 (D_\mu\phi)_a^{(-)}&\equiv&\displaystyle{\frac{1}{2}
                           \Big((D_\mu\phi)_a-{\epsilon_\mu}^\nu
                                              {\epsilon_a}^b(D_\nu\phi)_b
                           \Big)} = 0.                           
                                                       \label{eqn:sdgf2} 
\end{eqnarray}

We now derive the gauge fixed action with instanton relations as 
gauge fixing conditions together with the following Landau type gauge fixing 
condition for the one form gauge field and the ghost of the shift symmetry: 
\begin{equation}
  \del_\mu\omega^\mu = 0,~~~~~\del_\mu\widetilde{C}^\mu = 0.
\label{eqn:lgf}
\end{equation}

We introduce a set of antighost fields 
$\lambda$, $\chi_{\mu a}$, $\etab$ and $\overline{C}$,  
and Lagrange multipliers,
$\widetilde{\pi}$, $\pi_{\mu a}$, $\rho$ and $\pi$. 
These fields obey the standard BRST subalgebra
\begin{equation}
 \begin{array}{rclcrcl}
 s\lambda&=&\widetilde{\pi}, &\qquad& s\widetilde{\pi}&=&0, \\
 s\chi_{\mu a}&=&\pi_{\mu a}, &\qquad& s\pi_{\mu a}&=&0, \\
 s\etab&=&\rho, &\qquad& s\rho&=&0, \\
 s\overline{C}&=&\pi, &\qquad&s\pi&=&0,
 \label{eqn:ag-brs}
 \end{array}
\end{equation}
where anti-self-dual field $\chi_{\mu a}$ 
obeys the condition 
${\epsilon_\mu}^\nu{\epsilon_a}^b\chi_{\nu b}=-\chi_{\mu a}$ 
and $\pi_{\mu a}$ also obeys the similar condition.

We then obtain BRST invariant gauge fixed action
\begin{eqnarray}
 S_{\mbox{\scriptsize{g-f}}}
 &=& S_0+s\int d^2x\Bigg\{
                    +\lambda\Big(\frac{1}{2}\epsilon^{\mu\nu}F_{\mu\nu}
                               -|\phi|^2-\beta\widetilde{\pi}\Big) 
               -\chi_{\mu a}\Big((D^\mu\phi)^{a(-)}-\alpha\pi^{\mu a}\Big)
                                                              \nonumber \\
 & &\qquad\qquad\qquad\quad
                +\etab\del_\mu\widetilde{C}^\mu+\overline{C}
		\del_\mu\omega^\mu\Bigg\} 
                                           \nonumber \\
 &=&S_0+\int d^2x\Bigg\{\pi(\del_\mu\omega^\mu)
                       +\rho(\del_\mu\widetilde{C}^\mu)     \nonumber \\
 & &\qquad\qquad\qquad -\pi_{\mu a}\Big((D^\mu\phi)^{a(-)}
                                         -\alpha\pi^{\mu a}\Big)
                      +\widetilde{\pi}\Big(\frac{1}{2}
                                            \epsilon^{\mu\nu}F_{\mu\nu}
                                           -|\phi|^2-\beta\widetilde{\pi}\Big)
                                                                  \nonumber \\
 & &\qquad\qquad\qquad +(\del_\mu\overline{C})(\del^\mu C+\widetilde{C}^\mu)
                      -\chi_{\mu a}(D^\mu\widetilde{C})^a 
                      -\lambda\epsilon^{\mu\nu}\del_\mu\widetilde{C}_\nu
                      -\del_\mu\etab\del^\mu\eta                  \nonumber \\
 & &\qquad\qquad\qquad +2\chi_{\mu a}\epsilon^{ab}(D_\mu\phi)_bC
                      -2\chi_{\mu a}\epsilon^{ab}\widetilde{C}^\mu\phi_b
                      -2\lambda\phi_a\widetilde{C}^a\Bigg\},
\label{eqn:bs0}
\end{eqnarray}
where $\alpha$ and $\beta$ are arbitrary parameters.
In the second line of (\ref{eqn:bs0}) BRST transformations of the component 
fields from (\ref{eqn:ag-brs}), (\ref{eqn:gecu}) and (\ref{eqn:geBi}) are used.

\subsubsection{Twisted $N=2$ Super Yang-Mills Action with Dirac-
K$\ddot{\hbox{a}}$hler Fermions}

We now modify the gauge fixed action $S_{\mbox{\scriptsize{g-f}}}$ of 
(\ref{eqn:bs0}) by adding BRST exact terms and making the action Hermite 
and taking the particular choice of the parameters $\alpha$ and $\beta$ 
\begin{eqnarray}
 S=\int d^2x &\Big(&+\frac{1}{2}F_{\mu\nu}F^{\mu\nu} 
                      + (D_\mu\phi)_a(D^\mu\phi)^a + |\phi|^4  \nonumber \\
             &{}& +i\rho\partial_\mu\C^\mu 
                  -i\lambda\epsilon^{\mu\nu}\partial_\mu\C_\nu    \nonumber \\
             &{}& -i\chi_{\mu a}(D^\mu\C)^a 
                  +\partial_\mu\etab\partial^\mu\eta           \nonumber \\
             &{}& -2i\rho\epsilon^{ab}\phi_a\C_b 
                  -2i\lambda\phi^a\C_a
                  -2i\chi_{\mu a}\epsilon^{ab}\C^\mu\phi_b     \nonumber \\ 
             &{}& -\frac{i}{4}\epsilon^{\mu\nu}
                   \chi_{\mu a}{\chi_\nu}^a\eta 
                  +2i\etab\epsilon_{ab}\C^a\C^b 
                  +4\etab\eta|\phi|^2 \Big).
\label{eqn:wa}
\end{eqnarray}
It is easy to see that the kinetic terms of $\phi_a$, $\C_\mu$, $\rho$, 
$\lambda$, $\chi_{\mu a}$ and $\C_a$ are nondegenerate 
while that of $\omega_\mu$ is degenerate.
Indeed this action is invariant under the following $SO(2)$ 
gauge transformations with a gauge parameter $v$, 
\begin{equation}
 \begin{array}{rcl}
 \delta_{\hbox{\tiny gauge}}\omega_\mu &=& \partial_\mu v, \\
 \delta_{\hbox{\tiny gauge}}(\phi_a, \C_a, \chi_{\mu a}) 
                 &=& 2v{\epsilon_a}^b(\phi_b, \C_b, \chi_{\mu b}), \\
 \delta_{\hbox{\tiny gauge}}(\C_\mu, \eta, \lambda, \rho, \etab) &=& 0. \\
 \end{array}
\label{eqn:wgt}
\end{equation}

We claim that this action possesses BRST symmetry and $N=2$ twisted 
super algebra up to the above gauge symmetry. 
The algebra will be summarized for fermionic family of scalar BRST 
generator $s$ and the vector and pseudo-scalar counterpart $s_\mu$ and 
$\widetilde{s}$ as follows:
\begin{eqnarray}
 s^2&=&i{\delta_{\hbox{\tiny gauge}}}_{\eta},
\label{eqn:ss}\\
 \{ s, s_\mu \} &=& 2i\partial_\mu - 
 2i{\delta_{\hbox{\tiny gauge}}}_{\omega_{\mu}}, 
\label{eqn:ssm} \\
 \{ s_\mu, s_\nu \} &=& -2ig_{\mu\nu}{\delta_{\hbox{\tiny gauge}}}_{\etab},
\label{eqn:smsm}\\
\{ \st, s_\mu \} &=& -2i{\epsilon_\mu}^\nu\partial_\nu 
 + 2i{\delta_{\hbox{\tiny gauge}}}_{{\epsilon_\mu}^\nu\omega_{\nu}}, 
\label{eq:stsm} \\
 \{ \st, \st \} &=& 2i{\delta_{\hbox{\tiny gauge}}}_{\eta}, 
\label{eq:stst}\\
 \{ \st, s \} &=& 0.
\label{eqn:sts}
\end{eqnarray}
The more detailed confirmation of the component wise expressions for 
the twisted $N=2$ super algebra is given in (\cite{Kawamoto-Tsukioka}).

In this formulation we find out very interesting and nontrivial 
correspondence. 
We point out that the fermionic fields corresponding to ghosts 
and Lagrange multipliers can be interpreted as Dirac-K$\ddot{\hbox{a}}$hler 
fermion fields and thus the twisting procedure is nothing but the 
Dirac-K$\ddot{\hbox{a}}$hler interpretation of fermionic fields 
appearing in the quantization procedure.

The kinetic terms of multiplet $(\rho, \C_\mu, \lambda)$ and 
$(\C_a, \chi_{\mu a})$ in the action (\ref{eqn:wa}) can be 
expressed as 
\begin{eqnarray}
 S &=& \int d^2x(i\rho\partial_\mu\C^\mu-i\lambda\epsilon^{\mu\nu}
                 \partial_\mu\C_\nu
                 -i\chi^{\mu a}\partial_\mu\C_a)    \nonumber \\
   &=& \int d^2x\hbox{Tr}\Big(i\psi^\dagger\gamma^\mu\partial_\mu\psi
                 + i\chi^\dagger\gamma^\mu\partial_\mu\chi\Big),
\end{eqnarray}
where the Dirac-K$\ddot{\hbox{a}}$hler fields $\psi$ and $\chi$ are given by 
\begin{eqnarray}
 \psi &=& \frac{1}{2}(\rho+\C_\mu\gamma^\mu-\lambda\gamma_5), \nonumber \\
 \chi &=& \frac{1}{2}(-\C_{a=1} + \chi_{\mu a=1}\gamma^\mu - \C_{a=2}\gamma_5).
\label{eqn:dkf}
\end{eqnarray}

The final expression of the twisted $N=2$ super Yang-Mills action 
with Dirac-K$\ddot{\hbox{a}}$hler fermions is 
\begin{eqnarray}
 S=\int d^2x\Big(&+&\frac{1}{2}F_{\mu\nu}F^{\mu\nu} 
                 + (D_\mu\phi)_a(D_\mu\phi)^a + |\phi|^4  \nonumber \\
             &+&\frac{1}{2}\partial_\mu A\partial^\mu A 
	         - \frac{1}{2}\partial_\mu B\partial^\mu B \nonumber \\
             &+& \hbox{Tr}(i\psi^\dagger\gamma^\mu\partial_\mu\psi) 
	         \nonumber \\
             &+& \hbox{Tr}(i\chi^\dagger\gamma^\mu\partial_\mu\chi)
                 + 2i\omega_\mu\hbox{Tr}(\chi^\dagger\gamma^\mu\chi\gamma_5)
                 \nonumber \\
             &-& 4i\phi_1\hbox{Tr}(\psi^\dagger\gamma_5\chi)
	         + 4i\phi_2\hbox{Tr}(\psi^\dagger\gamma_5\chi\gamma_5)
		 \nonumber \\
             &-& i\sqrt{2}A\hbox{Tr}(\chi^\dagger\gamma_5 \chi)
	         +i\sqrt{2}B\hbox{Tr}(\chi^\dagger\chi\gamma_5 )
	         \nonumber \\
             &+& 2(A^2-B^2)|\phi|^2 \Big),
\label{eqn:dk}
\end{eqnarray}
where we denote $\eta\equiv \frac{2}{\sqrt{2}}(A+B)$ and 
$\overline{\eta}\equiv \frac{1}{2\sqrt{2}}(A-B)$. 
It should be noted that the gauge transformation on 
Dirac-K$\ddot{\hbox{a}}$hler field $\chi$ can be recognized as 
$SO(2)$ flavor group.

As we have seen in the formulation, the fermionic 
fields appearing in the quantization procedure such as ghosts and 
lagrange multiplier turns into the Dirac-K$\ddot{\hbox{a}}$hler matter 
fermion. 
It would be important to confirm algebraically that the 
Dirac-K$\ddot{\hbox{a}}$hler fermions transform as spinor fields and 
posses half integer spin unlike the ghost fields. 

Usually $N=2$ super algebra includes the generator $R$ corresponding 
to a conserved current $R_\mu$ associated with $R$-symmetry. 
The procedure of the topological twist is related with the redefinition 
of the stress tensor $T_{\mu\nu}$. 
We can define new stress tensor $T'_{\mu\nu}$ by the following relation 
without spoiling current conservation law of $R_\mu$,
\begin{equation}
 T'_{\mu\nu} = T_{\mu\nu} + \epsilon_{\mu\rho}\del^\rho R_\nu +
                          \epsilon_{\nu\rho}\del^\rho R_\mu.
\end{equation}
This modification of the stress tensor leads to a redefinition of 
the Lorentz rotation generator $J$ 
\begin{equation}
 J'=J+R.
\label{eqn:nj}
\end{equation}   
This rotation group is interpreted as the diagonal subgroup of 
$SO(2)\times SO(2)_I$. 
In other words the topological twist is essentially achieved by identifying 
spinor and isospinor indices. 

In the present model we can identify the transformation of the 
$R$-symmetry is the transformation of the flavor symmetry for 
the Dirac-K$\ddot{\hbox{a}}$hler fermion fields 
\begin{equation}
 \delta_{\scr R}\psi = \psi\Big(\frac{1}{2}\gamma_5\Big), ~~~~~~~~
 \delta_{\scr R}\chi = \chi\Big(\frac{1}{2}\gamma_5\Big). 
\end{equation}
Then the Lorentz transformation induced by the generator $J'$ is 
\begin{equation}
 \delta_{J'}\psi = -\frac{1}{2}[\gamma_5,\psi], 
\end{equation}
while the Lorentz transformation of $J$ is 
\begin{equation}
 \delta_{J}\psi = -\frac{1}{2}\gamma_5\psi. 
\end{equation}
We can obtain the same relations for $\chi$.
This implies that Dirac-K$\ddot{\hbox{a}}$hler fermion fields exactly 
transform as spinors and carry spin one half.

\subsection{Weinberg-Salam Model from Generalized Gauge Theory}

Connes has pointed out that the Weinberg-Salam model can be formulated as 
a particular case of the noncommutative geometry formulation of a gauge 
theory\cite{Connes}. 
Roughly speaking his idea is the following. 
Consider a manifold which is composed of a direct product of 
discrete two points and four dimensional flat space and 
define a connection and differential operator on this manifold. 
Due to the discrete nature of the two points the differential operator 
can be represented by two by two matrix. 
Thus the connection or equivalently the gauge field 
is now represented by two by two matrix and thus possesses diagonal 
and off-diagonal components. 
Then the weak and electromagnetic one form gauge fields 
are assigned to the diagonal component while the zero form Higgs fields 
are assigned to the off-diagonal components. 
Then the spontaneously broken Weinberg-Salam model comes 
out naturally from the pure Yang-Mills action by taking the group 
$SU(2)\times U(1)$\cite{Connes}\cite{Coq}.

\subsubsection{Generalized Gauge Theory with Dirac-K$\ddot{\hbox{a}}$hler 
Fermions}
In this subsection we show that the two by two matrix representation of 
the gauge fields are easily accommodated by the quaternions of the 
generalized gauge field since the quaternion algebra can be 
represented by two by two matrices.
We can, however, point out that our generalized gauge theory is more 
general formulation as noncommutative geometry since it includes not only 
zero and one form of gauge fields but also all the possible form degrees 
of gauge fields\cite{KTU}.

In formulating generalized Yang-Mills action we need to define 
the inner product to construct the action. 
In the sense of the generalized topological Yang-Mills theory the inner 
product has been defined by the naive wedge product.
Here we intend to incorporate fermionic form degrees of freedom in accordance 
with the generalized gauge field. 
We then formulate the fermionic form degrees by Dirac-K$\ddot{\hbox{a}}$hler 
fermion formulation to generate matter fermions. 
We thus need to define a new inner product to accommodate both the generalized 
gauge fields and Dirac-K$\ddot{\hbox{a}}$hler fields. 

We keep to use the generalized gauge field and parameter as defined above 
in (\ref{eqn:def-gf}) and (\ref{eqn:def-gp}). 
In addition we introduce fermionic matter generalized field 
\begin{eqnarray}
  \bm{$\psi$} &=& \mbox{\bf{1}}\psi + \mbox{\bf{i}}\hat{\psi} 
                  \in \Lambda_- \nonumber \\
 \overline{\bm{$\psi$}} &=& \mbox{\bf{j}}\hat{\overline{\psi}} 
                              + \mbox{\bf{k}}\overline{\psi} \in \Lambda_+,  
\end{eqnarray}
where $\psi$ and $\overline{\psi}$ fermionic odd forms while 
$\hat{\psi}$ and $\hat{\overline{\psi}}$ are fermionic even forms.
We now extend the generalized differential operator 
${\cal{Q}} = {\mbox{\bf{j}}}d$ into 
\begin{equation}
 {\cal D} = \mbox{\bf{j}}d + \mbox{\bf{k}}m,  \label{eqn:gy-ed}
\end{equation}
where $m$ is a bosonic constant matrix. 
In this subsection we consider the formulation only in four dimensions. 
It is straightforward to generalize the formulation into arbitrary 
even dimensions. 

This differential operator has the following grading structure:   
\begin{equation}
 {\cal DA} = \{ \ {\cal D}, \ {\cal A} \ \}, \quad {\cal A} \in \Lambda_{-}, 
\end{equation}
\begin{equation}
 {\cal DV} = [ \ {\cal D}, \ {\cal V} \ ], \quad {\cal V} \in \Lambda_{+}. 
\end{equation}
The graded Leibniz rule can be easily checked, 
but the nilpotency is not satisfied in general: 
${\cal{D}}^2 = -{\mbox{\bf{1}}}m^2$. 
The gauge transformation of the generalized curvature has extra factor due 
to the non-nilpotency of the differential operator 
\begin{eqnarray}
 \delta {\cal{F}} &=& [ \ {\cal{F}}, \ {\cal{V}} \ ] 
                      + [ \ {\cal{D}}^2, \ {\cal{V}} \ ] \nonumber \\
                  &=& [ \ {\cal{F}}, \ {\cal{V}} \ ] 
                      - [ \ m^2, \ {\cal{V}} \ ].
\label{eqn:delcurv}
\end{eqnarray}

In order to define generalized Yang-Mills action with 
Dirac-K$\ddot{\hbox{a}}$hler fermions as matter 
we introduce the notion of 
Clifford product or simply cup product $\vee$ which was introduced by 
K\"{a}hler and should be 
differentiated from the wedge product $\wedge$\cite{DK}\cite{BJ}.
We consider an element $u$ which is the direct sum of forms 
\begin{eqnarray}
   u= \sum^m_{p=0}\frac{1}{p!}u_{\mu_1\cdots\mu_p}
   dx^{\mu_1}\wedge\cdots\wedge dx^{\mu_p},
     \label{eqn:gy-gf}
\end{eqnarray}
where $m\le d$ with $d$ as spacetime dimension.      

We then define the linear operator $e_{\mu}$,
\begin{eqnarray}
  e_\mu u= \sum^{m-1}_{p=0}\frac{1}{p!}u_{\mu\mu_1\cdots\mu_p}
   dx^{\mu_1}\wedge\cdots\wedge dx^{\mu_p},
     \label{eqn:gy-cop}
\end{eqnarray}
which is understood as a differentiation of the polynomial of differential 
form with respect to $dx^\mu$.
In particular it has the role of contracting operator as  
\begin{eqnarray}
   e_\mu dx^{\alpha_1}&\wedge& dx^{\alpha_2}\wedge 
   dx^{\alpha_3}\wedge dx^{\alpha_4}
   \wedge\cdots \nonumber \\
   &=& g_\mu^{\alpha_1}dx^{\alpha_2}\wedge dx^{\alpha_3}\wedge dx^{\alpha_4}
   \wedge\cdots  - 
   g_\mu^{\alpha_2}dx^{\alpha_1}\wedge dx^{\alpha_3} \wedge dx^{\alpha_4}
   \wedge\cdots \nonumber \\
   &+& g_\mu^{\alpha_3}dx^{\alpha_1}\wedge dx^{\alpha_2} \wedge dx^{\alpha_4}
   \wedge\cdots  - 
   g_\mu^{\alpha_4}dx^{\alpha_1}\wedge dx^{\alpha_2} \wedge dx^{\alpha_3}
   \wedge\cdots + \cdots .
     \label{eqn:gy-copex}
\end{eqnarray}
We further define sign operators $\eta$ and $\zeta$ which generate the 
following sign factors: 
\begin{eqnarray}
  \zeta u_p &=& \zeta_p u_p \equiv (-1)^{\frac{p(p-1)}{2}} u_p \nonumber \\
  \eta  u_p &=& (-1)^p u_p, 
     \label{eqn:gy-sf}
\end{eqnarray}
where $u_p$ is a p-form variable.

We can now define the cup (Clifford) product of $u$ and $v$
\begin{eqnarray}
  u\vee v= \sum^m_{p=0} \frac{1}{p!} \zeta_p 
  \{ \eta^p(e_{\mu_1}\cdots e_{\mu_p} u) \} \wedge e^{\mu_1}\cdots 
  e^{\mu_p} v,      \label{eqn:gy-cp}
\end{eqnarray}
where $v$ has the similar form as $u$ in (\ref{eqn:gy-gf}).
The sign factor $\zeta_p$ and the operator $\eta^p$ are necessary to confirm 
the associativity of the cup product
\begin{eqnarray}
  (u\vee v)\vee w = u\vee (v\vee w) .
   \label{eqn:gy-associativity}
\end{eqnarray}
We introduce the generalized Yang-Mills action 
\begin{equation}
  S_G = \int \hbox{Tr}[\overline{\bm{$\psi$}}\vee({\cal D}
  +{\cal A}){\bm{$\psi$}} + {\cal{F}}\vee {\cal{F}}]_{\hbox{{\bf 1}}} * 1 
   \label{eqn:gy-gga}
\end{equation}
where $*$ is Hodge star operator and in particular 
$* 1 = \sqrt{g}\frac{1}{4!}\epsilon_{\alpha_1\alpha_2\alpha_3\alpha_4}
dx^{\alpha_1}\wedge dx^{\alpha_2}\wedge dx^{\alpha_3}\wedge dx^{\alpha_4}
$ in four dimensions with $g$ as the metric determinant.
The simbol ${\bf 1}$ denotes to pick up 
the coefficient of ${\bf 1}$ for the quaternion expansion in the trace. 

We then define the generalized gauge parameter
\begin{eqnarray}
 { \cal{V}}= \hbox{{\bf 1}} v,
   \label{eqn:gy-ggp}
\end{eqnarray}
where $v$ is 0-form gauge parameter and thus ${\cal{V}} \in \Lambda_+$.
This has a clear contrast with the generalized gauge parameter of generalized 
Chern-Simons formulation where all the degrees of differential forms were 
introduced. 

We now claim that the generalized Yang-Mills action with 
Dirac-K$\ddot{\hbox{a}}$hler matter fermion is gauge invariant 
under the following gauge transformation: 
\begin{eqnarray}
  \delta {\cal A} &=& [{\cal{D}}+{\cal{A}},{\cal{V}}]_{\vee}, \nonumber \\
  \delta {\bm{$\psi$}} &=& - {\cal V}\vee {\cal \psi}, \nonumber \\
  \delta \overline{\bm{$\psi$}} &=& \overline {\cal \psi}\vee {\cal V},
  \label{ym-ggt}
\end{eqnarray}
where we have used the notation 
$[\cal{A},\cal{B}]_\vee = \cal{A}\vee\cal{B}-\cal{B}\vee\cal{A}$. 
Here we should mention that in the quaternion algebra we should 
choose the sign choice $(\epsilon_1,\epsilon_2)=(1,1)$ to prove 
the generalized gauge invariance of the generalized Yang-Mills action 
(\ref{eqn:gy-gga}).

\subsubsection{Weinberg-Salam Model from Generalized Gauge Theory 
as Non-Commutative Geometry Formulation}
In the formulation of previous subsection we have introduced all the 
degrees of differential 
forms but only zero form for the generalized gauge parameter. 
The reason why we have introduced only zero form gauge parameter 
is that the action is not gauge invariant under the higher form 
gauge parameters. 
Hereafter we introduce only zero form and one form gauge fields for the 
generalized gauge field and do not consider the 
Dirac-K$\ddot{\hbox{a}}$hler fermionic matter for simplicity. 

We take the algebra of super group $SU(2|1)$ as the graded Lie algebra.
$SU(2|1)$ generators can be represented by $3\times3$ 
matrices~\cite{Coq}
$$
 T_{i}=\left(
             \begin{array}{@{\,}cc|c@{\,}}
                & \mbox{ \hspace{-7.6mm} \raisebox{-2.5mm}[0pt][0pt]
                  { $ \displaystyle \frac{\sigma_{i}}{2} $ } } & 0 \\ 
                &   & 0 \\ \hline 
              0 & 0 & 0
             \end{array}
            \right), \quad
  Y   =\left(
             \begin{array}{@{\,}cc|c@{\,}}
              1 & 0 & 0 \\
              0 & 1 & 0 \\ \hline
              0 & 0 & 2    
             \end{array}
          \right),
$$
$$
  \Sigma_{+}=\left(
                   \begin{array}{@{\,}cc|c@{\,}}
                   0 & 0 & 1 \\
                   0 & 0 & 0 \\ \hline
                   0 & 0 & 0 
                   \end{array}
             \right), \
  \Sigma_{-}=\left(
                   \begin{array}{@{\,}cc|c@{\,}}
                   0 & 0 & 0 \\
                   0 & 0 & 1 \\ \hline
                   0 & 0 & 0
                   \end{array}
             \right), \
  \Sigma'_{+}=\left(
                   \begin{array}{@{\,}cc|c@{\,}}
                   0 & 0 & 0 \\
                   0 & 0 & 0 \\ \hline
                   0 & 1 & 0
                   \end{array}
             \right), \
  \Sigma'_{-}=\left(
                   \begin{array}{@{\,}cc|c@{\,}}
                   0 & 0 & 0 \\
                   0 & 0 & 0 \\ \hline
                   1 & 0 & 0
                   \end{array}
             \right), \
$$
where $\sigma_{i}$'s are Pauli matrices.
They satisfy the following graded Lie algebra: 
$$
 [ \ T_{i}, \ T_{j} \ ]=i\epsilon_{ijk}T_{k}, \quad
 [ \ Y, \ T_{i} \ ]=0,
$$
$$
 [ \ T_{\pm}, \ \Sigma_{\pm} \ ]=0, \quad
 [ \ T_{\pm}, \ \Sigma_{\mp} \ ]=\frac{1}{\sqrt{2}}\Sigma_{\pm},
$$
$$
 [ \ T_{\pm}, \ \Sigma'_{\pm} \ ]=0, \quad
 [ \ T_{\pm}, \ \Sigma'_{\mp} \ ]=-\frac{1}{\sqrt{2}}\Sigma'_{\pm},
$$
$$
 [ \ T_{3}, \ \Sigma_{\pm} \ ]=\pm\frac{1}{2}\Sigma_{\pm}, \quad
 [ \ T_{3}, \ \Sigma'_{\pm} \ ]=\pm\frac{1}{2}\Sigma'_{\pm},
$$
$$
 [ \ Y, \ \Sigma_{\pm} \ ]=-\Sigma_{\pm}, \quad 
 [ \ Y, \ \Sigma'_{\pm} \ ]=\Sigma'_{\pm},
$$
$$
 \{ \ \Sigma_{\pm}, \ \Sigma_{\pm} \ \} 
 = \{ \ \Sigma_{\pm}, \ \Sigma_{\mp} \ \}=0, \quad
 \{ \ \Sigma'_{\pm}, \ \Sigma'_{\pm} \ \}
 = \{ \ \Sigma'_{\pm}, \ \Sigma'_{\mp} \ \}=0,
$$
$$
 \{ \ \Sigma_{\pm}, \ \Sigma'_{\pm} \ \}=\sqrt{2}T_{\pm}, \quad
 \{ \ \Sigma_{\pm}, \ \Sigma'_{\mp} \ \}=\pm T_{3}+\frac{1}{2}Y,
$$
with $T_{\pm}=\frac{1}{\sqrt{2}}(T_{1}\pm iT_{2})$.
$T_{3}, Y$ correspond to a generator of a weak isospin 
and a weak hypercharge, respectively. 
This algebra contains $SU(2)\times U(1)_{Y}$ 
in the even graded parts $T_{i}, Y$, 
and $SU(2)$ doublet in the odd graded parts $\Sigma_{\pm}, \Sigma'_{\pm}$ 
whose subscripts $\pm$ correspond to the generators with eigenvalues 
$\pm\frac{1}{2}$ 
of the generator $T_{3}$. 
Indeed Higgs doublet 
which we have introduced as Lie algebra valued gauge fields 
corresponds to these odd parts.
It is interesting to note that the super symmetric algebra of $SU(2|1)$ 
can be accommodated in the generalized gauge theory even without fermionic 
fields. 

The gauge field is now expanded by corresponding fields of the generators  
\begin{equation}
 {\cal{A}}={\mbox{\bf{j}}}
             \Big(iA^{i}T_{i}+\frac{i}{2\sqrt{3}}BY\Big) 
             +{\mbox{\bf{k}}}
             \Big(\frac{i}{\sqrt{2}}(\phi_{0}\Sigma_{+}+\phi_{+}\Sigma_{-}
                  +\phi_{0}^{*}\Sigma'_{-} 
                  +\phi_{+}^{*}\Sigma'_{+})\Big),
\end{equation}
where $A^{i}, B$ and $\phi_{0}, \phi_{+}$ are 
real $SU(2)\times U(1)_{Y}$ gauge fields 
and complex Higgs scalar fields respectively.
We can choose a particular form of the constant matrix $m$ of the generalized 
differential operator $m=\frac{i}{\sqrt{2}}(v\Sigma_{+}+v^*\Sigma'_{-})$ 
which leads 
\begin{equation}
 m^2=-\frac{|v|^2}{2}\Big(T_{3}+\frac{1}{2}Y\Big). \label{em}
\end{equation}
Then eq. (\ref{em}) exactly represents 
the electro-magnetic charge.
As we can see from the gauge transformation of the curvature in 
(\ref{eqn:delcurv}) the gauge symmetries $v$ spontaneously break down 
to the special direction which is required from the consistency of the 
extended differential algebras. 

The curvature is given with a suitable redefinition of fields 
${\cal{A}}\rightarrow g{\cal{A}}$,
\begin{equation}
 {\cal{F}}={\mbox{\bf{1}}} 
            \bigg(g^2{\cal{F}}^{(0)} 
                  -\frac{i}{2}g{\cal{F}}_{\mu\nu}^{(2)}
                                        dx^{\mu}\wedge dx^{\nu} \bigg)
          +{\mbox{\bf{i}}}ig{\cal{F}}_{\mu}^{(1)}dx^{\mu}.
\end{equation}
The kinetic terms of $SU(2)\times U(1)_{Y}$ gauge fields are 
\begin{equation}
 {\cal{F}}_{\mu\nu}^{(2)}=F_{\mu\nu}^kT_{k}+\frac{1}{2\sqrt{3}}G_{\mu\nu}Y, 
\end{equation}
where
$$
 F_{\mu\nu}^{k}=\partial_{\mu}A_{\nu}^{k}-\partial_{\nu}A_{\mu}^{k}
                 -2g\epsilon_{ij}{}^{k}A_{\mu}^{i}A_{\nu}^{j},
$$
$$
 G_{\mu\nu}=\partial_{\mu}B_{\nu}-\partial_{\nu}B_{\mu}.
$$
The kinetic terms of Higgs fields and interaction terms are
\begin{eqnarray}
 {\cal{F}}_{\mu}^{(1)}=\frac{1}{\sqrt{2}}\Bigg\{
                         &\bigg(&\partial_{\mu}\phi_{0}
                                +\frac{i}{\sqrt{2}}gW_{\mu}\phi_{+}
                                +\frac{i}{\sqrt{3}}gZ_{\mu}
                                            \Big(\phi_{0}+\frac{v}{g}\Big) 
                         %
                         %
                        \bigg)\Sigma_{+} \nonumber \\
                      + &\bigg(&\partial_{\mu}\phi_{+}
                                 +\frac{i}{\sqrt{2}}gW_{\mu}^{\dagger}
                                  \Big(\phi_{0}+\frac{v}{g}\Big)
                                 -i\frac{\sqrt{3}}{6}gZ_{\mu}\phi_{+}
                                 -\frac{i}{2}gA_{\mu}\phi_{+}
                         %
                         %
                         \bigg)\Sigma_{-} \nonumber \\
                      +&{}&{\mbox{h.c.}} \ \Bigg\},
\end{eqnarray}
where
$$
 W_{\mu}=\frac{1}{\sqrt{2}}\Big( A_{\mu}^{1}-iA_{\mu}^{2} \Big),
$$
and
\begin{eqnarray}
  Z_{\mu} &=& \frac{\sqrt{3}}{2}A_{\mu}^3-\frac{1}{2}B_{\mu}, \nonumber \\
  A_{\mu} &=& \frac{1}{2}A_{\mu}^3+\frac{\sqrt{3}}{2}B_{\mu}.  \label{az} 
\end{eqnarray}
These identifications (\ref{az}) fix the 
Weinberg angle to be $\theta_{W}=\frac{\pi}{6}$ 
which is an arbitrary parameter in the Weinberg-Salam model. 
Thus the direction of spontaneous breaking is particularly chosen by the 
model itself. 
The Higgs potential term is given by 
\begin{eqnarray}
 {\cal{F}}^{(0)}=\frac{1}{2}
               &&\Bigg\{\bigg(\Big|\phi_{0}+\frac{v}{g}\Big|^2 
                              -\Big|\frac{v}{g}\Big|^2\bigg)
                        \Big(T_{3}+\frac{1}{2}Y\Big)
                        +\bigg(\phi_{0}+\frac{v}{g}\bigg)
                                    \phi_{+}^{*}\sqrt{2}T_{+} \nonumber \\
               &&       +\phi_{+}\bigg(\phi_{0}^{*}+\frac{v^{*}}{g}\bigg)
                                    \sqrt{2}T_{-}
                        +|\phi_{+}|^2\Big(-T_{3}+\frac{1}{2}Y\Big)\Bigg\}.
                                                              \label{c0}
\end{eqnarray}

Then the full expression of the action in Minkowski space-time is given by
\begin{eqnarray}
 S = \int d^4x \Bigg\{
      &-& \frac{1}{4}F_{\mu\nu}F^{\mu\nu} 
          - \frac{1}{4}Z_{\mu\nu}Z^{\mu\nu}                    \nonumber \\
      &-&\frac{1}{2}(D^{\mu\dagger}W^{\nu\dagger}-D^{\nu\dagger}W^{\mu\dagger})
                    (D_{\mu}W_{\nu}-D_{\nu}W_{\mu})            \nonumber \\
      &-&2ig\Big(\frac{\sqrt{3}}{2}Z_{\mu\nu}
                 +\frac{1}{2}F_{\mu\nu}\Big)W^{\mu}W^{\nu\dagger} \nonumber \\
      &+&2g^2\Big(|W_{\mu}W^{\mu}|^2-(W_{\mu}W^{\mu\dagger})^2\Big) 
                                                                 \nonumber \\
      &+&\Big|{\partial}_{\mu}\phi_{0}+\frac{i}{\sqrt{2}}gW_{\mu}\phi_{+} 
                          +\frac{i}{\sqrt{3}}Z_{\mu}
                             \Big(\phi_{0}+\frac{v}{g}\Big)\Big|^2
                                                                 \nonumber \\
      &+&\Big|{\partial}_{\mu}\phi_{+}
                          +\frac{i}{\sqrt{2}}gW_{\mu}^{\dagger}
                                      \Big(\phi_{0}+\frac{v}{g}\Big)
                          -i\frac{\sqrt{3}}{6}gZ_{\mu}\phi_{+}
                          -\frac{i}{2}gA_{\mu}\phi_{+} \Big|^2    \nonumber \\
      &-&\frac{g^2}{2}\bigg( \Big|\phi_{0}+\frac{v}{g}\Big|^2 
                               + |\phi_{+}|^2 \bigg)^2            \nonumber \\
      &+&\frac{3}{4}g^2\Big|\frac{v}{g}\Big|^2(1-2y)
          \bigg( \Big|\phi_{0}+\frac{v}{g}\Big|^2+|\phi_{+}|^2 \bigg)
                                                                 \nonumber \\
      &-&\frac{3}{8}g^2\Big|\frac{v}{g}\Big|^2(1-2y)^2  \Bigg\},
\end{eqnarray}
where
\begin{eqnarray*}
 F_{\mu\nu} &=& \partial_{\mu}A_{\nu}-\partial_{\nu}A_{\mu}, \\
 Z_{\mu\nu} &=& \partial_{\mu}Z_{\nu}-\partial_{\nu}Z_{\mu}, \\
 D_{\mu}    &=& \partial_{\mu}+i2g\Big(\frac{\sqrt{3}}{2}Z_{\mu}
                                              +\frac{1}{2}A_{\mu} \Big).
\end{eqnarray*} 
Higgs potential term can be minimized 
by shifting the field with $y=-\frac{1}{6}$.
This leads to
the ratio of the mass of physical Higgs $M_{\phi}$ 
and weak boson $M_{W}$ to $\frac{M_{\phi}}{M_{W}}=\sqrt{2}$.
 
The predictions from this scheme agree 
with noncommutative geometric models~\cite{Connes}\cite{Coq}.
This model has fewer parameters than the usual Weinberg-Salam model 
and has predictive power on the Higgs mass.

\section{Possible Scenario and Conjectures for Unified Model on the Lattice}

I would like to discuss a possible scenario 
and then propose conjectures of unified model on the lattice 
based on the analyses of previous sections and suggestive known results. 

Constructive and reguralized quantum theory of gravity will be given by 
dynamically triangulated simplicial lattice manifold. 
In two dimensional quantum gravity we have enough evidences that this is 
the fact and we have the analytical and numerical formulations to describe 
it.
As we have seen in the three dimensional example of Chern-Simons gravity 
the lattice manifold may be interpreted as spin graph where links carry 
the quantum numbers of angular momentum. 
Two dimensional topological gravity has similar nature as the Chern-Simons 
gravity where links may carry conformal spin quantum 
numbers\cite{Verlinde-Verlinde}. 
In four dimensional lattice gravity formulations there are similar 
evidences\cite{Ooguri}\cite{FDgravity}\cite{Smolin}.

Gauge theory formulation of the lattice gravity and matter will be formulated 
by the generalized gauge theory. 
In other words the generalized gauge theory formulated by differential forms 
may play a crucial role in formulating a lattice gravity 
since the diffeomorphism invariance is automatic in the formulation of 
differential forms. 
Furthermore differential forms and simplexes have one to one correspondence 
and thus the fields on the forms can naturally be put on the corresponding 
simplexes. 

Quaternion algebra may play a crucial role in the formulation of the lattice 
gravity formulated by the generalized gauge theory. 
In other words even forms, odd forms, fermions and bosons are the fundamental 
algebraic ingredients which are nicely classified by the quaternion algebra. 
Furthermore the graded Lie algebra will be another fundamental algebra 
which governs the gauge structure of the generalized gauge theory, 
in particular, the super symmetry. 
Thus the semi-direct product of the graded Lie algebra and the quaternion 
algebra is the fundamental structure of the generalized gauge theory and 
may play an important role in formulating lattice gravity as a gauge theory. 

Fermionic matter will be generated from the ghosts of quantization by the 
twist. 
The twisting procedure is essentially related with the 
Dirac K$\ddot{\hbox{a}}$hler fermion formulation as we have shown in the 
quantization of topological Yang-Mills action. 
The fermionic ghosts appear as fermionic differential 
forms and can be put on the 
simplicial lattice similar as the bosonic fields. 
We need to understand this twisting mechanism from lattice point of view. 

As we have seen in the quantization of the generalized Chern-Simons actions 
we needed to introduce infinite number of ghosts due to the infinite 
reducibility of the system\cite{KSTU}. 
It is interesting to consider possible physical aspects of the introduction 
of an infinite number of the ghost fields. 
An immediate consequence is a democracy of ghosts and classical fields, 
{\it i.e.}, the classical fields are simply the zero ghost number sector 
among infinitely many ghost fields and thus the classical gauge fields 
and ghost fields have no essential difference in the minimal action.

Furthermore fermionic and bosonic gauge fields are treated in an 
equal footing and the series of infinite ghosts originated from the 
classical fermionic and bosonic fields are complementary.
In other words if we only introduce bosonic classical fields in the 
starting action we need to introduce fermionic 
fields with odd integer ghost number 
and bosonic fields with even integer ghost number 
as we have shown in section 5. 
If we introduce the classical fermionic gauge fields, 
the odd and even nature 
should be reversed for the ghost numbers when introducing 
the corresponding ghost fields to the fermionic gauge 
fields.
It seems to mean that even the fermionic and bosonic fields have no essential 
difference in the generalized Chern-Simons theory. 
In other words fermionic fields, bosonic fields, classical fields and 
ghost fields 
are mutually inter-related via the quantization procedure.

Super symmetry is the natural consequence of topological symmetry with 
partial gauge fixing of instanton conditions.
Thus the introduction of super symmetry is realized via quantization 
procedure. 
In the generalized gauge theory graded Lie algebra plays an important role 
leading to the super symmetry. 

The lessons from the lattice Chern-Simons gravity of the section 4 
tell us that the drei bein and spin connection can be simply put 
on the original and dual lattice, respectively. 
This means that the fields of differential forms from the 
generalized gauge theory will 
be put on both of the original and dual lattice.

The standard model including Weinberg-Salam model will be realized by 
the generalized gauge theory via non-commutative geometry mechanism. 
In particular the quaternion algebra plays an essential role here 
again to formulate the gauge theory of all the differential forms. 
It should be noted that Connes's non-commutative geometry formulation 
of Weinberg-Salam model is a particular example of our generalized gauge 
theory formulation. 
Connes has speculated that the discrete two points which he introduced 
in the non-commutative geometry formulation can be interpreted as the 
discrete distance of the space time structure which is physically 
related with the Higgs vacuum expectation value\cite{Connes}.

In our generalized gauge theory the quaternion algebra is playing 
a fundamental role and should have some physical or geometrical 
interpretation. 
I would like to give conjectures on the possible interpretation of 
the quaternion and the fermion-boson correspondence following to 
the suggestive examples of analyses given in this summary.

\begin{center}
{\bf Conjectures}
\end{center}

The generalized gauge theory will be formulated on the simplicial lattice
manifold and may play a fundamental role in formulating a unified model 
on the lattice. 
Quaternion algebra is responsible for differentiating even simplexes, 
odd simplexes corresponding to even forms and odd forms, 
and original simplicial lattice and the dual simplicial lattice reflecting 
boson and fermion nature.
In particular discrete two points in terms of non-commutative geometry 
would correspond to the original lattice and dual lattice.

Matter fermions are generated from ghosts via twisting mechanism which is 
nothing but the Dirac K$\ddot{\hbox{a}}$hler fermion formulation. 
Bosons and fermions are dual to each other in the lattice simlex sense.
In other words bosons are sitting on the original lattice while fermions 
are sitting on the dual lattice and {\it vice versa}. 
Chiral fermion problem will be understood via 
Dirac K$\ddot{\hbox{a}}$hler fermion formulation.
This duality relation would be analogous to the Ising model where the 
disorder variable defined on the dual lattice is related to the 
fermion field as shown in the subsection 2.1. 
Strong coupling and weak coupling duality of Krammers and Wannier type 
is working here and the fermion will be a soliton on the dual 
lattice. 
The origin of the Seiberg and Witten formulation\cite{Seiberg-Witten} will 
be related to the duality of original and dual lattice. 

Due to the complete duality of the original lattice and the dual lattice 
where fermions and bosons are located on the mutually dual lattices 
the super symmetry is the natural consequence for the whole system.
The super symmetry transformation and the gauge transformation 
are the change of the fields from one simplex to the other on the 
original lattice and the dual lattice. 
This is analogous to the lattice gauge theory where even sites and odd sites 
exchange corresponds to the chiral transformation. 

Instanton will play an essential role in this bosonization mechanism 
and also in the breaking of the higher super symmetry or the topological 
symmetry.   

The phenomenological results like the group structure of the standard model 
and the number of 
the generations mentioned in the introduction will be understood from 
this point of view.

\begin{center}
{\bf Questions}
\end{center}
There are still many fundamental questions. 
In this summary we have mainly discussed topological field theories 
which do not include dynamical degrees of freedom. 
Then the question arises: \\
\lq\lq How do the dynamical degrees of freedom appear starting from the 
theory with higher symmetry such as $N=2$ super symmetry and topological 
symmetry ?'' \\
We can guess that the classical solution like instanton may 
play an important role to resolve the higher symmetry into lower symmetry 
but the mechanism is not yet clear. 
The topological symmetry or $N=2$ super symmetry may play a similar role 
as the kinetic term in the lagrangian formulation of field theory. 
The field theories keeping higher symmetry such as topological symmetry 
or higher super symmetry may not yet have genuine 
non-perturbative interactions which break the symmetry dynamically.

There are other questions concerning to the number of dimensions of 
our space-time and Minkowskian nature. 
String approach may be able to understand this question\cite{Kawai1}.
I believe that the string approach and our approach is not 
incompatible but rather complementary.

\vskip 1cm

\noindent{\Large{\bf Acknowledgements}} \\
I would like to thank my collaborators J.Ambj\o rn, K.N.Anagnostopoulos, 
T.Ichihara, L.Jensen, H. Kawai, V.A. Kazakov, T. Mogami, H.B.Nielsen, E.Ozawa, 
Y. Saeki, N.Sato, J.Smit, K.Suehiro, T.Tsukioka, H.Umetsu, Y. Watabiki and K.Yotsuji for fruitful discussions and collaborations .

 
\end{document}